\documentclass[preprint,authoryear,12pt]{elsarticle}

\usepackage{graphicx}

\usepackage{amssymb}

\usepackage{multirow}
\usepackage{rotating}

\newcommand{\kms}{km~s$^{-1}$\,}

\newcommand {\sm}{\rm\,M$_\odot$}
\newcommand {\sL}{\rm\,L$_\odot$}
\newcommand {\mgi} {Mg\,{\small I}\,}
\newcommand {\disp} {$\sigma_{\rm l.o.s}$\,}
\newcommand {\dispr} {$\sigma_{\rm l.o.s}(R)$\,}

\newcommand{\amevscl}{AmE12}
\newcommand{\bred}{B12}
\newcommand{\bredfour}{BH13}
\newcommand{\giusscl}{B08}
\newcommand{\jardfnx}{JG12}
\newcommand{\jardeldraco}{J13}
\newcommand{\longdraco}{LM10}
\newcommand{\kleynaone}{Kl01}
\newcommand{\kleynatwo}{Kl02}
\newcommand{\lokasfnx}{L01}
\newcommand{\lokasdraco}{L05}
\newcommand{\lokasinter}{L09}
\newcommand{\kochleotwo}{Ko07a}
\newcommand{\kochleoone}{Ko07b}
\newcommand{\mateoleoone}{M08}
\newcommand{\walkerseven}{Wa07}
\newcommand{\walkeruni}{Wa09}
\newcommand{\battsxt}{B11}
\newcommand{\hayashi}{HC12}
\newcommand{\gilmore}{G07}

\journal{New Astronomy Reviews}

\begin{document}

\begin{frontmatter}



\title{Internal kinematics and dynamical models of dwarf spheroidal galaxies 
around the Milky Way}


\author[1]{Giuseppina Battaglia\corref{cor1}}
\ead{gbattaglia@oabo.inaf.it}
\cortext[cor1]{Corresponding author}
\address[1]{INAF - Astronomical Observatory of Bologna, via Ranzani 1, 40127 Bologna, I}

\author[2]{Amina Helmi}
\author[2]{Maarten Breddels}

\address[2]{Kapteyn Astronomical Institute, P.O. Box 800, 9700 AV Groningen, NL}

\begin{abstract}
We review our current understanding of the internal dynamical
properties of the dwarf spheroidal galaxies surrounding the Milky Way. These are the most
dark matter dominated galaxies, and as such may be considered ideal
laboratories to test the current concordance cosmological model, and
in particular provide constraints on the nature of the dominant form of dark matter. 
We discuss the latest observations of the kinematics of stars in
these systems, and how these may be used to derive their mass
distribution. We tour through the various dynamical techniques
used, with emphasis on the complementarity and limitations, and discuss what the results imply 
also in the context of cosmological models. Finally we provide an
outlook on exciting developments in this field. 

\end{abstract}

\begin{keyword}

Galaxies: kinematics and dynamics \sep Galaxies: dwarf \sep dark matter \sep 
Local Group \sep Galaxies: stellar content.


\end{keyword}

\end{frontmatter}
\let\clearpage\relax

\section{Introduction}
\label{sec:intro}

With absolute magnitudes ranging from $M_V \sim -9$ to $\sim -13.5$
and central surface brightness between $\mu_{0,V} \sim 22.5 -27$ mag
arcsec$^{-2}$, the ``dwarf spheroidals''  (dSphs) are the faintest and
lowest surface brightness galaxies known to date, beaten only by the
relatively recently discovered ultra-faint dwarf galaxies (UFDs).

Although of dull appearance, dSphs reveal an unexpectedly complex
stellar populations mix \citep[for a recent review see][]{tht09} that
makes them very useful laboratories for understanding star formation
and chemical enrichment processes at the faint end of the galaxy
luminosity function. In terms of their internal dynamics, they might
well be key in constraining the nature of dark matter. Even though the very 
first measurement of the line-of-sight velocity dispersion of a dSph 
was based on just 3 carbon stars in Draco \citep[][]{aaronson1983},
it already hinted at a dynamical 
mass-to-light ratio about one order of magnitude larger than for globular clusters. 
Subsequent works have confirmed this result using larger samples that included red giant stars 
\citep[e.g.][see also Sec. \ref{sec:obs} and \ref{sec:dyn-models}]{armandroff1986, aaronson1987, hargreaves1994sext, hargreaves1994umi}. If in
dynamical equilibrium, dSphs have the
highest mass-to-light ratios known to date\footnote{While at face value UFDs exhibit even
  larger dynamical mass-to-light ratios, these values are subject to numerous important caveats, as discussed later in this review.}, with $M/L \sim 100$s \sm/\sL . 

In the remainder of the Introduction we describe the latest
observational surveys of the kinematics of dSphs, place these
systems in a cosmological context and briefly discuss why most of these systems may be considered to be in dynamical equilibrium. 
In Sec.~\ref{sec:obs} we describe 
the determination of the kinematic properties of dSphs from spectroscopic 
samples of individual stars. 
In Sec.~\ref{sec:dyn-models} we review the methods used
to model the internal dynamics of spheroidal systems
and discuss their application to dSphs around the Milky Way.  We discuss 
possible future developments in Sec.~\ref{sec:disc}, and
in Sec.~\ref{sec:concl} we briefly summarize the current status of the field.

\subsection{Surveys of dwarf spheroidals around the Milky Way} 

Determining the mass content of a system requires 
observations of the kinematics of suitable tracers. Since  
dSphs are devoid of a neutral interstellar 
medium, the only tracers available are stars. Because of their distance, to-date all
measurements of their internal kinematics are based on line-of-sight velocities. 
The stars accessible for spectroscopic observations with current 
facilities are resolved for systems within the Local Group, since there is
no crowding because of 
the low surface brightness of these galaxies.  In this review, we concentrate on the dwarf galaxies that are satellites of the Milky Way \citep[MW, hereafter. We refer the reader to][for a nice and comprehensive 
historical excursus on the growing kinematic samples for MW dSphs]{Walker2012}. 

The first attempt to go beyond the determination of a global l.o.s. velocity dispersion of a dSph 
was made by \citet{mateo1991fnx} using a 2.5m telescope. These authors 
measured the kinematics of $\sim 30$ stars in the Fornax dSph, in the center and 
in a field located at about two core radii. This first l.o.s. 
velocity dispersion ``profile'' turned out to be approximately flat, and this led the authors to suggest
that it could be due to a dark halo spatially more extended than the visible matter. 
These results opened a whole line of investigation to measure l.o.s. velocity dispersion profiles of dSphs around the Milky Way, and to use these to determine their dark matter (hereafter, DM) distribution,  orbital structure and dynamical state. 

The samples of l.o.s. velocities collected in the 90s contained a 
few dozens of individual members per dSph 
\citep[e.g.][]{mateo1991fnx, hargreaves1994sext, hargreaves1994umi, olszewski1996, queloz1995, mateo1998leo1}. An increase in sample size became possible
with  multi-object spectrographs such as the KPNO/4m Hydra multi-fiber positioner \citep[100 members in Draco and Ursa~Minor,][]{armandroff1995}, and the AF2/Wide Field Fibre Optical Spectrograph on the WHT \citep[150 members in Draco,][]{Kleyna2001}.

In the second half of the 2000s several large spectroscopic surveys of the classical MW dSphs were carried out. In broad terms we can distinguish them in 3 main ``streams'':
\begin{enumerate}
\item Surveys that obtained l.o.s.\ velocities for 
typically $\sim$100-150 members per dSph, with a large success ratio of dSph  
members/target stars thanks to an optimized target selection using Washington photometry 
\citep[M, T2, and DDO51 filters, e.g.][]{majewski2005, munoz2005, munoz2006, westfall2006, sohn2007}. These have made use of  Keck/HIRES, Magellan/MIKE, CTIO/Hydra and Keck/DEIMOS. 
\item Surveys to obtain several 100s of stars per dSph to determine 
both the internal kinematics and the metallicity distribution from Ca~II triplet lines using intermediate resolution spectroscopy. This includes the Dwarf Abundances and Radial velocities Team (DART, PI: Tolstoy) 
\citep[$\sim$570, 800, 170 members for the Sculptor, Fornax, Sextans dSphs, respectively, at 
R$\sim$6500 over the wavelength range 8200\AA\, - 9400\AA\,,][]{
tolstoy2004, tolstoy2006, battaglia2006, helmi2006, battaglia2008cat, starkenburg2010, battaglia2011}; and program 171.B-0520 (PI: Gilmore) ``Towards the Temperature of Cold Dark Matter'' 
\citep[$\sim$500, 170 members for the Carina and Leo~II dSphs with the same set-up as for the DART data-set,][]{koch2006, koch2007leo2}. These have taken advantage of 
the VLT's large collecting area coupled to the wide-field, multi-object capability and stability of the FLAMES-GIRAFFE spectrograph \citep{pasquini2002} and, also of  Keck/DEIMOS and GeminiN/GMOS 
\citep{koch2007leo1}. 
\item Surveys to obtain several 100s to 1000s of l.o.s.\ velocities and 
spectral indices (providing estimates of the {\it relative} 
metallicity of red giants) on a restricted wavelength range (5140 \AA\,-5180\AA\,) at resolution R$\sim$20000 
\citep[PI: M.Mateo, e.g. $\sim$800, 2500, 1400, 400 members for Carina, Fornax, Sculptor and Sextans, 
respectively][]{walker2007data, walker2009}. These have been mainly carried out with the 
Michigan/MIKE Fiber System (MMFS) at the Magellan/Clay (6.5m) telescope 
and with MMT/Hectochelle \citep[see][for Leo~I]{mateo2008}. With a comparable field-of-view to FLAMES 
(20 arcmin), MMFS has the advantage of almost double the number of fibres (equally shared between the 
blue and red channel of the MIKE spectrograph).  
\end{enumerate}

Therefore, to-date the combined data-sets for the best studied dSphs have impressive sizes 
($\sim$2900 and 1700 probable members for Fornax and Sculptor, respectively), permitting 
studies of their internal properties to a level of detail that was unthinkable a little more
than a decade ago. 

The low luminosities of UFDs imply that very few RGB stars (the most luminous targets available for galaxies with old stellar populations) are available for spectroscopy. The size and spatial coverage of existing kinematic samples resemble those in the early days of the``classical'' MW dSphs, even when targeting fainter stars (on the horizontal branch, and/or close to the main-sequence turn-off).  Given our interest in 
exploiting the full l.o.s.\ velocity distribution, in what follows we concentrate 
on the ``classical'' dSphs and discuss only briefly results on the internal kinematics of UFDs.

\subsection{dSphs in a cosmological context} 

In our current understanding of the Universe, a mere 5\% 
of the total mass/energy density budget consists of baryons, atoms essentially, with the remaining 95\% comprising about 24\% non-baryonic ``dark matter'' and 71\% ``dark energy'' 
\citep[see][for the 9-years WMAP results]{hinshaw2012}. This has become known as the 
$\Lambda$ cold-dark matter ($\Lambda$CDM) model. As the evocative naming suggests, we are ignorant of the nature of the great majority of constituents of 
the Universe. 

There are several DM candidates such as weakly interacting massive 
particles, axions, sterile neutrinos, light gravitinos etc., whose existence 
is also motivated to solve problems in the Standard Model of particle physics \citep[for a review see][]{feng2010}. 
Some of these behave as cold and some as warm dark matter, where e.g. ``cold'' is defined as being non-relativistic at the time of 
structure formation. A wealth of experiments and strategies for direct and indirect detections of DM particles 
are underway \citep[e.g. for reviews see][]{bertone2005, hooper2008, feng2010}, 
but at present the evidence for the existence of DM (based on the validity of Newton's law of gravity on all gravitational acceleration regimes) is provided by astrophysical observations on a variety of scales, from the smallest galaxies such as the dSphs up to the largest structures in the Universe\footnote{Alternative theories of gravity, or modifications of Newton's law have also been presented in the literature. We decided not to discuss these here because their application to model
the dynamics of dSphs has been very limited.}. 

Potentially, astrophysical observations can provide important constraints on the dominant form of DM, as the characteristics of the DM particle are expected to influence the growth of structures, the substructure content and internal properties of DM halos. Rather than reviewing the extensive literature on the topic, we proceed to discuss results that are most directly related to this review, highlighting the crucial role of dwarf galaxies. 

Cosmological pure DM N-body simulations, carried out in the $\Lambda$CDM framework,
show that the halos formed follow very specific functional forms, such as 
the Navarro, Frenk \& White profile \citep[NFW,][]{Navarro1996,Navarro1997}
\begin{equation}
\label{eq:nfw}
\rho(r) = \frac{\rho_0}{r/r_s (1 + r/r_s)^2}
\end{equation}
where $\rho_0$ and $r_s$ are a characteristic density and radius.
More recently the Einasto form has been found to provide better fits 
\citep[e.g.][]{Springel2008, navarro2010}
\begin{equation}
\label{eq:einasto}
\rho(r) = \rho_{-2} \exp \left\{ -\frac{2}{\alpha_E} \left[ \left(\frac{r}{r_{-2}}\right)^{\alpha_E} -1 \right] \right\},
\end{equation}
where $\rho_{-2}$ and $r_{-2}$ are the density and radius where the logarithmic slope $d \log \rho/d \log r = - \gamma_{DM} = -2$, and $\alpha_E$ is a shape parameter\footnote{For $\alpha_E \sim 0.2$ the resulting profile resembles an NFW.}. These density profiles are rather steep near the centre, with the NFW being cuspy with $\gamma_{DM} = 1$, while the Einasto profile has $\gamma_{DM} = 0$ at the centre.  

Although not necessarily theoretically motivated, other density profiles are also often employed in the literature. Typically they have the form 
\begin{equation}
\rho(r) = \frac{\rho_0}{(r/r_s)^\gamma (1 + (r/r_s)^\kappa)^{(\alpha - \gamma)/\kappa}},
\label{eq:rho_general}
\end{equation}
where $\alpha, \gamma, \kappa \ge 0$. Note that $\gamma$ and $\alpha$ correspond to the inner and outer slopes respectively.  The sharpness of the transition between these two regimes is thus given by $\kappa$. A cuspy profile has $\gamma > 0$, while for a 
cored one $\gamma = 0$ and
$\kappa > 1$. This is because in the cored case, the profile must have a flat shape
at the centre, i.e. $d \rho/dr = 0$. A profile that has $\gamma = 0$ and 
$\kappa \le 1$ has at the centre $d \log \rho/d\log r = 0$ and a finite density, but in
this case $d \rho/dr$
is non-zero, and hence this profile should not be confused with a core.

In the $\Lambda$CDM high-resolution cosmological N-body simulations described above the 
sub-halo mass function of MW-sized main halos is $dN/dM \propto M^{-\alpha}$, with  $\alpha =1.9$ down to the simulations' resolution limit \citep{Springel2008}, which is smaller than the mass estimates for the faintest dSphs (see Sec.~\ref{sec:dyn-models}). These simulations predict that MW halos contain 20\% of the mass in subhalos, which results in a very large number of (mostly extremely low mass) satellites.

A comparison between the results of these pure DM N-body simulations with 
observations on galactic scales is not straightforward. Part of the issue
lies in making the link between a luminous satellite to what should be its corresponding sub-halo in a DM simulation \citep[e.g. of what mass? how dense?, see][]{Strigari2010}. This is particularly difficult because such simulations do not include baryons. This has motivated
numerous theoretical efforts to provide a realistic treatment of baryonic effects using semi-analytical models and hydrodynamical simulations of dwarf galaxies
\citep[e.g.][]{revaz2009, li2010, Font2011, sawala2012, starkenburg2013}. 
Observationally, it is clearly important to obtain reliable estimates of the 
mass content and its distribution in dwarf galaxies. 

For example, there is a debate about the inner shape of the density profiles of the DM halos hosting galaxies. For dSphs, this issue is still very open (see Sec.~\ref{sec:dyn-models}). On the other hand, for isolated late-type dwarfs and low surface brightness 
galaxies, the rotation curves seem to favor cored rather than cusped DM distributions \cite[e.g.][and references therein]{deblok2010}. It has been suggested that feedback from supernovae explosions for these more massive systems could transform a cuspy halo into a cored one  \citep[e.g.][]{navarro1996cores, read2005, Governato2010, pontzen2012, teyssier2013}. Note that in the case of an UFD, a single SN event releases an amount of energy comparable to the binding energy of the whole system. On the other hand, it is still to be assessed whether this mechanism is important or relevant on the scales of the MW dSphs, also given
their low star formation rates \citep[see][]{penarrubia2012}. 

The ``missing satellites'' problem refers to the large mismatch 
between the observed number of dwarf galaxies satellites of the MW and M31 
and the predicted number of DM subhalos
 \citep{klypin1999, moore1999}. The discovery of dozens low-luminosity dwarf galaxies in the Local Group, mainly by SDSS around the MW \citep[e.g.][to mention a few]{willman2005, zucker2006, belokurov2006, belokurov2007}  
and the PandAS survey for M31 \citep[e.g.][]{mcconnachie2009, martin2009}, has mitigated somewhat the ``missing satellite'' problem, after taking into 
account the surveys' coverage and selection function 
\citep{koposov2009}. The most appealing solution to reconcile predictions and observations 
is to suppress star formation, or gas accretion, in low-mass halos because of the joint effects of feedback and of a photo-ionizing background due to re-ionization \citep[e.g.][]{bullock2000, benson2002, somerville2002}.  

Another interesting issue was the recently reported ``too big too fail problem'' pointed out by \citet{Boylan2011}, who used the Aquarius suite of DM simulations to argue that there exists a population of subhalos that are too massive and too dense 
to be consistent with the internal kinematics of the MW dSphs, and yet they do not 
have an observed stellar counterpart. However, as argued by \cite{Wang2012} and \cite{Vera2012}, the number of massive satellites is a stochastic quantity
that also depends on the mass of the host. For example, if the mass of the MW is around 8$\times 10^{11}$\sm, i.e. the least massive MW-like halos of the Aquarius suite
\citep[which reproduces well the observed MW satellite luminosity function, see][]{koposov2008,starkenburg2013}, the mismatch disappears. Furthermore, \citet{Vera2012} show that M31, if assumed to be more massive than the Milky Way, does not miss such a population. 

A plausible alternative to CDM is warm dark matter (WDM). The warm component has the effect of reducing the small-scale power in the primordial fluctuations spectrum, yielding fewer subhalos and of lower central densities \citep{colin2000, colin2008,lovell2012}. Specifically, in the numerical simulations of \citet{maccio2012,maccio2013err}, which explore a range of masses for the WDM particles, cored density profiles arise naturally. However, either the core sizes 
are too small to be consistent with those suggested in some studies of the internal kinematics of MW dSphs (see Sec.~\ref{sec:dyn-models}) or if large enough, they would be due to particles whose masses are inconsistent with the limits 
imposed by  observations of the Lyman-$\alpha$ forest 
\citep[e.g.][]{viel2005, seljak2006, viel2008}. Note however that e.g.\ \citet{Busha2007} find in their WDM simulations that the halos are well described by an NFW form (i.e. cuspy) while \citet{Wang2009} find this even holds for halos in hot dark matter simulations. Given that the state-of-the-art 
of WDM simulations is not as extensive and developed as for CDM, we await future 
developments. 
 
From the above it is clear that there are numerous reasons to try and pin down the DM content and its distribution in the dSphs. Given that the overall evolution of small systems like dwarf galaxies will most likely be 
sensitive to their relatively small potential well \citep[e.g.][]{revaz2012, sawala2012}, obtaining such measurements 
will also allow us to make sense of the variety of star formation and chemical enrichment histories of these galaxies, 
in particular in conjunction with the information on the dSphs orbital history that the Gaia satellite mission \citep{Prusti2011} will provide. 

\subsection{Are the stellar components of dwarf spheroidals 
affected by tides or are they in dynamical equilibrium?}
\label{sec:tides}

An assumption in dynamical modeling of dSphs is that these objects are 
in dynamical equilibrium, while if they were significantly affected 
by tidal interactions with the MW this would need to be taken into account.

The possibility that dSphs are fully tidally disrupted dark-matter
free galaxies has been excluded on the basis of their observed
internal kinematic and structural properties \citep[see for
example][]{klessen2003, munoz2008}, the large distances of some of
these galaxies (up to 250~kpc from the MW) and a well-established
luminosity-metallicity relation. It would also be difficult to explain
the dSphs' extended SFHs and broad metallicity distributions \citep[see
e.g.][]{tolstoy2004, battaglia2006, koch2006, starkenburg2010,
  battaglia2011} if the potential well would be due solely to the
dSph stars (amounting to typically 10$^5$-10$^6$ M$_{\odot}$,
e.g. McConnachie 2012).

Partly because of the lack of knowledge of the orbits of dSphs around
the MW, the importance of tides on the stellar components of dSphs is
largely unknown. This also depends on the degree of embedding of this
component in its dark matter halo, as well as on the average density
of the system. \citet{mayer2001} propose that dSph galaxies are what
results when a disky dwarf is tidally stirred by the MW. For this
process to be effective, the stellar component of the dSph today has
to be tidally limited, in which case tidal tails are expected. 
However, \citet{penarrubia2008tides} find that the stars are
very resilient to tides in their simulations where the stellar
component follows a King-profile and is deeply embedded in an NFW
halo. In any case, there is general consensus that the central
velocity dispersion (or the dispersion at the half-light radius)
continues being a good indicator of the present maximum circular
velocity and bound mass, as long as the objects retain a bound core
\citep[e.g.][]{munoz2008, penarrubia2008tides, klimentowski2009,
  kazantzidis2011}.

Besides the obvious case of Sagittarius, the only classical dSph presenting
unambiguous signs of tidal disturbance such as tails and isophote twists is Carina \citep{battaglia2012car}.
This object has been a candidate for tidal disturbance for a long time, with convincing arguments
given by the presence of spectroscopically confirmed RGB stars, probable members, out to very large distances
from its center
(4.5 times the central King limiting radius), observed together with a break in the surface brightness profile,
a velocity shear with turn-around, and
a rising line-of-sight velocity dispersion profile \citep[e.g.][]{munoz2006}.
Among the classical dSphs, other candidates for tidal disruption are Leo~I \citep[e.g.][]{sohn2007, mateo2008}
and Ursa Minor
\citep[e.g.][]{martinez-delgado2001, palma2003, munoz2005},
although the observational evidence is not as strong as for Carina. Note that even for Carina, the
N-body simulations by \citet{munoz2008} show that large amounts of DM ($M/L \sim 40$~\sm/\sL) within the remaining
bound core are still needed to explain its characteristics.

N-body simulations of tidally perturbed dSphs agree in predicting rising l.o.s. velocity dispersion profiles
in the majority of cases, while only Carina and perhaps Draco \citep{Walker2012} are observed to show
such feature. Together with the fact that most classical dSphs show no tidal streams, this 
may be taken as indicative 
that the outer parts of the stellar components of dSphs have
not been significantly affected by tides. All these arguments provide some justification for the assumptions made in this review, namely that we may consider the dSphs to be in dynamical equilibrium.

However, it would be well-worth the effort to carry out observational campaigns
designed to maximize the chances of detecting the smoking-gun signature of tidal disruption, i.e. tidal tails.
Detection of these low-surface brightness features needs
deep and spatially extended photometric data-sets.
Instruments like CTIO/DECam and the forthcoming Subaru/HyperCam, but also the proper motion information from the Gaia mission, are excellent matches to this type of problem.

\section{Observed kinematics} 
\label{sec:obs}

The heliocentric distances to MW dSphs, ranging from 75 to 250~kpc, have made it unfeasible 
to obtain accurate proper motions of individual stars in these galaxies with current facilities. This 
 implies that we only have access to one component of their velocity vector, namely that
along line-of-sight (l.o.s.) $v_{\rm l.o.s}$. Therefore, all current studies of the 
internal kinematics of dwarf galaxies are based on their line-of-sight velocity distributions (LOSVD) and 
their moments.

In this Section we start by describing how to derive a reliable LOSVD from the
measurements of the velocities of individual stars. In Sec.~\ref{sec:losvd_err} we consider the effects of velocity
errors, the presence of binary stars and the contamination introduced by objects that do not belong to the dSph, such as Milky Way stars (hereafter ``contaminants'' or ``interlopers''). 
In Sec.~\ref{sec:losvd_char} we discuss how the main characteristics of an LOSVD 
are related to the internal dynamics of the system. In Sec.~\ref{sec:losvd_obs} we describe
how to measure the moments of the LOSVD, and present the latest results for the
MW dSphs.

\subsection{Derivation of a reliable LOSVD}
\label{sec:losvd_err}

\subsubsection{Velocity errors} 
\label{sec:errors}

In general l.o.s. velocities accurate to a few \kms are essential to kinematic studies of dSphs, given their low internal velocity dispersion and small velocity gradients (see Sec.~\ref{sec:losvd_obs}). 

\begin{figure}
\vspace*{7cm}
\centering
\includegraphics[scale=0.75]{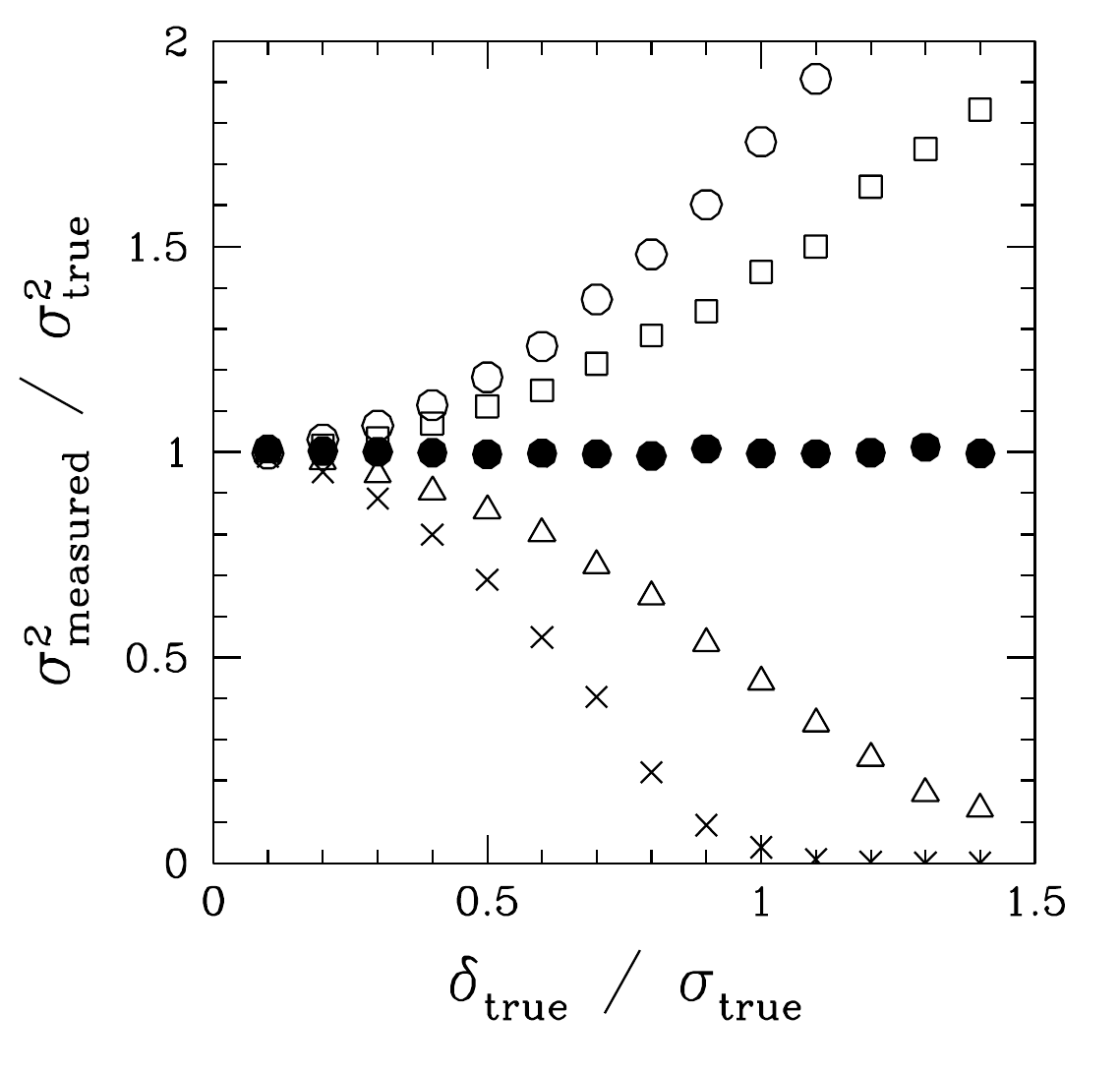}
\caption{This figure \citep[from][reproduced by permission of the American Astronomical Society]{koposov2011} illustrates the accuracy of velocity dispersion estimates, 
expressed as the ratio of the measured variance over the true one, as a function 
of the ratio of velocity error $\delta_{\rm true}$ to the true velocity dispersion $\sigma_{\rm true}$. The filled circles 
show the case of perfectly known velocity errors; open squares and circles when the errors used in the analysis are underestimated by a factor of 0.5 and 0.75; 
the triangles and the crosses for an overestimation of 1.25 and 1.5, respectively.  
} \label{fig:koposov2011}
\end{figure}

\citet{koposov2011} have shown that it is crucial to obtain a reliable estimation of the velocity errors to provide a realistic value of the dSph \disp, and hence of its dynamical mass (see Sec.~\ref{sec:dyn-models}). Fig.~\ref{fig:koposov2011} shows that
if the velocity errors are perfectly known, the intrinsic (true) \disp  
can be accurately recovered even if the measurement errors are larger than the intrinsic \disp. 
On the other hand, overestimated (pessimistic) errors will yield a smaller estimate for the dispersion, while the opposite happens for underestimated errors. This effect 
becomes particularly strong when the velocity error is $\gtrsim 0.5$ the true \disp.  


In the case of the classical dSphs, with \disp$\sim 6-11$ \kms,  this is not an issue.
With current fiber-feb spectrographs, it is possible to obtain very accurate $v_{\rm l.o.s}$ for
their red giant branch stars. For example, with 1h observing time at VLT/FLAMES+GIRAFFE at intermediate resolution (R$\sim$6500) one can achieve velocity errors to better than 
$\sim$5 \kms on RGB stars with V$=$19.5 in the region of the nIR Ca~II triplet lines \citep[e.g.][]{battaglia2008cat}. This 
corresponds to $\sim$0.5mag below the tip of the RGB of the most distant MW dSph, Leo~I. 
The RGB of MW classical dSphs is well-populated, making it unnecessary to observe 
fainter stars to gather a large sample of 
targets. The situation is different for the UFDs, both because of the need to target fainter stars but also because of their intrinsically smaller \disp.

An example of the importance of well-determined velocity errors is given by \citet{koposov2011}, who developed a sophisticated data-reduction procedure for VLT/FLAMES+GIRAFFE data. In their application to the Bo\"{o}tes~I UFD, 
and after the removal of radial velocity variable stars,  this allowed \citet{koposov2011} to ``resolve'' the LOSVD into  
two Gaussians with very different velocity dispersions. As highlighted by 
the authors, this result shows that  
current determination of masses for UFDs need to be taken with care 
\citep{martin2007, Wolf2010, koposov2011}.

\subsubsection{Binaries}  
\label{sec:binaries}

In this review, we are mostly concerned  
with the effect that binaries may have on the LOSVD and its moments, 
for example by inflating the measured velocity dispersion of dSphs. This concern is as old as the first measurements of 
a dynamical mass-to-light ratio of these objects \citep{olszewski1996}. \citet{hargreaves1996} concluded that the highest velocity dispersion that could stem from the orbital motions of binaries alone is 3 \kms if the binary population is similar to that of the MW.  The authors state ``To produce larger dispersions, more binary orbits with a mixture of lower periods, 
higher mass secondaries or primaries with radii smaller than $10 R_\odot$ are required''. 
More recently, \citet{minor2010} showed that, if the measured velocity dispersion of a dSph ranges between 4 and 10 \kms, 
the inflation from binaries should not be more than 20\% in samples of RGB stars with 
absolute magnitude M$_V \lesssim$1 and older than 1 Gyr (again assuming similar binary populations as in the MW), 
and at most 30\% when including fainter stars. The authors also 
devise a method to correct the measured velocity dispersion for the contribution of binaries that is 
expected to yield intrinsic velocity dispersions accurate to a few percent. They do this 
by measuring the fraction of stars that exhibit a change in velocity exceeding a certain threshold 
over a time between measurements.

Therefore, the general conclusion is that the effect of 
binaries  for the \disp measured for dSphs is minor. 
Unidentified binaries can instead have a strong impact on the \disp measured for UFDs. 
\citet{mcConnachie2010} explore whether it is plausible to explain the velocity dispersions of UFDs with $\mathrm{M}_V \gtrsim -7$ 
as the result of inflation by unidentified binaries from an intrinsic dispersion of  $\sim 0.1-0.3$ \kms (i.e. they test whether UFDs could be devoid of DM).  The authors consider only binaries with periods smaller than 10-100~yr \citep[for comparison, the mean period of binaries in the solar neighborhood is 180~yr,][]{Duquennoy1991}. \citet{mcConnachie2010} find that binaries cannot account for observed dispersions much in excess of $\sim$4.5 \kms, but contamination by binaries might still be a concern for some UFDs. 

In their analysis of the Segue~1 UFD, \citet{martinez2011} introduce a comprehensive Bayesian method 
to analyze multi-epoch data, including foreground contamination from the MW and the presence of binaries. 
Their analysis shows that, if in dynamical equilibrium, then Segue~1 has an intrinsic velocity dispersion of 
3.7$_{-1.1}^{+1.4}$ \kms at 1$\sigma$. They estimate that without a binary correction, the most likely 
dispersion would only have been 12\% larger. However, the authors point out the importance of multi-epoch data 
to disfavor the possibility that most of Segue~1 observed dispersion is due to binary stars. 

The determination of the binary fraction and distribution of orbital parameters in dSphs is also of intrinsic interest to
establish dependences on different galactic environments \citep[for a nice introduction see][]{minor2013}, but no
such dedicated campaigns have been carried out to-date. The recent work of \citet{minor2013} constitutes
the best effort in determining the binary fraction in MW dSphs by exploiting the information on 
repeated observations from the large spectroscopic survey of 
\citet{walker2009}. The binary fraction in 3 dSphs is found to be consistent 
with the one of the MW (around 0.5), while it appears to be much lower for Carina (Car). Unfortunately, 
the binary fraction and period distribution parameters  are strongly degenerate:  
a small binary fraction with short mean period produces very similar observed 
velocity variations (within the measurement errors) to a large binary fraction with long mean periods.

To constrain the period distribution independently from the binary fraction, the measurement errors 
on the individual velocities need to be $<$0.1 \kms \citep[achievable with high resolution spectrographs, see Eq.~(19) of][]{minor2013}, unless the measurements are taken over time-scales much longer than 5 years. This is therefore a difficult task, and larger amplitude velocity variations would only be measurable if the mean period in dSphs is considerably 
shorter than for MW field stars. 

\subsubsection{Weeding out Galactic contaminants}  
\label{sec:contaminants}

Contamination by MW stars on the LOSVD needs to be modeled carefully. An example is 
the case of Willman~1, whose nature as a galaxy or stellar cluster 
(as inferred by the presence/absence of an [Fe/H] spread among its stars) is 
still debated due to 
the difficulty of gathering interloper-free samples of members \citep[][]{martin2007, siegel2008, willman2011}. 

Methods for removal of contaminants have different levels of sophistication and 
effectiveness, and make use of different sets of observables, such as the magnitude and colors of the objects, 
their velocities and their spectral features. We describe these in detail in what follows, 
and leave for Sec.~\ref{sec:secondm} how to model the LOSVD including a foreground component. 

\begin{figure}
\centering
\includegraphics[scale=0.5]{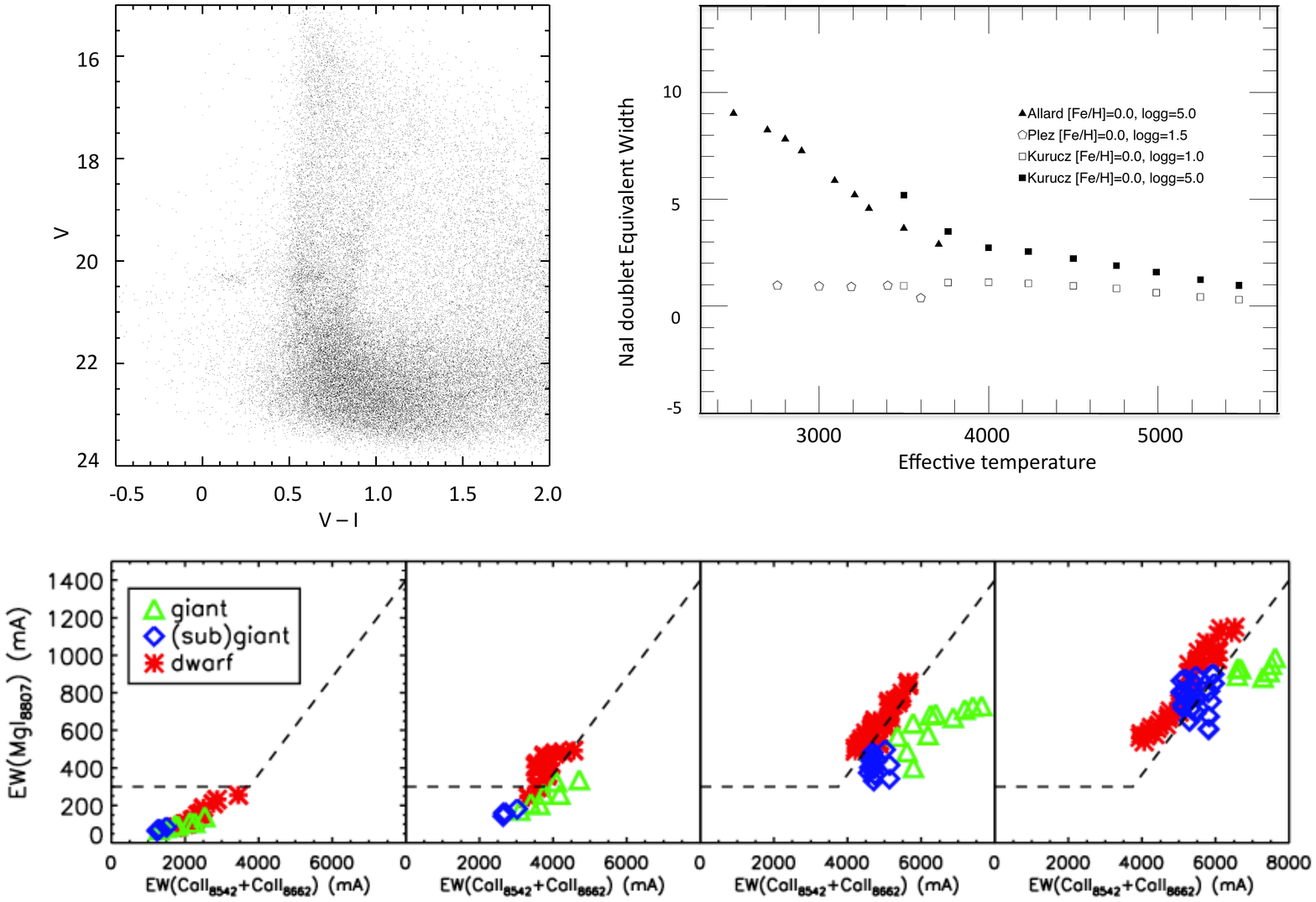}
\caption{Top left: 
Color-magnitude diagram of stellar objects along the l.o.s. to the Sextans dSph from INT/WFC and ESO/WFI data 
\citep[see][]{battaglia2011} showing how heavily contaminated by MW foreground stars is red giant branch of this dSph. 
Top right: 
Equivalent width of the Na~I feature, for giant stars (open symbols) and dwarf stars (filled symbols) from \citet[][reproduced by permission of the American Astronomical Society]{schiavon1997}. Bottom: 
Summed EW of the two strongest CaT line versus the \mgi 8806.8 line. 
Giants above the horizontal branch are shown as green triangles, (sub)giants below the horizontal branch as blue diamonds and
dwarfs as red asterisks \citep[for definition see][]{bs12}.  
Overplotted in each panel is the line separating dwarf and giant stars as given by Eq.~(\ref{eq:line}). 
Figure from Battaglia \& Starkenburg 2012, A\&A, 539, 123, reproduced with permission \copyright ESO.} 
\label{fig:contamination}
\end{figure}

The photometric data provide the first aid in minimizing the presence of interlopers: resolved 
galaxies can be weeded out using information on shape parameters 
(unresolved galaxies usually are not an issue at these magnitudes); while a fraction of MW stars can be 
excluded by restricting target selection to a range of magnitude and colors consistent 
with the locus of specific stellar evolutionary phases 
at the distance of the dwarf galaxy (e.g. the RGB, main sequence turn-off, horizontal branch etc). 
Even then, the large solid angles subtended by MW dwarf galaxies can 
result in severe contamination from MW stars, depending on the Galactic coordinates of the object. 
This is the case for example, of Sextans (Sext), as shown in Figure~\ref{fig:contamination}(top-left panel), 
whose RGB is barely visible, being buried by MW stars. Fortunately, 
spectroscopic observations come to our aid.


\paragraph{ Information from the stars kinematics} 
Assuming that the velocities of the stars belonging to the dwarf follow a Gaussian distribution, 
the simplest approach to remove MW contaminants is to adopt a hard cut in 
$v_{\rm l.o.s.}$ by iteratively clipping the sample at a 
given number of \disp from the systemic velocity $v_{\rm sys}$ \citep[where \disp is an estimate of the width of the LOSVD, 
such as the standard deviation or robust bi-weight estimator, e.g.\ ][]{mateo1991, walker2006}. 
Although in general the outer parts need to be treated more carefully, this method may suffice if there is not much overlap between the LOSVD of the MW and of the dSph, otherwise more sophisticated methods are to be preferred. 
Such methods simultaneously model the expected LOSVD 
of MW and dSph stars, and take into account that the probability that a star is a member of the dSph decreases as a function of projected radius $R$. 
This follows from the fact that the projected number density of dSph stars decreases with $R$, while 
the projected number density of MW stars is roughly constant over the area subtended by the dwarf (see Sec.~\ref{sec:secondm}). 

\paragraph{Diagnostics sensitive to gravity from spectroscopy} 
Since the great majority of spectroscopic studies on MW dSphs target RGB stars, and most
contaminants are dwarf stars in the Galactic disk(s), diagnostics sensitive to surface gravity
can strengthen membership determination.

A popular indicator is 
the equivalent width (EW) of the Na~I doublet lines at $\lambda=$8183, 8195\AA\,, which \citet{schiavon1997} showed 
is a useful dwarf/giant stars discriminator as dwarf stars exhibit larger Na~I EWs than giant stars. 
As shown in Fig.~\ref{fig:contamination} (top-right panel),  the Na~I EW 
of dwarf stars declines with increasing temperature: at $T_{eff} < $ 4000 K there is a very marked difference in Na~I EW between 
dwarf and giant stars, making the discriminator very powerful, while at $T_{eff} > $ 4000 K the Na~I EW 
of dwarf stars almost approaches the values of giants (as shown in the top-right panel of 
Fig.~\ref{fig:contamination}). Although some studies have used hard-cuts on the Na~I EW to separate dwarfs from giants 
\citep[e.g. for  UFDs,][]{martin2007, simon2007}, other works have considered the location of the  
stars in the Na~I EW vs de-reddened (V-I)$_0$ color plane, to take into account 
the aforementioned dependence of the Na~I EW of dwarf stars on $T_{eff}$
\citep[e.g.][for applications to the M31 system using training sets]{gilbert2006, guhathakurta2006, geha2010}. 

\citet{walker2007data} introduced the use of the \mgi index 
to assign membership of stars to their samples of MW dSphs. This is similar in spirit to the reduced CaT EW 
\citep[e.g.][]{martinez2011} or equivalently the [Fe/H] values derived from CaT calibration 
\citep[e.g.][]{helmi2006}. There is some degree of overlap between the distribution of Mg~I index or CaT [Fe/H] 
of MW and dSph stars \citep[see e.g. Figs.~1 and 2 of][and Fig.~1 of Helmi et al. 2006]{walker2009algo}. 
This also happens for the EW of the nIR \mgi  line at $\lambda \sim $8807 \AA\,, as shown by \citet{bs12} (see their Fig.~7).  
\citet{bs12} have suggested that the position of a star in the \mgi EW versus CaT EW plane, rather than just its \mgi EW, 
is a much more efficient dwarfs/giants discriminator over the range $-2 \le$ [Fe/H] $\le$0 
(see bottom panel in Fig.~\ref{fig:contamination}). The relation to discriminate giants and dwarfs by \citet{bs12} is:
\begin{equation} \label{eq:line} 
{\rm EW}_{\rm Mg} (m\AA\,) = \left\{ \begin{array}{ll}
 300 &\mbox{ if $\lambda \le$ 3750 m\AA\,} \\
 0.26 \times \Sigma {\rm W}_{\rm CaT} -670.6  &\mbox{ if $ \lambda >$ 3750 m\AA\,}
       \end{array} \right. 
\end{equation}
where $\Sigma {\rm W}_{\rm CaT}$ is the summed EW of the two strongest CaT lines. 

Other diagnostics such as the EWs of the K7665 and K7699 absorption features and 
the strengths of the TiO bands at 7100, 7600, and 8500\AA\,, are effective at dereddened (V-I)$_0 > 2.5$ 
\citep{gilbert2006}.

\paragraph{\it Diagnostics sensitive to gravity from photometry} 
Photometric indices sensitive to gravity (and luminosity, effective temperature, etc.) can be derived from combinations of particular filters. 
For example, the intermediate-band DDO~51 filter measures the strength of the Mg~b/MgH feature near 5170 \AA\,, which is strong in
late-type dwarfs and weak in giants \citep[e.g.][]{morrison2001}, and is in general used in combination to 
the Washington $M$ and $T_2$ filters \citep[e.g.][]{majewski2000, palma2003}. However, verification of membership with velocity information is crucial if the area considered is large enough to contain a significant number of 
metal-poor halo dwarfs  -- which cannot be isolated with this method either at these faint magnitudes -- 
and/or if low photometric precision moves stars from the dwarfs to the giants locus  in the M$-$T$_2$ vs M$-$51 plane 
\citep[e.g.][]{morrison2001}. This method allowed 
\citet{munoz2005} to find members of the Car dSph out to very large distances from its center. 
Str\"{o}mgren photometry (filters $u$, $v$, $b$ and $y$, where the first two and partially $b$ are covered by the broader 
Washington C- filter) has been successfully used to determine membership in the Draco dSph 
\citep{faria2007}, and to weed out non-members in the sparsely populated Hercules, leading to a decrease of its estimated dynamical mass \citep{aden2009a}. 
\\

Choosing what membership method to adopt probably implies a trade-off between available facilities, 
observing times involved, permitted tolerance to the presence of contaminants, and expected 
stellar population mix both in the target galaxy and 
foreground population along the specific line-of-sight. It is important to keep in mind that, even if the 
inclusion of the kinematic information often is crucial in distinguishing contaminants from member stars, the methods 
that use such information need to make assumptions on the intrinsic LOSVD of the dwarf, which is instead 
what we want to determine. It may be worth to assess the effectiveness 
of cleaning up samples only on the basis of diagnostics independent of kinematics or when adopting less strict 
kinematic criteria; for example, \citet{bs12} show that the \mgi line method is still very effective when relaxing the 
kinematic criterion from the more standard 3$\sigma$ clipping to a larger 4$\sigma$, which would include the high velocity tails of the dSph LOSVD.

The situation will improve dramatically for MW dSphs when parallax and proper motion information becomes available from the Gaia mission. Most foreground dwarf stars will be identified thanks to their large parallaxes and proper motions, 
while even metal-poor MW halo giants (nowadays the most difficult contaminants because they have similar spectral features as RGB stars in the dwarf galaxy), could be identified through their different tangential velocities, and possibly 
also their parallaxes. In either case, it will be important to develop models that include all populations of contaminants,
as suggested by \citet{walker2009algo}.

\subsection{Characterization of the line-of-sight velocity distribution} 
\label{sec:losvd_char}

The shape of the LOSVD depends both on the distribution function (DF) of the stars as well as on the gravitational potential in which these orbit. Here we focus specifically on  {\it spherical systems}  
and present some general results on what can be learned from the characteristics of the LOSVD. 
There is extensive literature on the topic from the dynamical modeling of elliptical galaxies, 
but we choose here to briefly review a few results that may be particularly useful for dSphs. It is important to stress that there are two important differences between elliptical galaxies and dSphs. Firstly, baryons are not negligible in ellipticals. Secondly, in contrast to the distant ellipticals, in principle, the 
individual kinematic measurements for the stars in a dSph could be used to characterize the LOSVD, however
most works use its moments, as we shall see in Sec.~\ref{sec:dyn-models}.

\cite{Dejonghe1987} has studied the LOSVDs and its moments for  
self-consistent Plummer anisotropic models. He has shown that in the
central regions the \disp  has a stronger contribution of radial
orbits even if the system is tangentially anisotropic. This is known as the ``complementarity'' property,
and is simply the result of projection effects. Moreover, a more peaked LOSVD results for a tangentially biased system than for a 
radial one when looking in the central regions; however, a 
tangentially anisotropic ellipsoid can be recognized in the outer parts, 
as a flat-topped LOSVD. 

\begin{figure}
\includegraphics[scale=0.6]{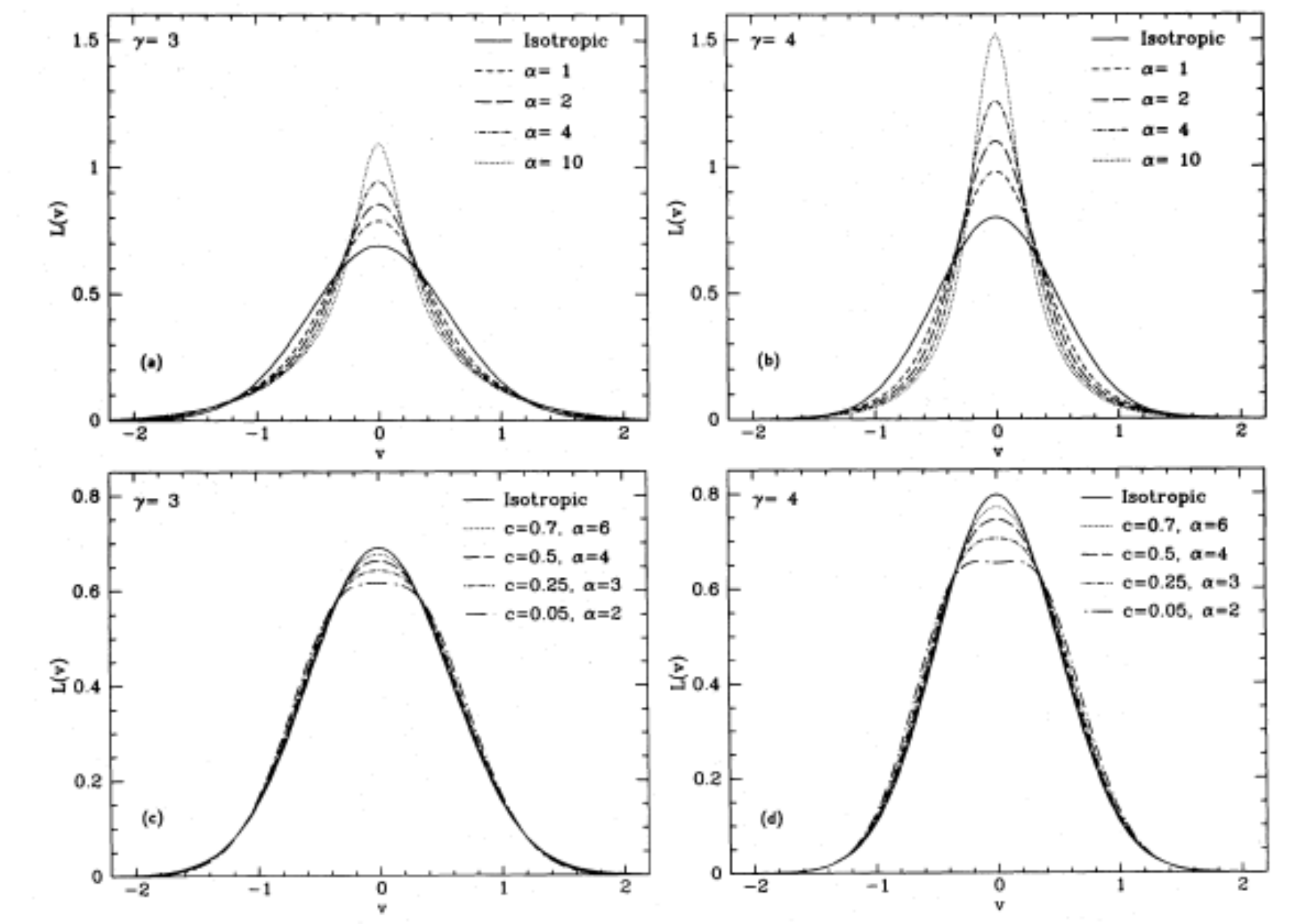}
\caption{L.o.s. velocity profiles for quasi-separable distribution functions in the potential of a 
singular isothermal sphere from Gerhard 1993, MNRAS, 265, 213, reproduced by permission of Oxford University Press. Radially anisotropic and tangentially anisotropic models are shown in the top and bottom panels, respectively. 
The stellar density is $\rho \propto r^{-3}$ (left) and $\rho \propto r^{-4}$ (right). See \citet{Gerhard1993} for more details.} \label{fig:gerhard}
\end{figure}

\cite{Gerhard1993} has explored the shape of the LOSVD in the case of 
different gravitational potentials (Kepler and the isothermal
sphere), for power-law light distributions for the tracers, and for a
family of DFs that are quasi-separable in energy and total angular momentum.  Note that this study is particularly
relevant for dSphs since it also deals with non self-consistent systems.  
\cite{Gerhard1991,Gerhard1993} show that 
the velocity profiles (VPs)\footnote{Here we use the terms velocity profiles VPs and LOSVD, interchangeably.} 
for self-consistent radially anisotropic models are strongly non-Gaussian 
(the isotropic case is Gaussian), with a two component 
structure: a narrow inner core and strong extended outer wings. As the stellar density profile steepens,  
the VP becomes more dominated by the inner core and the relative amplitude of the 
extended wings with respect to the core decreases, as shown in Fig.~\ref{fig:gerhard}. When the system is embedded in an isothermal potential, tangentially anisotropic models produce nearly Gaussian VPs with flattened tops. These  
become more noticeable for steeply falling stellar density profiles. 
In the Keplerian potential instead, the VP in the isotropic case is not Gaussian, and it becomes 
more Gaussian-like for radially anisotropic systems. Nonetheless also here, 
the VPs in the tangential case are flat-topped. {\it Overall, strongly tangentially anisotropic DFs in the outer parts lead to flat-topped VPs that are easily 
recognizable, and this effect does not depend strongly on the potential, but it becomes more enhanced for 
steeper stellar density profiles}, as shown in Fig.~\ref{fig:gerhard}. This result is also supported by the
work of \citet{wilkinson2002} who have introduced a family of constant anisotropy models where the stars follow a Plummer sphere, 
while for the DM halo they discuss the mass-follows-light case, an isothermal sphere and 
an extended halo with a harmonic core.  \citet{wilkinson2002} reach a similar conclusion, namely that it is 
the VP in the outer parts that holds most information on the velocity anisotropy. 


The results above indicate that shape of the LOSVD, mainly how peaked or flat-topped it is especially at
large radii, could be directly be used to infer the orbital structure of a dSph in this regime. This shape information 
is encoded in its 4th moment,  be it the kurtosis or the Gauss-Hermite $h_4$ moment, see Sec.~\ref{sec:higherm}.
In the outskirts of a dSph, its stellar density profile may be assumed to fall off as a power-law, hence in this
region the conclusions drawn by \cite{Gerhard1993} would be most
relevant.  In that case, the LOSVDs should be more flat-topped for a tangentially anisotropic ellipsoid. 
Furthermore, this result also ought to be valid for systems with multiple
stellar components, as this outer region is generally dominated by a single (the most metal-poor) component (see Sect.~\ref{sec:multiple}).

\subsection{Results}
\label{sec:losvd_obs}

In this section, we discuss step by the step how to measure the first four moments of the LOSVD
for dSphs, as well as present some first results.

\subsubsection{Surface brightness profiles}
Even though not a moment of the LOSVD, we briefly describe the dSphs surface brightness profile (as derived from stellar number counts) as 
this is an ingredient for the dynamical modeling. 

The results from the literature are rather heterogeneous as the surface stellar number count profiles are 
derived from samples of different photometric depths/completeness, spatial extent, and available regions for accurate determination of the 
(foreground) density. Studies in the literature have mostly
focused on the overall fit at all radii and in the performance of various functional forms in the outer parts as a possible way of 
detecting signs of tidal stripping from the MW. 

Typically, empirical King models \citep{king1962} fit better the MW dSphs surface stellar number counts profile with respect to
 exponential profiles \citep{ih95}. However, 
it is now well-established that King models are not necessarily good representations of the outer parts, 
as they typically underestimate the observed densities 
\citep[e.g.][]{ih95, martinez-delgado2001, wilkinson2004, coleman2005scl, battaglia2006, battaglia2008}. Note that, if the stellar 
component of MW dSphs is embedded in massive and extended DM halos, such an excess does not need to be interpreted as evidence for 
tidal disruption. 
 
Although not tested for all MW dSphs, in some cases better representations at all radii are given by 
Sersic profiles \citep{sersic1968} that decline faster than exponentials 
\citep[e.g.][]{odenkirchen2001, battaglia2006} or by Plummer profiles \citep[e.g.][]{battaglia2008}. Fits to the light distribution in the very central regions have not received much attention thus far. The variety of profiles used allow
for the presence of cores as well as small (logarithmic) cusps in the intrinsic 3D luminosity density profile. 

\subsubsection{The first moment: velocity gradients}  
\label{sec:firstm}

A variation of the mean (or median) l.o.s. velocity across a dSph (hereafter ``velocity gradients'') 
potentially allows us to gain information on the physical mechanisms shaping the dSphs,  such as 
intrinsic rotation or tidal disruption. However, velocity gradients also result from mere geometrical effects due to the perspective orbital motion of the dwarf galaxy around the Milky Way\footnote{The mean velocity of a star in an object with a large angular extent such as MW dSphs 
is expected to vary as a function of position of the star because of geometrical projection effects due the proper motion of the 
object (e.g. Feast, Thackeray \& Wesselink 1961). Such ``spurious'' velocity gradients mimic solid body rotation.} 
and we refer to these as ``spurious'' gradients. 
 
A variety of methods have been used in the past to measure the variation of the LOSVD first moment 
across the object:  
$i$) considering how the average/median velocity (in the heliocentric or Galactocentric Standard of Rest [GSR] frame) varies as a function of distance along some direction \citep[e.g.][]{munoz2006}; $ii$) as before, but considering stars falling within ``slits'' of varying position angle \citep[PA, e.g.][]{battaglia2008}; $iii$) calculating the mean velocity differences, $\Delta v$,  on either side of a bisector line passing through the galaxy center 
and having a range of position angles, $\theta$ \citep[see][e.g.]{walker2006, mateo2008}; $iv$) exploring the variation of the mean velocity in angular sectors \citep{amorisco2012lp}. 

Until a few years ago, no statistically significant velocity gradients had been 
detected from spectroscopic samples limited to the central regions of dSphs, except for a 3$\sigma$ detection along the 
minor axis of Ursa Minor but from a sample of only 35 stars (Hargreaves et al. 1994).  
In the past decade this has changed considerably thanks to the large increase in the size and spatial coverage 
of kinematic samples of MW dSphs brought by wide-area multi-object spectrographs. 
At present, statistically significant velocity gradients have been reported for Sculptor (Scl), Car, Fornax (Fnx), 
Sext and Leo~I \citep{munoz2006, battaglia2008, mateo2008, walker2008, battaglia2011}. 
The weak velocity gradient along the projected major axis of Leo~II is significant only at the 
0.17$\sigma$ level \citep{koch2007leo2}, and 
Sext could certainly benefit from a larger spatial coverage. 
The velocity gradients measured are of only a few km s$^{-1}$ deg$^{-1}$.

\begin{figure}
\centering
\includegraphics[scale=0.5]{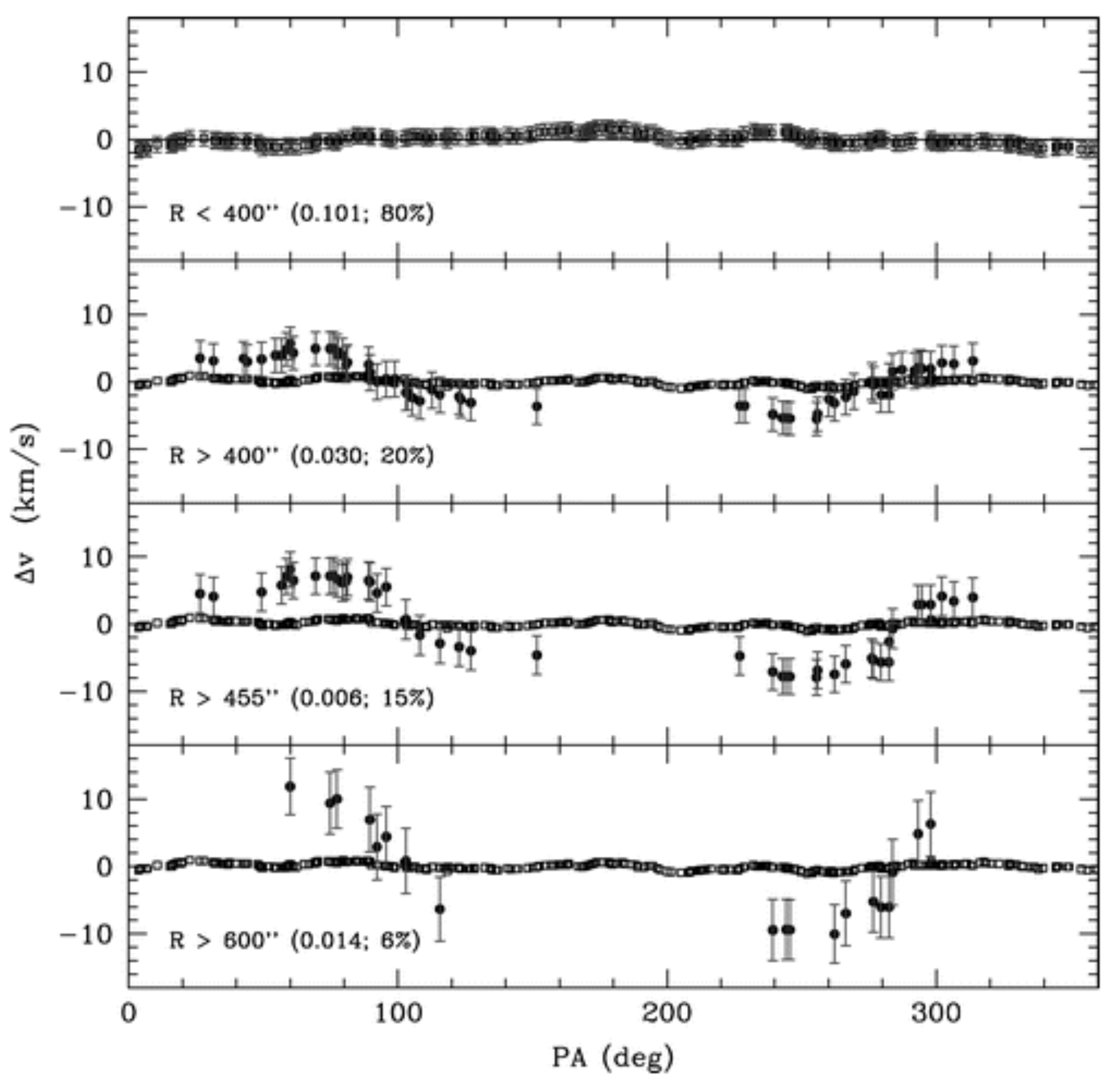}
\caption{Difference of mean velocities of Leo~I members on both 
sides of a bisector oriented along a given position angle (P.A.) from \citet[][reproduced by permission of the American Astronomical Society]{mateo2008}. The labels indicate the radial range of the data-set plotted in each panel 
(points with error-bars), the probability of exceeding  $\delta v_{\rm max}$ and the fraction of Leo~I members used to produce each plot. 
In the panels with $R>$400'', the symbols without error bars are the data-set from the full sample (N=328).
} \label{fig:leoI}
\end{figure}

The importance of exploring the outer regions of dSphs to detect such gradients has become increasingly evident. 
For some dSphs the mean velocity is found to slowly rise\footnote{From current samples it appears difficult to pin down the exact shape of the velocity gradient, whether it can be approximated by a straight-line or 
other functional forms.} with projected radius $R$ \citep[see Fig.~1 of][for the case of Scl]{battaglia2008}. 
It is then in the outer regions that the mean and systemic velocities will likely differ the most, 
while at the center the system is basically entirely dominated by velocity dispersion. 
The case of Leo~I, where the presence of a gradient was investigated via the bisector method, 
illustrates this clearly as shown in Fig.~\ref{fig:leoI}: no signs of a velocity gradient are visible 
at $R < 400"$, while the detection becomes statistically significant when 
considering stars at projected radii $>600"$. On the other hand, application of the bisector method to the whole sample, i.e. 
not including the projected distance information, yields no compelling evidence for a significant velocity gradient, 
indicating that, at least in this case, such method washes out the larger signal present in the outer parts. 
Another, although indirect, example is given by Car, 
for which \citet{munoz2006} detect a clear velocity gradient with a velocity amplitude of $\sim$5 \kms over a scale of 1.2~deg, while a much lower value is 
found by \citet{walker2008} for a sample restricted to a smaller spatial region. Homogeneous spatial coverage 
is also essential to recover the direction of the strongest gradient. 

The origin of the detected velocity gradients could in principle impact the mass modeling. If the detected 
velocity gradients are entirely due to perspective effects (``spurious''), 
the only action required would be to place the 
velocities in a system at rest with the dwarf galaxy. 
However, for ratios of maximum apparent versus intrinsic velocity dispersion $v/\sigma \sim 0.5-0.6$, 
as observed in dSphs, \citet{amorisco2012lp} show that it may not even be necessary to subtract such velocity gradients as the impact is merely an increase of 10\% in the observed dispersion compared to the intrinsic one, while the change in the higher moments is less than what can be detected with sample size as large as 800 stars.

\begin{figure}
\includegraphics[scale=0.35]{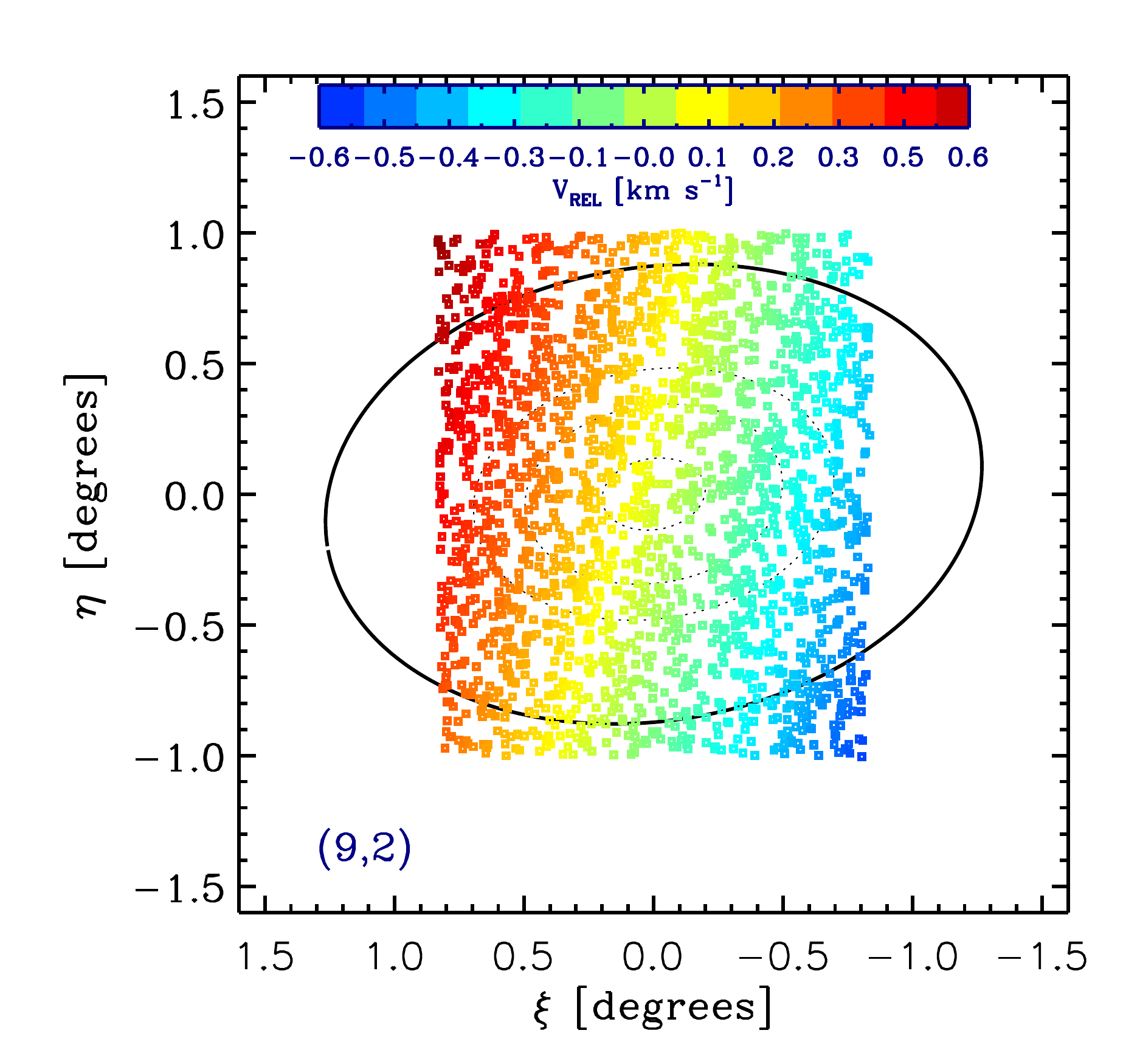}
\includegraphics[scale=0.35]{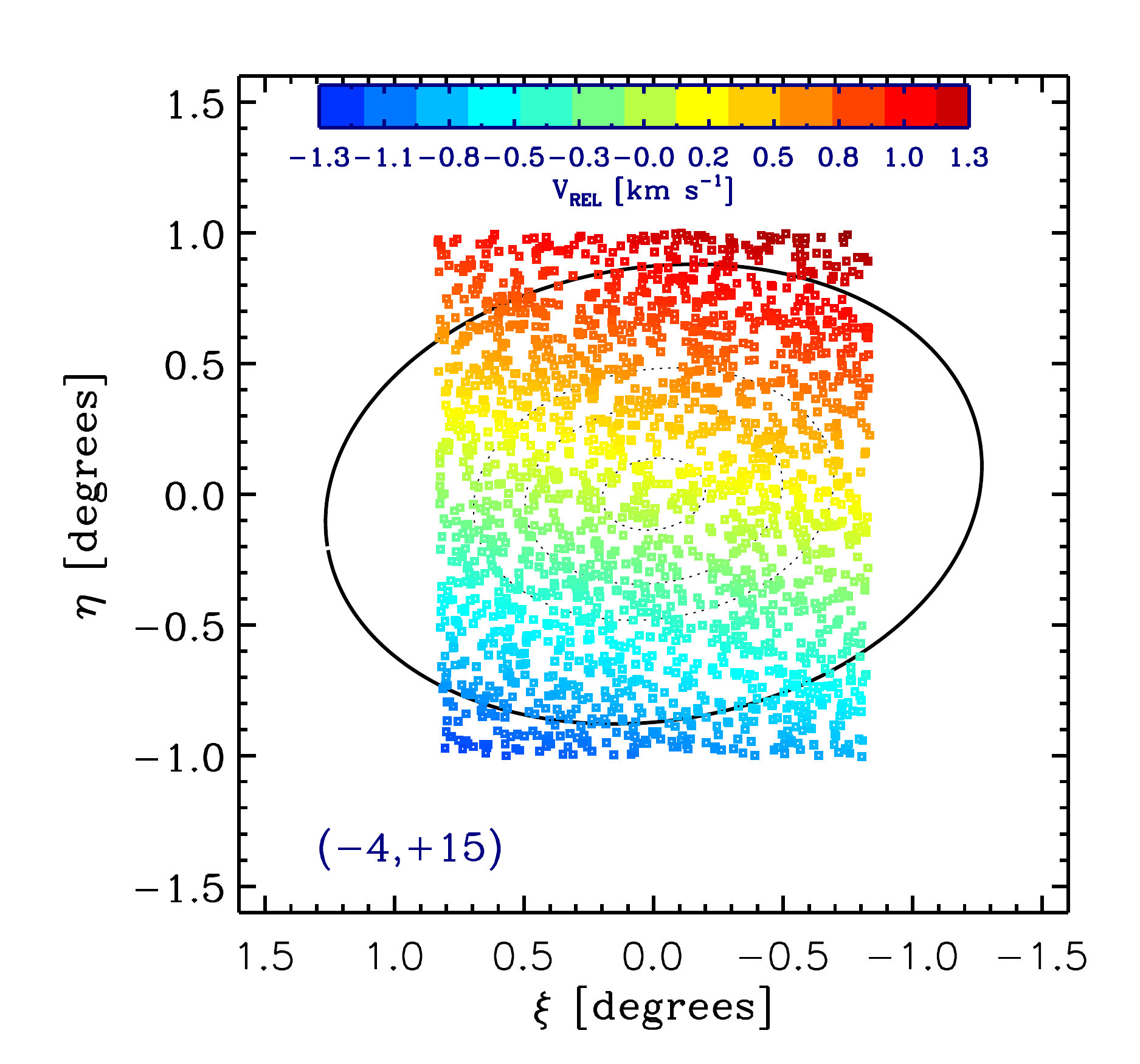}
\includegraphics[scale=0.35]{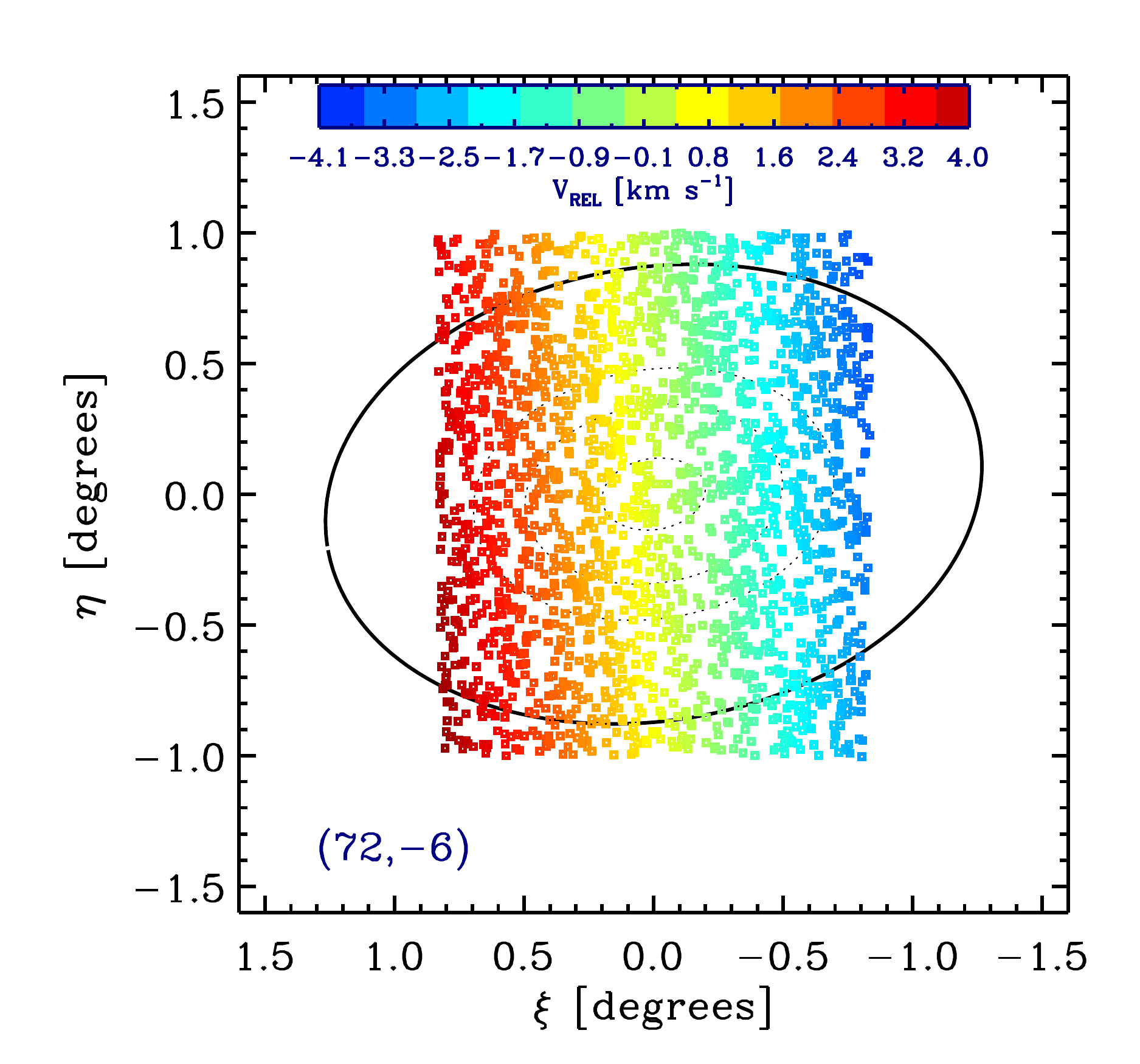}
\includegraphics[scale=0.35]{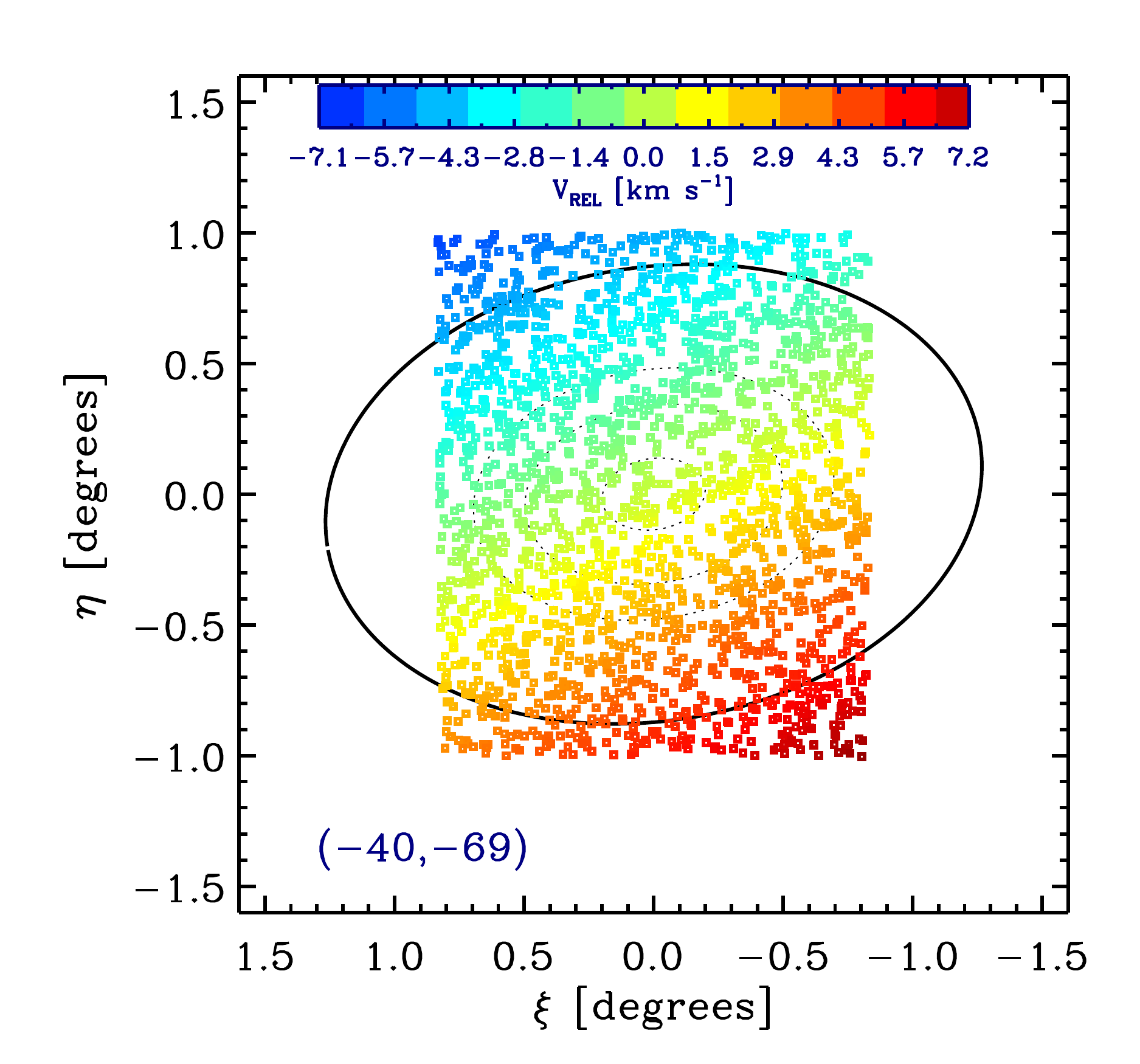}
\caption{Velocity field of the projection of the relative motion between the Sun and the Sculptor 
dSph along the line of sight defined by equatorial coordinates ($\alpha$,$\delta$). 
The $v_{rm rel} (\alpha,\delta)$ has been derived applying the formulae in the Appendix of \citet{walker2008} and 
subtracting Sculptor systemic heliocentric velocity. 
The coordinates of the mock data-set (squares) 
were randomly extracted from a uniform distribution in $\alpha$ and $\delta$ of $\pm$2deg around 
Sculptor's center. The four panels assume different values for the Sculptor proper motion 
($\mu_{\alpha}$, $\mu_{\delta}$) [mas/century] {\it in the heliocentric reference frame}, as indicated by the labels: top left:  
from \citet{piatek2006}; 
top right: value from \citet{piatek2006}, but taking ($\mu_{\alpha} - 1 \sigma$, $\mu_{\delta} + 1 \sigma$); 
bottom left: from \citet{schweitzer1995}; bottom right: from \citet{walker2008}.  
} \label{fig:fov_vel}
\end{figure}

Although desirable, the correction for the 3D motion of the dwarf galaxy around the MW is made uncertain by the 
large errors on the current measurements of their proper motion  
(in reality, for some dSphs entirely lack astrometric proper motion measurements).  This is illustrated in Fig.~\ref{fig:fov_vel}, where we see 
how the amplitude and direction of the ``spurious'' gradients may vary according to the systemic proper motion measurements existing 
in the literature, taking the Scl dSph as an example. Here we have randomly generated a uniform distribution of stars in right ascension and declination, with no intrinsic velocity assigned. As shown in the top left panel of this figure, when adopting \citet{piatek2006} proper motion the amplitude 
(at the last measured point) is very small, less than 1 \kms and the gradient points at PA $\sim 75$~deg. However it is enough to consider $\mu_{\alpha} - 1\sigma$ and $\mu_{\delta} + 1\sigma$, as shown in the top right panel, to make the gradient direction change considerably, with the vector now pointing at P.A.$\sim\, - $15deg (the amplitude is still $\sim$1 \kms). On the other hand,  
the amplitude obtained when adopting the proper motion from \citet{schweitzer1995} is larger, up to 4 \kms (bottom left). We note that the ``spurious'' solid-body rotation results in velocity fields with parallel iso-velocity contours, perpendicular to the proper motion direction; however, the intrinsic velocity dispersion of dSphs makes it challenging to detect such a signature with current data-sets.

It is also possible to assume that dSphs do not have intrinsic velocity gradients, and 
derive a systemic proper motion from the measured stellar redshifts \citep{walker2008}. 
For Scl, there is disagreement with the direct proper motion measurements by \citet{schweitzer1995,piatek2006} and those derived by \citet{walker2008}, as evidenced in the comparison of the four panels of Fig.~\ref{fig:fov_vel}. 
This tells us that this dSph has an intrinsic residual velocity gradient, be it due to rotation 
(Battaglia et al. 2008) or tidal streaming, or that both proper motion measurements are incorrect.  On the other hand, for Fnx and Car there is good agreement between the proper motions from stellar redshifts and those 
directly measured \citep[see][]{walker2008}, so that the gradients detected {\it over the region sampled by the data} are most likely due just to perspective effects. We look forward to the accurate proper motions that the Gaia satellite will deliver, because they will determine the amplitude and directions of ``spurious'' velocity gradients, and herewith also help unravel intrinsic gradients due to rotation or tidal effects.

Even if the origin of the velocity gradients is intrinsic rotation, 
the relatively low $v/\sigma$ values suggest that the internal kinematics of dSphs is dominated 
by random motions; we note that, in the hypothesis of internal rotation, this is the ``projected'' value, 
which should be corrected for the -unknown- inclination of the angular momentum vector. 

\subsubsection{The second moment: l.o.s. velocity dispersion profiles} \label{sec:secondm}

Much care is invested in deriving robustly the l.o.s. velocity dispersion profiles
\dispr\, of dSphs, not only because of the 
sensitivity of mass estimates to the value of the dispersion, 
but also because of the importance of the velocity dispersion profile {\it shape} in determining the 
DM mass distribution (see also Sec.~\ref{sec:dyn-models}) and providing 
possible hints on the dynamical status of the dSph. 

There are two main schools of thought for deriving $\sigma_{\rm l.o.s.}(R)$, which assume that: 
$i$)  the dSph is in {\it dynamical equilibrium}; $ii$) the kinematic samples contain {\it unbound 
dSph stars} lost as a consequence of tides induced by the MW. These need to be removed as 
they are no longer tracers of the dSph galaxy potential. We will 
discuss the methodologies used in the literature for deriving $\sigma_{\rm l.o.s.}(R)$ within either of the 
assumptions. 

\paragraph{Dynamical equilibrium} 

The way the l.o.s. velocity dispersion profile is derived is very much related to the 
way of dealing with the foreground contamination.  \citet{hargreaves1994sext} introduced a simple 
maximum likelihood procedure to evaluate simultaneously the systemic velocity $\overline{v}_{\rm dSph}$ and 
l.o.s. velocity dispersion $\sigma_{\rm dSph}$, 
assuming that the measurement errors and the velocities are Gaussianly distributed:
\begin{equation} 
L(v_1, v_2, \dots, v_N) = \prod_{i} P(v_i) = \prod_{i} \frac{1}{\sqrt{2 \pi (\sigma_i^2 + \sigma_{\rm dSph}^2)}} 
\mathrm{exp} \left[ \frac{ -(v_i - \overline{v}_{\rm dSph})^2}{2 (\sigma_i^2 + \sigma_{\rm dSph}^2)} \right], 
\end{equation} 
where $L(v_1, v_2, \dots, v_N) $ is the likelihood of observing a set of $N$ velocities $v_1, v_2, \dots, v_N$, $i$ goes from 1 to $N$, 
and $\sigma_i$ are the errors on the individual velocity measurements. The error ellipses on $\overline{v}_{\rm dSph}$ and $\sigma_{\rm dSph}^2$ 
are then given by steps down the peak of ln($L$), i.e. ln($L$)-1/2 corresponds to the 68\% confidence limit. They also provide the analytic formulas to use in an iterative scheme: 
\begin{equation}  \label{eq:barv_har}
\hat{\overline{v}}_{\rm dSph} = \frac{\sum_{i} w_i v_i}{\sum_{i} w_i},
\end{equation}
and 
\begin{equation} \label{eq:sigma_har}
\hat{\sigma}_{\rm dSph}^2 = \frac{1}{N-1} \frac{\sum_{i} [(v_i - \overline{v}_{\rm dSph})^2 - \sigma_i^2 ] w_i^2}{\sum_i w_i^2},
\end{equation}
where the weights $w_i = 1/(\sigma_i^2 + \sigma_{\rm dSph}^2)$ are updated in each iteration and the factor $1/(N-1)$ is needed to 
make the estimated dispersion unbiased since the systemic velocity is evaluated at the same time. 
The $\sigma_{\rm l.o.s.} (R)$ is then derived for each distance bin, where in most studies $\overline{v}_{\rm dSph}$ is assumed to be constant with $R$. 
While in absence of streaming motions this is a reasonable assumption, care should be taken when dealing with samples 
with asymmetric spatial coverage in presence of velocity gradients as this can result in $\overline{v}_{\rm dSph}$ varying with $R$. 

This simple iterative approach suffices when the 
number of MW contaminants expected within the velocity range used for membership is 
small in comparison to the number of genuine members, for example in the central regions (note that this is not 
necessarily the case for UFD). For samples extending also to the outer parts, it 
is more appropriate to explicitly take into account the contribution of MW stars when deriving the internal kinematic properties of the dSph.  

For example \citet{battaglia2008} propose to derive $\overline{v}_{\rm dSph}$ and $\sigma_{\rm dSph}$ 
as a function of projected radius $R$ by maximizing the likelihood 
$L(v_1, \dots, v_N) = \prod_{i} P(v_i)$, with the individual probabilities given by: 
\begin{eqnarray}
P(v_i \mid \overline{v}_{\rm dSph}, \sigma_{\rm dSph})& = & \frac{N_{\rm MW}}{N_{\rm TOT}} f_{MW}(v_i) + \nonumber \\
 & + & 
\frac{N_{\rm dSph}}{N_{\rm TOT} \sqrt{2 \pi (\sigma_{\rm dSph}^2 + \sigma_i^2)}} e^{- \frac{(v_i- \overline{v}_{\rm dSph})^2}{2 ( \sigma_{\rm dSph}^2+\sigma_i^2)}}.
\end{eqnarray}
The dSph LOSVD is assumed to be a Gaussian, while the MW LOSVD, $f_{\rm MW}(v)$ the sum of two Gaussians. 
This provides a good fit to the LOSVD of MW stars predicted by the Besan\c{c}on model \citep{robin2003}, in the direction to Scl and in the magnitude and color range of Scl target RGB stars. The surface number count profile of the target RGB stars in the dSph and the 
number density of MW contaminants in the same CMD region of the targets is derived from the wide-area ESO/WFI DART 
photometric data. This allows to determine the fraction of MW and Scl RGB 
stars per distance bin in elliptical annuli ($\frac{N_{\rm MW}}{N_{\rm TOT}}$ and $\frac{N_{\rm dSph}}{N_{\rm TOT}}$, respectively). This method was also 
applied to the chemo-dynamical components of Scl (for all the details see Battaglia 2007).

\citet{walker2009algo} present an expectation maximization (EM) algorithm with applications to dSphs. 
The equations that identify this method are the probability 
that $M=1$ (i.e. that the star is a member), subject to the data $X_{i=1}^N$ and the prior constraints $p(a)$ 
(a priori membership fraction):
\begin{eqnarray}
P_{M_i} & \equiv & P(M_I = 1 | X_i, a_i) \nonumber \\
 & = & 
\frac{p_{\rm mem}(X_i) p(a_i)}{p_{\rm mem}(X_i) p(a_i) + p_{\rm non}(X_i) [1-p(a_i)]}
\end{eqnarray}
and the expected (log) likelihood given the data-set $S \equiv {X}_{i=1}^N$:
\begin{eqnarray}
E(\ln L(\xi_{\rm mem}, \xi_{\rm non} | S) & = & \sum_{i=1}^N P_{M_i} \mathrm{ln}[p_{\rm mem}(X_i) p(a_i)]  \nonumber \\
 & + & \sum_{i=1}^N (1 - P_{M_i})  \mathrm{ln}[p_{\rm non}(X_i) [1- p(a_i)]]
\end{eqnarray}
where $\xi_{\rm mem}, \xi_{\rm non}$ are the parameter set to evaluate, $p_{\rm mem}$ and 
$p_{\rm non}$ the probability distribution of members and non-members in the variable $X$. Before maximizing 
the expected ln($L$), the  $P_{M_i}$ are evaluated for all $i$. 

In their formulation of the EM analysis applied to dSphs, the data-set 
consists of $S \equiv \{V_i, W_i, a_i\}_{i=1}^N$, where beside the individual 
l.o.s. velocities $V_i$, the variables are the Mg-index values of the stars $W_i$ and $a_i$ the elliptical 
distances. Except for the velocity probability distribution of non-members, which is estimated from the Besan\c{c}on model, for members and the probability distributions in W are assumed to be Gaussian; finally, the only assumption made on $p(a)$ is that it decreases with 
radius. While in the 
approach of \citet{battaglia2008} the fraction of members as a function of projected distance 
is fully determined from the photometric data, without free parameters, here the $p(a_i)$ need to be evaluated for each $i$. 
The approach from Walker et al. has the advantage of being more flexible in the inclusion of new variables and mathematically more rigorous. 

The membership probabilities obtained from the above likelihood approaches can be used to retain 
in the sample only stars with high probability values or as weights for subsequent calculations; 
for example, for the velocity dispersion profiles in \citet{walker2009uni}  
only stars with probability of membership larger than 0.95 are retained and then 
the \dispr is found using a maximum likelihood procedure analogous to 
that of \citet{hargreaves1994sext}. In Bayesian extensions of the method, for example by \citet{martinez2011} and \citet{walker2011}, the probability of membership does not need to be directly evaluated. 

While previous works obtained flattish velocity dispersion profiles with cold points at the last measured 
radius for Sext, Draco and Ursa Minor \citep{kleyna2004, wilkinson2004}, now the general consensus is that dSphs have flat l.o.s. velocity dispersion profiles, with only gentle hints of a decline or a rise, as shown in Fig.~\ref{fig:walker2009}. The approximately flat l.o.s. velocity dispersion profiles of dSphs are considered as the best evidence that  the DM has a different spatial distribution than the stars, so that mass-follows-light models can be excluded, {\it in the hypothesis of dynamical equilibrium}.  

\paragraph{Presence of unbound stars} 
The concern is that these might affect the inferred DM mass distributions and dynamical mass-to-light ratios.
 For example, N-body simulations show that inclusion of unbound stars typically causes  
rising \dispr  \citep[or less often flat, e.g.][]{read2006, klimentowski2007, munoz2008}, 
even though the \dispr derived considering only bound stars is declining. 

In the literature, the methodology used to get rid of unbound dSph stars is based on the virial theorem and 
was originally proposed to eliminate interlopers from clusters of galaxies 
\citep[e.g.][]{denhartog1996} and tested using dark-matter only N-body simulations of 
galaxy clusters in which random sub-sets of particles where chosen to represent galaxies 
\citep[e.g.][]{wojtak2007}; in other words in the situation where the number density 
of the tracers is proportional to that of the overall mass density 
(i.e. a ``mass-follows-light'' model). Subsequently \citet{klimentowski2007} tested the performance of the method using 
a high-resolution N-body simulation of a dSph formed within the 
tidal stirring framework, for which strong tidal 
stripping of the stellar component and a final (approximately) mass-follows-light configuration were found. 
Using mock samples of l.o.s. velocities, they showed that 70-80\% of 
the unbound stars from the tidal tails could be removed by the \citet{denhartog1996} method, while only a 
very small fraction ($\sim$1\%) of bound stars was mistaken for interlopers. When including foreground contamination 
from the MW, the fraction of correctly identified interlopers (both from the MW and from the tidal tails) decreases 
only of a few percent. 

The above iterative procedure rejects objects whose velocity exceeds the maximum velocity available to a tracer particle 
(here a star) at a certain projected radius $R$. This maximum velocity
 is determined under the assumption that a star is either on a circular orbit with velocity $v_{\rm circ}= \sqrt{ G M(r)/r}$ 
or in free fall under the pull of the 
galaxy's potential with velocity $v_{\rm inf}= \sqrt{2} \, v_{\rm circ}$, 
given the mass profile $M(r)$ determined from the virial mass estimator:
\begin{equation}
M(r_j) \equiv M_{\rm VT}(R_j) = \frac{3 \pi N}{2 G} \, \frac{\sum_i (v_i - \overline{v})^2}{\sum_{i < j} 1/R_{ij}}
\end{equation}
where - adapting to our case the definitions from \citet{denhartog1996} - 
$M(r_j)$ is the mass enclosed within the distance to the star $r_j$, $R_j$ is the projection of $r_j$, $N$ is the number of stars in the sample, $R_{ij}$ 
are the projected distances of pairs of stars within a cylinder of radius $R$ around the center, $v_i$ and $\overline{v}$ have been 
defined above. Here  $v_{\rm inf}$ can be seen as the escape velocity from the mass interior to $r$ \citep{wojtak2007}.

By applying this method for rejecting (putative) unbound stars, \citet{lokas2008} and \citet{lokas2009} 
rederive the l.o.s. velocity dispersion 
profiles for Leo~I, Car, Fnx, Scl and Sext using data from \citet{mateo2008} and \citet{walker2009}; they  
find gentle declines for all these galaxies and are able to explain the data with ``mass-follows-light models'' close 
to isotropic or mildly tangential. 
Note that the best fits still require considerable amounts of DM (see Sect.~\ref{sec:dyn-models}). 

\begin{figure}
\centering
\includegraphics[width=0.35\linewidth]{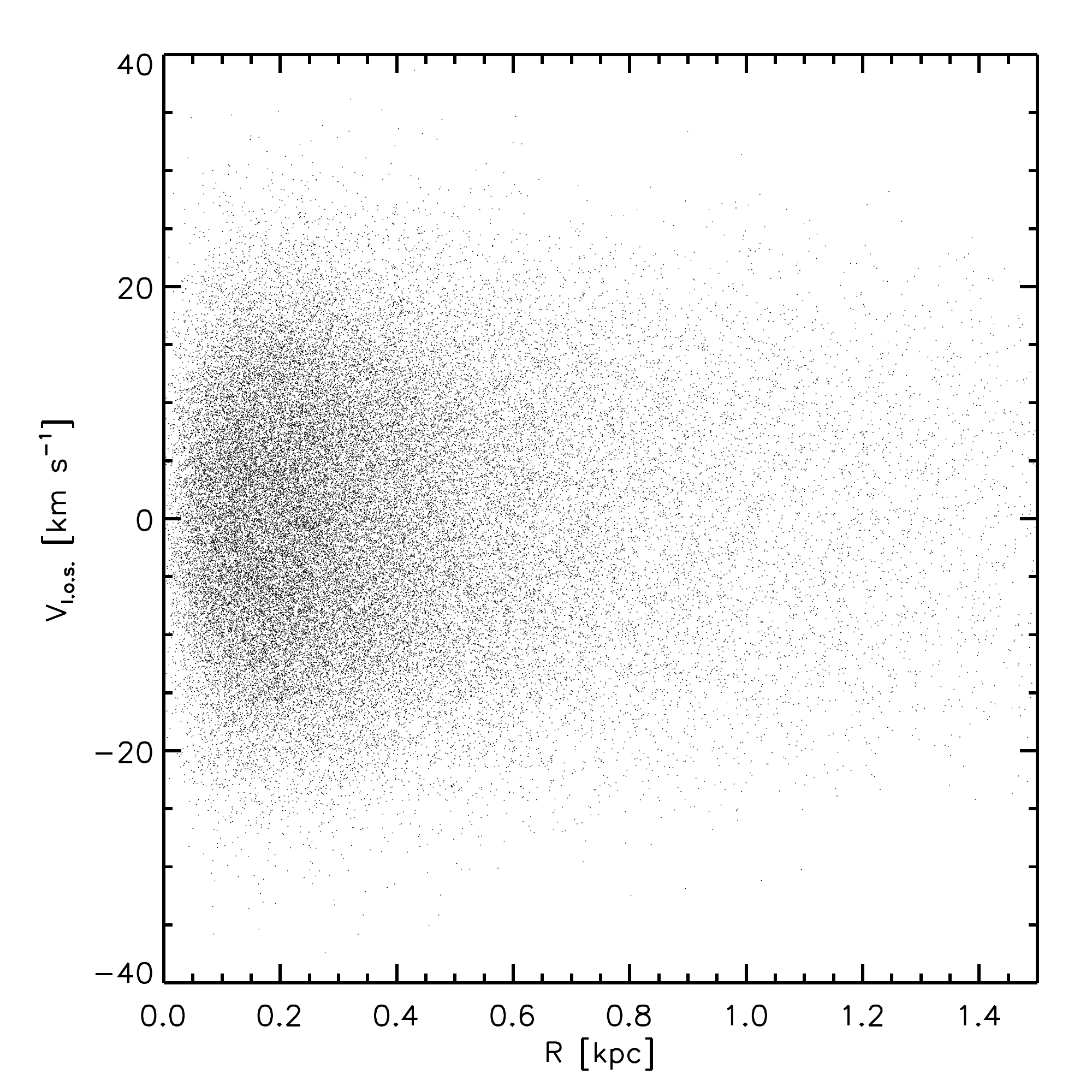}
\includegraphics[width=0.35\linewidth]{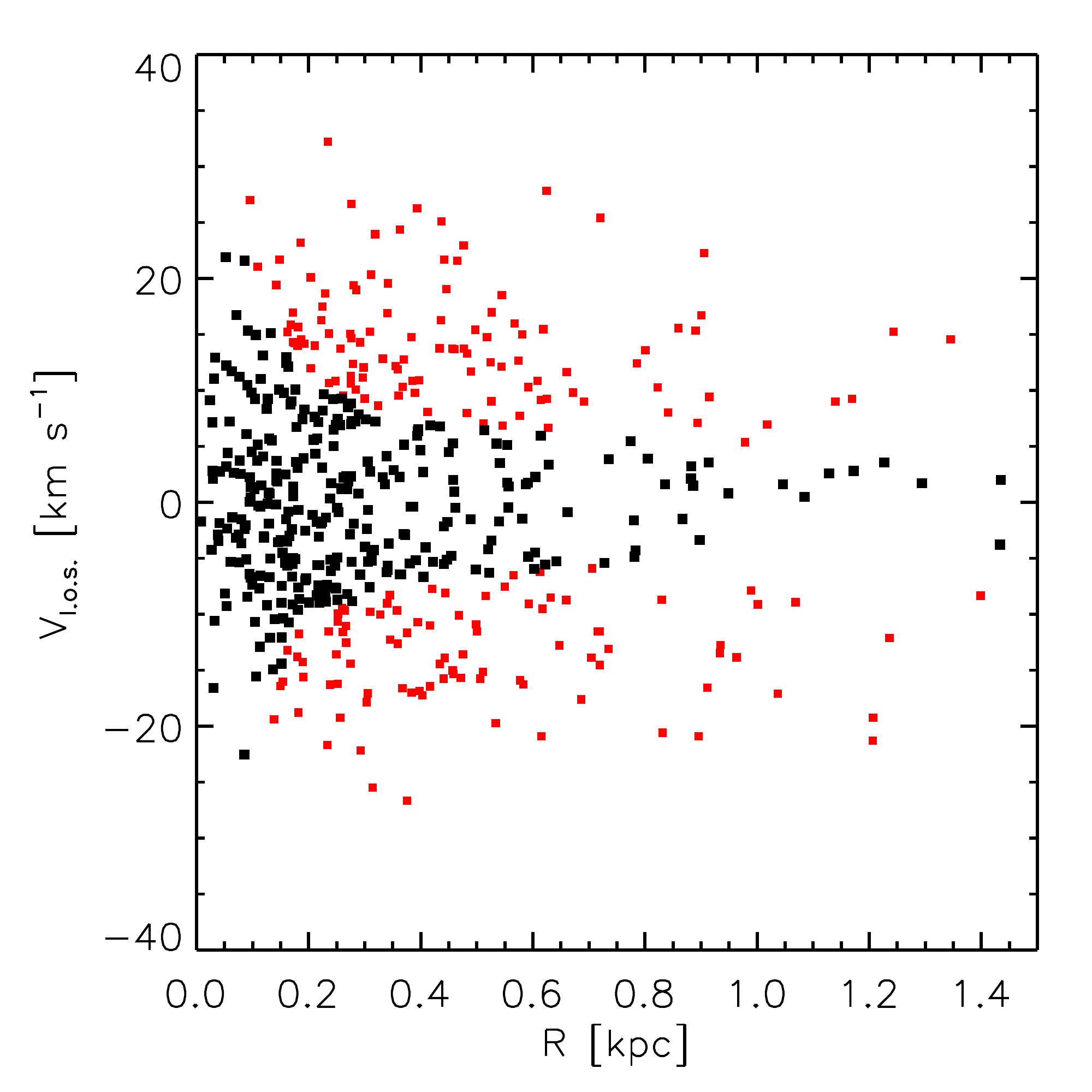}
\includegraphics[width=0.35\linewidth]{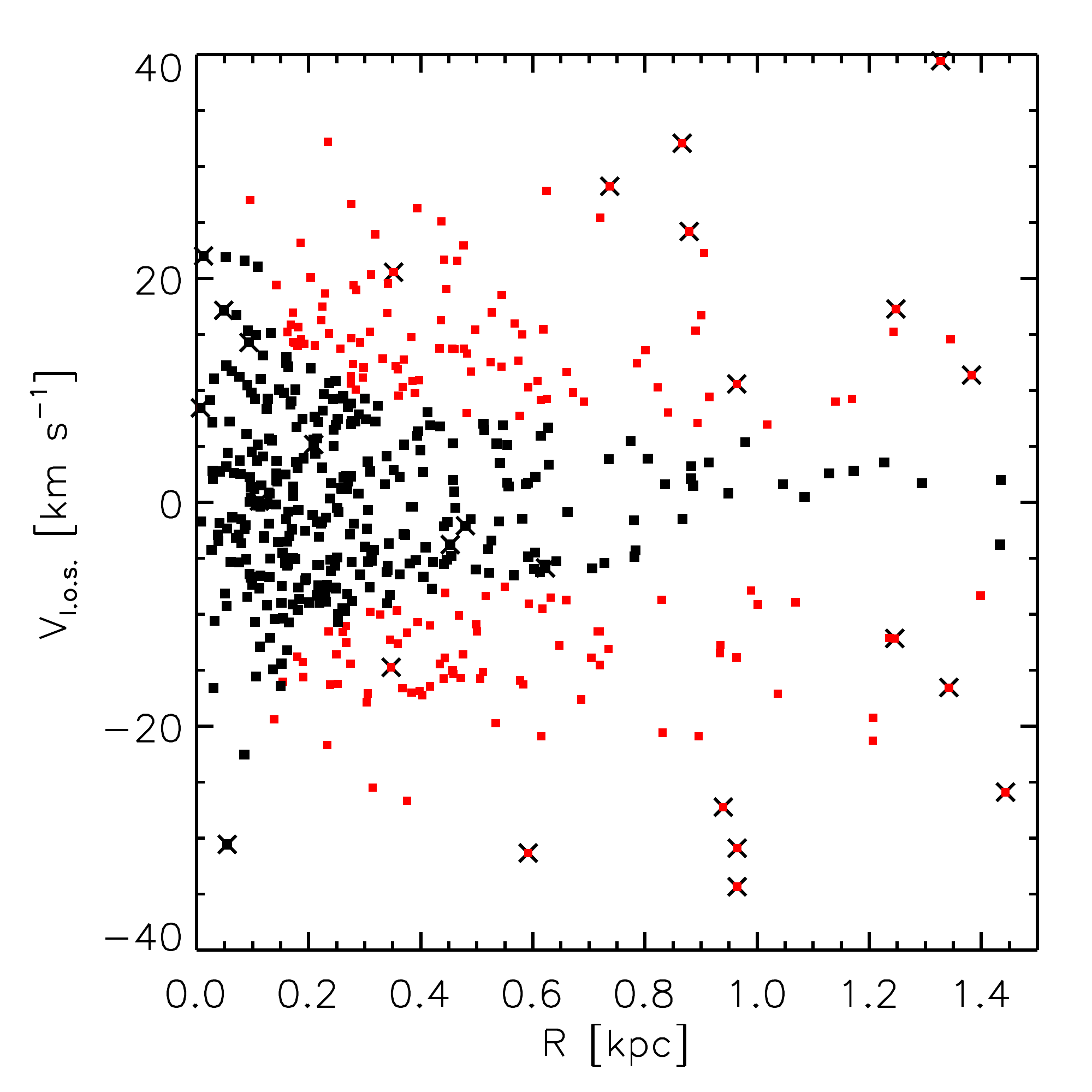}
\includegraphics[width=0.35\linewidth]{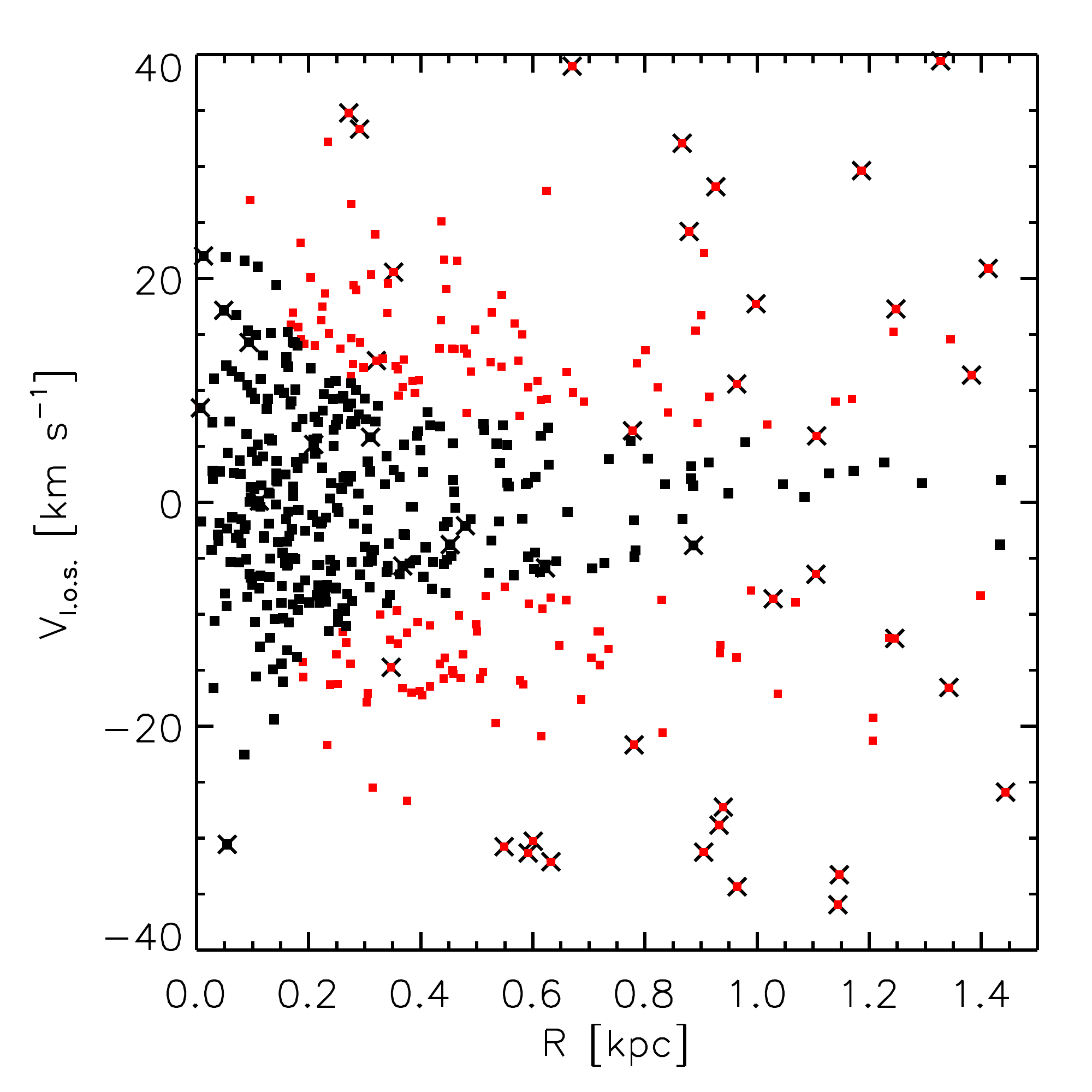}
\includegraphics[width=0.35\linewidth]{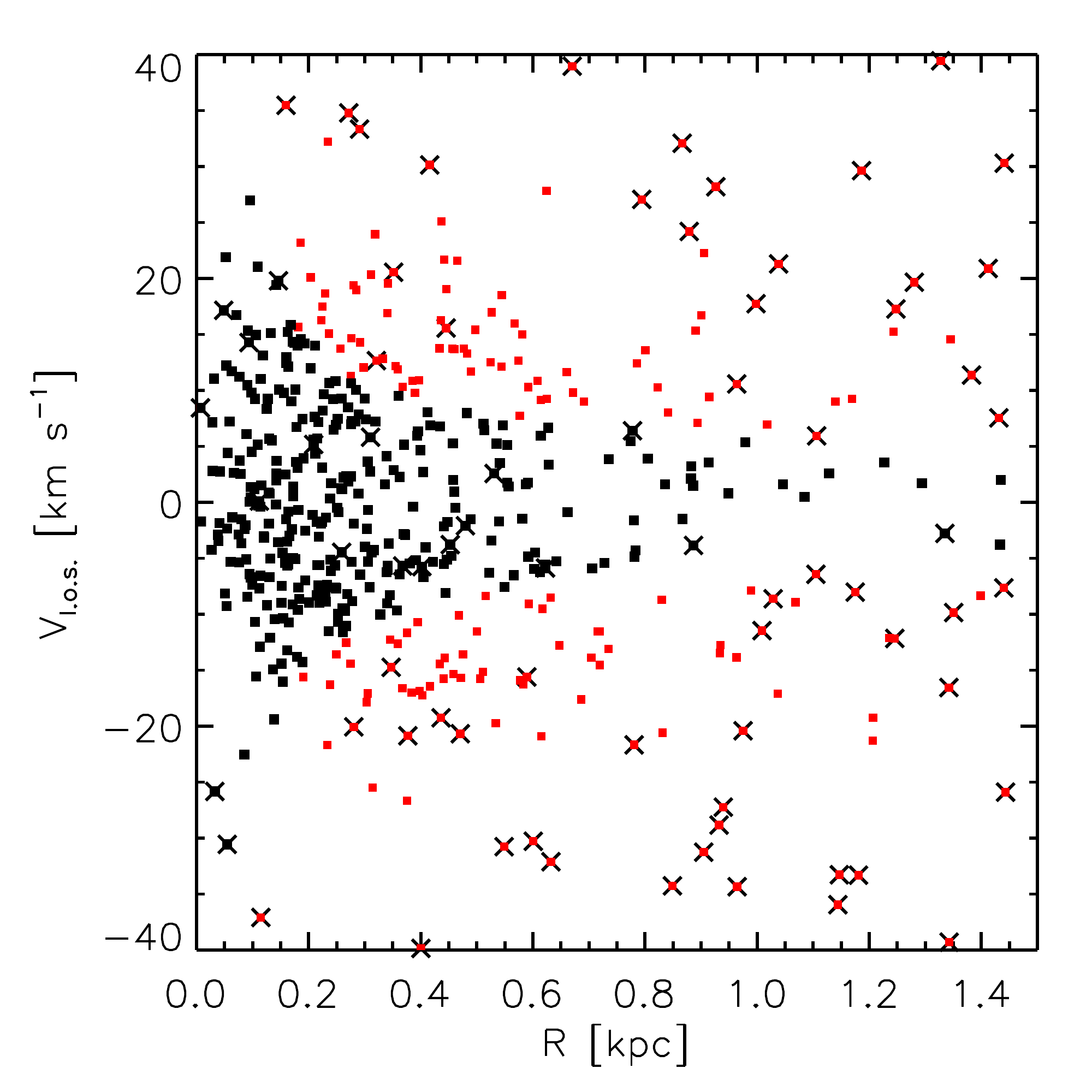}
\includegraphics[width=0.35\linewidth]{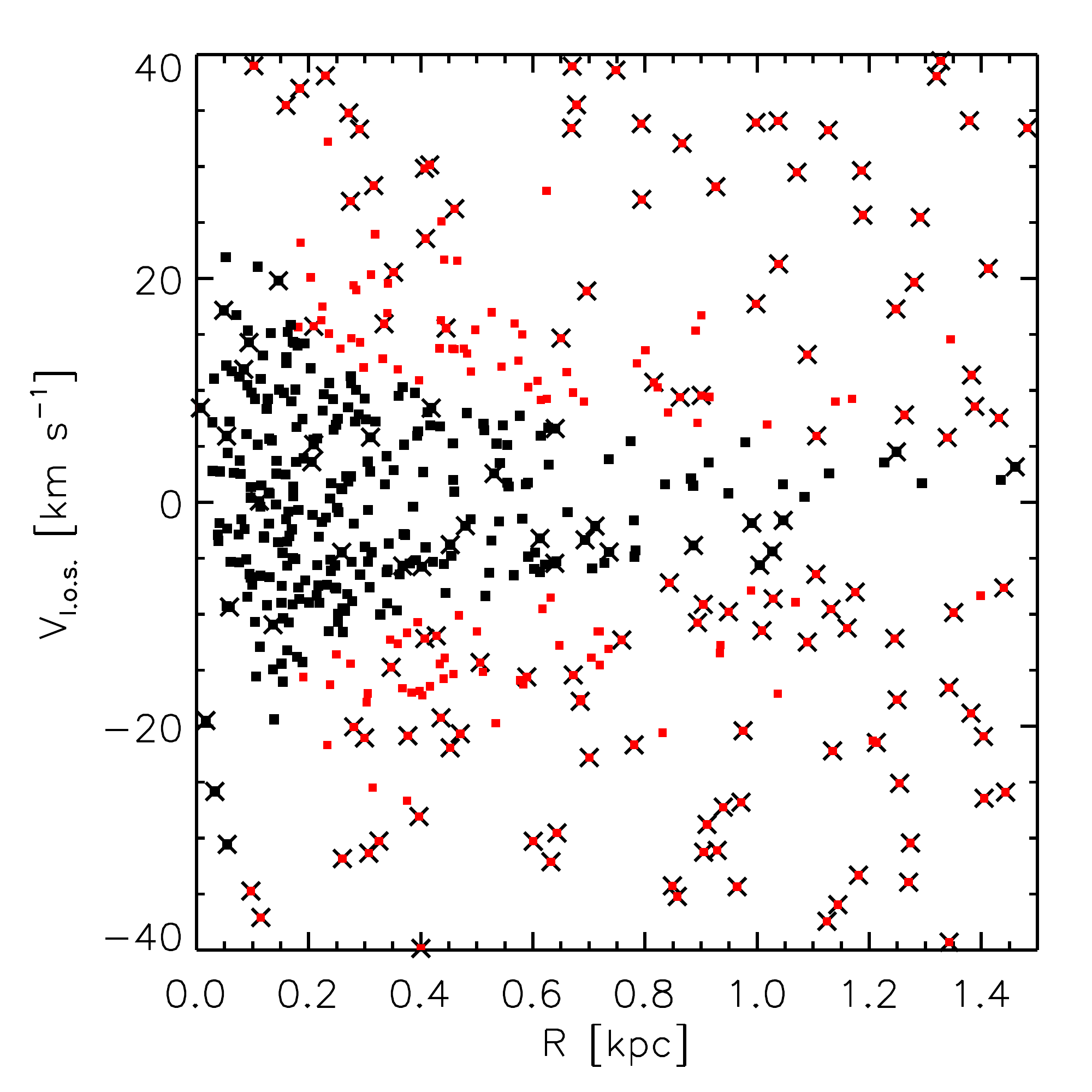}
\caption{Top left: L.o.s. velocity versus projected radius $R$ diagram for a mock Sculptor galaxy \citep{Breddels2012} 
(50000 points). Top right: application of the virial theorem interloper rejection procedure applied to a randomly 
extracted sub-sample of 500 stars from the mock Sculptor (see text). 
The red squares show the objects rejected after three iterations and the black squares those retained in the sample. 
Middle and bottom panel: as in the previous panel, but adding an increasing 
amount of mock MW interlopers, shown by the crosses (see text). The interloper rejection procedure correctly identifies the mock 
MW interlopers found outside of the velocity envelope, but also rejects an important fraction of genuine members: initially, the genuine members are 95\%, 91\%, 85\% and 70\% of the 
sample of 500 objects, while after the 3rd iteration only the 60-65\% are accepted as members 
(therefore the remaining  32\%, 27\%, 21\% and 13\% are mistaken for interlopers).  
} \label{fig:mockscl}
\end{figure}

\begin{figure}
\centerline{\includegraphics[scale=0.7]{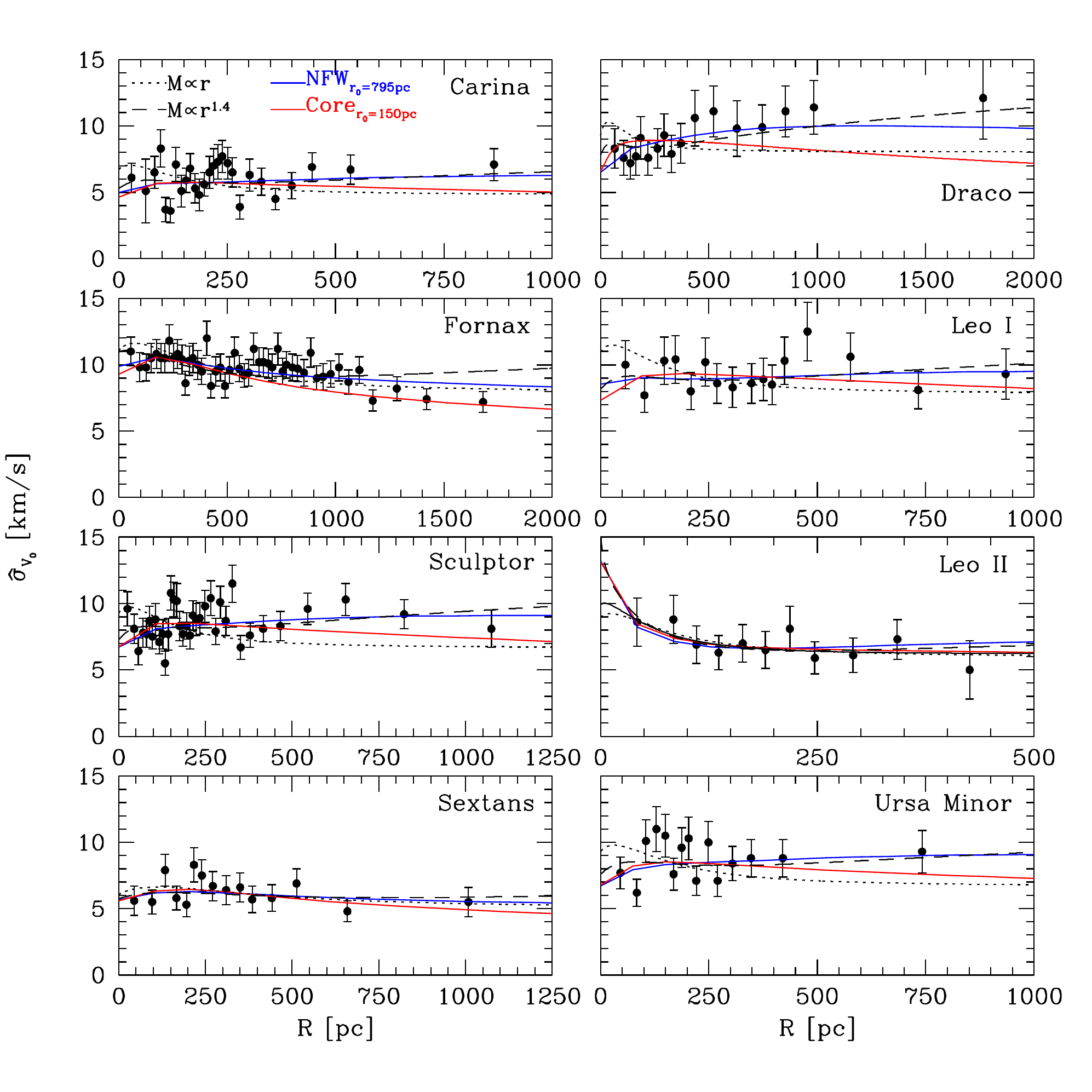}}
\caption{Projected velocity dispersion profiles for eight bright dSphs obtained by \citet[][reproduced by permission of the American Astronomical Society]{walker2009uni}. The profiles calculated from isothermal, power-law, NFW and cored halos (with $M(r) \propto r^{1.4}$ are also shown. These fits have been obtained using the spherical Jeans equations, and in particular Eqs.~(\ref{eq:sigma_r-beta-cst}) and (\ref{eq:sigma_los-beta}). For each type of halo these authors fit only for the anisotropy (assumed to be constant) and mass normalization (at a given scale/distance). The scale radius of the system is fixed for the various models as indicated in the panels. 
\label{fig:walker2009}}
\end{figure}

At present it is unclear how such method would perform for a system in which the mass-to-light 
ratio strongly increases with radius, for example if dSphs were embedded in an NFW or cored DM halo
in dynamical equilibrium. To explore this, here we apply such a rejection procedure on the mock Scl 
model from \citet{Breddels2012} in which the stars are distributed as a Plummer sphere with $b =$0.3~kpc, stellar mass $M_* =10^6 \mathrm{M}_{\odot}$, 
velocity anisotropy $\beta = -0.5$,  
embedded within a NFW halo with scale radius 0.5~kpc and $10^8 \mathrm{M}_{\odot}$ mass enclosed within 1 kpc radius. 
To the 50000 Scl mock stars, we add a mock MW foreground contamination, amounting to 5\%, 10\%, 20\% and 50\% of these stars, 
uniformly distributed in projected radius and $\pm$40 \kms around the systemic velocity. From the overall sample, we randomly extract 500 
objects (genuine members and contaminants) and determine the 
velocity envelope using the infall velocity at $R$, $v_{\rm inf} = \sqrt{2} v_{\rm circ}$ evaluated at $r=R$. 
As shown in Fig.~\ref{fig:mockscl}, after 3 iterations of the algorithm, several genuine members are identified as contaminants and 
the stars retained in the sample would lead to a declining $\sigma_{\rm l.o.s.}(R)$.  
It still remains to be extensively tested how the rejection method based on the virial theorem 
would perform for a system in equilibrium for different functional forms of the DM halo and light distributions; here we 
have shown that a particular example of a dSph in equilibrium within extended and massive DM halos for which this rejection method 
would remove an important fraction of genuine members. 

\subsubsection{The higher moments: third and fourth}  
\label{sec:higherm}

The third moment has barely been used in the study of MW dSphs. The most detailed analysis has been carried out by 
\citet{amorisco2012lp} by analyzing the angular behavior of asymmetric deviations from Gaussianity and
 found what appears to be an intrinsic rotation signal of 1 \kms about the minor axis in the Fnx dSph. 
No conclusions could be drawn for Scl, Car and Sext given the large error-bars and more limited sample sizes. 

Therefore, in this section we focus mostly on the fourth moment because of the  
information it carries about the velocity anisotropy of the stars in dSphs. 
Knowledge of the velocity anisotropy allows to test possible formation scenarios of dSphs and 
helps break modeling degeneracies. For example, \cite{Merrifield1990} show explicitly how two
systems with the same surface density profile and l.o.s. 
velocity dispersion can have very different distribution
functions and be embedded in different gravitational potentials. The 
only way to tell them apart is through a measure of the 4th moment of the LOSVD. 

The 4th moment of the LOSVD has been quantified in various ways in the literature. \cite{Merrifield1990} use the kurtosis, defined as 
\begin{equation} \label{eq:merr}
\kappa_{\rm l.o.s}(R) = \overline{v}_{\rm l.o.s}^4/\overline{v^2}^2_{\rm l.o.s}.
\end{equation} 
A Gaussian function has a $\kappa_{\rm l.o.s}=3$, 
while boxier LOSVD distributions have kurtosis $<$3 and more peaked LOSVD with longer tails have kurtosis $>$3.
\cite{Merrifield1990} state that to measure the kurtosis with an error $<$ 10\% a sample of 750 stars is needed, while
if only 150 stars are available, then the error is $\sim 20$\%, both with
90\% confidence, which ought to allow the detection of 
gross variations in the velocity anisotropy. 

\citet{lokas2003} define $K'$, an unbiased estimator of the kurtosis for a sample of $n$ l.o.s.\
velocity measurements $v_i$  
\begin{equation}
K' = \frac{3}{C} K 
\end{equation}
with $K$ being 
\begin{equation} \label{eq:K}
K = \frac{ \frac{1}{n} \sum_{i=1}^{n} (v_i - \overline{v})^4}{(S^2)^2}
\end{equation}
and 
\begin{equation}
\overline{v} = \frac{1}{n} \sum_{i=1}^{n} v_i  \: \mathrm{,} \: S^2 = \frac{1}{n} \sum_{i=1}^{n} (v_i- \overline{v})^2.
\end{equation}
$K'$ is introduced because $K$ was found to underestimate the kurtosis in Monte Carlo realizations of a Gaussian distribution. Since the sampling distribution of $K$ is strongly skewed, \citet{lokas2003} also suggest to use the 
function $k = \left[ \mathrm{log} K' \right]^{1/10}$, which follows a Gaussian sampling distribution, valid also 
for weakly non-Gaussians LOSVD \citep{lokas2005}. 
The value of $C$ and the standard error on $k$ vary with the number of objects (per bin) and become 
$C=$2.75, 2.89, 2.93 for $N =$40, 100, 200, while the corresponding sampling error for $k$ is 0.02, 0.0124, 0.009 \citep[see also][]{lokas2009}. For sample sizes with a few hundreds objects per bin, the value of the kurtosis as given by 
Eq.~(\ref{eq:K}) remains underestimated by a few percent.  $k = $0.93 corresponds to a Gaussian distribution, while flatter (more peaked) distributions have 
$k < 0.93$ ($> 0.93$). 

More recently, \citet{Breddels2012} defined a 4-th moment estimator as:
\begin{equation} \label{eq:bred}
\hat{\mu_4} = \sum_{i=1}^{n} (v_i + \epsilon_i)^4 - 3s_2^2 + 6 \mu_2 s_2 
\end{equation}
where $\epsilon_i$ is the velocity error for star $i$ (this term accounts for the noise of
the measured velocity $v_i$), 
$s_2$ is the average of the measured squared velocity errors, and the second moment of the LOSVD $\mu_2$ is approximated by the estimator 
$\hat{\mu_2}$:
\begin{equation}
\hat{\mu_2} = \frac{1}{n}\sum_{i=1}^{n} (v_i + \epsilon_i)^2 - s_2.
\end{equation}
This estimator has the same distribution of values as the kurtosis.

One problem of the kurtosis, in its various definitions, is that it heavily weighs the tails of the LOSVD (since it is
proportional to $v^4$), and so for example, contamination by
foreground/unbound stars can be an issue. In this sense it has been argued that the use of Gauss-Hermite
moments is generally preferable as these are less dependent 
on the wings of the distribution, and because they can be derived more accurately. 

Mainly in the context of studying the properties of elliptical galaxies, \cite{Gerhard1993} and  \cite{vdMarel1993} showed that the LOSVD ($L(v)$) from absorption line spectroscopy could be well described with just a few
terms of the expansion in Gauss-Hermite functions, typically just up to order 4. Since Gaussians provide good first-order 
approximations to the $L(v)$, \cite{vdMarel1993} introduced the following definition for the Gauss-Hermite serie:   
\begin{equation}
L(v)= [ \gamma \alpha(w) / \sigma ] \left \{1 + \sum_{j=3}^{N} h_j H_j(w) \right \} \, , \mathrm{with} \: w= (v-V)/\sigma.
\end{equation}
where $\gamma$ , V, $\sigma$ are the line strength, mean radial velocity and velocity dispersion, respectively, of the 
best-fitting Gaussian to the observed $L(v)$; the functions $H_j(w)$ are Hermite polynomials \citep[for the definition see][]{vdMarel1993}; $h_0$, $h_1$, $h_2$ are set to (1,0,0) and $h_3$, ..., $h_N$ contain the shape information;  
specifically, $h_3$ and $h_4$ measure asymmetric and symmetric deviations from Gaussianity. Here $h_4 > 0 $ corresponds to a peaked distribution, while $h_4 < 0$ to a more
flat-topped distribution. They also show that negative $h_4$ corresponds typically
to tangential anisotropy, for a similar set of gravitational
potentials and distribution functions as explored by \cite{Gerhard1993}. 

\cite{amorisco2012lp} have tested how the discrete nature of dSph data-sets affects the reliable
determination of the Gauss-Hermite moments of the LOSVD. They estimate that 200 stars are 
necessary not to be dominated by noise
due to limited sampling: in this case, the shot noise on $h_3$ and $h_4$ is $\sim$0.05. 
If one wishes to study how the moments vary as a function of distance, of course the error refers 
to the number of stars per bin. 

These authors also argue that the Gauss-Hermite formalism cannot be readily applied to the kinematic
data available for the dSphs because of the difficulties 
in accounting for the heterogeneous observational uncertainties and probability of membership 
associated to each star, and for the fact that the LOSVD is a convolution of the intrinsic $L(v)$ with 
the measurement errors. 
They also note that this expansion does not define a proper probability density function since the Gauss-Hermite series is not always positive definite everywhere. Therefore they introduce
a set of quantities to characterize asymmetric
and symmetric deviations from a Gaussian distribution, and which they
argue are less sensitive to shot noise than the Gauss-Hermite moments
(roughly a factor 2 smaller for $h_4$ for a sample of 800 stars). 

The fourth moment of the LOSVD of MW dSphs has been derived and analyzed for most classical dSphs, 
although not always used in the mass modeling. At present the results for Leo~I are contrasting: \citet{sohn2007} detect a broad and skewed LOSVD, in 
particular for stars at large angular separations on the West side of the galaxy; this has not been confirmed by 
\citet{mateo2008}, who find that the skewness and kurtosis are consistent with a Gaussian distribution. It is unclear 
if the different results may be due to the systematic velocity differences between the data-sets used in the two works. \citet{lokas2008} have re-analyzed the sample of \citet{mateo2008} finding that the LOSVD 
departs from Gaussian and becomes rather irregular at $R >$ 6 arcmin, but consistent with a Gaussian at smaller $R$. 

For Car, Fnx, Scl and Sext, \citet{lokas2009} 
apply the interloper removal scheme based on the virial estimator to the \citet{walker2009} sample and find either flat $k(R)$ very close to the value expected for a Gaussian distribution or 
slightly declining $k(R)$ with values in the range 0.90-0.95; the latter imply very mild 
tangential anisotropy, consistent within 1$\sigma$ with an isotropic velocity ellipsoid. Fourth moment values 
corresponding to mild tangential anisotropies 
are found by \citet{Breddels2013} for Scl and by \citet{amorisco2012lp} for Fnx, assuming
that the dSphs are in dynamical equilibrium.

On the other hand, \citet{amorisco2012lp} find a preference for small positive values of $h_4$ (and of their new parameter measuring symmetric deviations from Gaussianity) for Scl, and also Sext.  The contrasting results may be attributed to 
the different ways of dealing with membership, as both works used similar data-sets as \citet{Breddels2012}. Also \cite{lokas2005} found that the values (and profile shapes) of the unbiased kurtosis estimator 
vary according to the treatment of interlopers, going from flat around 
the value expected for a Gaussian distribution, to slightly declining or slightly rising. 

From the above it is clear that it is important to model carefully the stars' membership, as this can impact  
the conclusions on the orbital structure of stars in dSphs. The situation could likely be improved by using the probability of membership only as a weight when deriving the kinematic properties, rather than to use it for eliminating stars from the sample. 

Notwithstanding the above issues, the current measurements suggest that the LOSVD of stars in dSphs are not dramatically different from Gaussians in most cases, so that one can conclude that the velocity ellipsoid is neither strongly radial nor strongly tangential.

\subsubsection{Chemo-dynamical components} \label{sec:multiple}
Several classical dSphs exhibit spatial variations in their metallicity distribution functions, 
with the mean metallicity being higher in the inner regions and lower in the outer parts 
(e.g. Tolstoy et al. 2004, Battaglia et al. 2006, Koch et al. 2006, Faria et al. 2007, Battaglia et al. 2011). 

\begin{figure*}
\centering
\vspace*{7cm}
\includegraphics[width=0.5\linewidth]{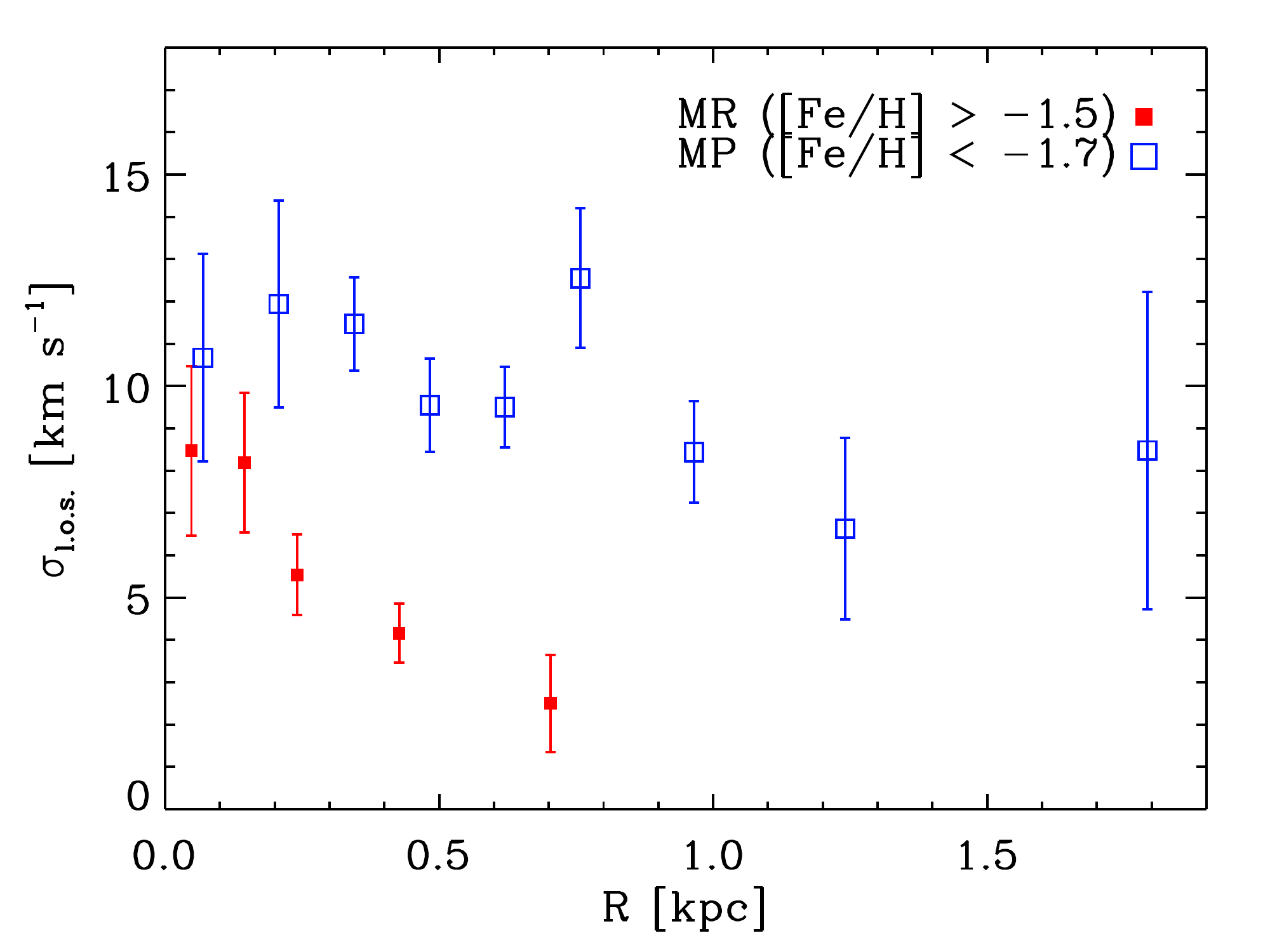}
\caption{L.o.s. velocity dispersion profiles for the Sculptor members more metal-rich than [Fe/H]$=-1.5$ (filled squares) and 
more metal-poor than [Fe/H]$=-1.7$ (open squares), 
from rotation-subtracted velocities in the Galactocentric Standard of Rest system \citep[see][]{battaglia2008}. 
} \label{fig:sigmascl}
\end{figure*}

For Scl, Fnx and Sext, the internal kinematics are found to be linked to the metallicity, 
in that the ``metal-rich'' stars are more centrally concentrated and exhibit a lower $\sigma_{\rm l.o.s}$ than the metal-poor stars which form a more extended population
(Tolstoy et al. 2004, Battaglia et al. 2006, Battaglia et al. 2011). 
This is clearly seen in Fig.~\ref{fig:sigmascl} for the Scl dSph, where the $\sigma_{\rm l.o.s}$ of  
stars with [Fe/H]$ > -1.5$ is also found to decline with $R$, while for those with 
 [Fe/H]$< -1.7$ it stays approximately constant \citep[see also][]{walker2011}. 

The analysis of the kinematic properties of these ``chemo-dynamical'' stellar components paints in some cases a very complex evolutionary picture, as for the Fnx dSph. The analysis of \citet{amorisco2012fnx} suggests 
 that this system is best described by the superposition of 3 chemo-dynamical components, increasingly more metal-rich, 
more spatially concentrated and with colder kinematics. The authors detect 
what appears to be a misalignment of the angular momentum of intermediate-metallicity and 
metal-poor stars. Such a detailed analysis would not have become possible without the large spectroscopic data-sets with metallicity and kinematical information available nowadays. 
Furthermore, from a dynamical modeling perspective, the presence of different ``chemo-dynamical'' stellar components have provided interesting new constraints on the DM mass distribution of dSphs  (see Sect.~\ref{sec:dyn-multiple}).

\section{Dynamical modeling}
\label{sec:dyn-models}

The techniques to model the internal dynamics of spheroidal systems have long been in place. 
However, their application to nearby dwarf spheroidals has only really taken off in the last decade,
with the need for more sophisticated approaches thanks to the manifold increase in data samples. 
In this section we review the 
methods used, briefly discuss their limitations and the results obtained thus far for these systems. 
Table \ref{tab:models} gives an overview of the various modeling 
techniques applied to the MW dSphs.

We divide this Section according to the groups of methods that have
been used so far. In general, we can broadly classify methods on
whether they are {\it parametric}, i.e. they assume a family of
models, or {\it non-parametric}, in which the distribution function is
expressed in more general terms, for example as an expansion of basis
functions. Most works attempt to fit the moments of the velocity
distributions, while the use of the velocities and positions of
individual stars to determine the likelihood of a given model (also
known as discrete modeling) has been explored to a lesser extent in
the literature.

\subsection{Modeling with the Jeans Equations}

To a very good approximation, a dwarf galaxy may be considered a
collisionless system. The internal structure of such a dynamical
system can be described through its distribution function $f({\bf x},
{\bf v},t)$, which in the collisionless case, obeys the Boltzmann
equation
\begin{equation}\label{eq:boltzmann}
\frac{\partial f}{\partial t} + {\bf v} {\bf .} \nabla_x f - \nabla_x \Phi {\bf .} \nabla_v f = 0,
\end {equation}
where $\Phi({\bf x})$ is the total gravitational potential of the
system \citep[including stars and dark matter contributions][]{BT2008}. For our
purposes $f({\bf x}, {\bf v},t)$ describes the probability of finding
a star with a given position {\bf x}, and velocity {\bf v} at time $t$.

In general we assume that the distribution function is time-independent (see Sec.~\ref{sec:tides}), so that the first term in this equation may be dropped.
Deriving the distribution function from Eq.~(\ref{eq:boltzmann}) by comparison to observations is not straightforward
(see below), so a commonly used approach is to take moments of this equation, and compare these
moments to observables, since also low-order moments are easily measured from observations.

The zero-th moment corresponds to the continuity equation in
hydrodynamics, and it is generally not used in data-model
comparison. The first moment is obtained by multiplying
Eq.(\ref{eq:boltzmann}) by $v_j$ and integrating over all
velocities. The resulting equation is
\begin{equation}
\frac{\partial \nu \langle v_i v_j \rangle }{\partial x_i} + \nu \frac{\partial \Phi}{\partial x_j} = 0,
\label{eq:first_mom_cbe}
\end{equation}
where $\nu({\bf x})$ is the stellar density, i.e. $\nu({\bf x}) = \int
d^3 v f$, and the brackets $\langle\rangle$ denote moments, e.g. here 
$\langle v_i v_j \rangle = \int
d^3 v v_i v_j f$. Eq.(\ref{eq:first_mom_cbe}) represents a set of 3 equations known as the Jeans
equations. These are useful because they relate to observables,
however, it should be born in mind that this is not a closed set of
equations, in the sense that even if we knew the potential and the
density, to derive the streaming (mean) velocities (3 components) and
the full velocity ellipsoid (6 independent quantities), we only have 4
equations, i.e. the continuity and the Jeans equations. Although it is possible to use higher moments of the
Boltzmann equation, this tends to be more cumbersome. Higher moments are 
also difficult to measure observationally reliably, and nonetheless 
the use of closure relations would still be necessary. Therefore, typically, as we
shall see below, certain assumptions are made, regarding for example the form of the 
velocity ellipsoid, to find a solution to the system.

The distribution function of a steady state system depends on the
integrals of motion.  If the potential is time-independent, then the
energy $E$ is an integral of motion. For a spherical system, all
components of the angular momentum $ {\bf L}$ are conserved, while if
the system is axisymmetric, then only $L_z$ will be, but a third
integral $I_3$ might exist. Therefore, in non-rotating spherical
systems, the distribution function can be a function $f(E)$ or
$f(E,L)$. Although it is possible for a spherical system to rotate
\citep{Lynden-Bell1960}, in which case the distribution function will be
of the form $f(E, {\bf L})$ this is not the most general
configuration. Rotation would be more natural in the axisymmetric
case, when $f(E,L_z)$, i.e. there is a preferred axis (that about which
the system rotates). As discussed in Sec.~\ref{sec:firstm}, there is evidence
of small velocity gradients in the dSphs, however, their origin is
unclear, and in many cases these can be explained by projection
effects. Therefore, in the rest of this review we assume that our
systems to do not rotate. In that case, the second moment and the variance of the
velocity distribution are equal (after subtraction of the gradient), and we refer to these
interchangeably. 

\subsubsection{Jeans equations for spherical systems}

In the case of a spherical system, only one of the Jeans equations is non-trivially zero, and it relates
the 2nd moment of the radial velocity  $\langle v_r^2\rangle$, the stellar density $\nu(r)$, the velocity anisotropy
$\beta(r) = 1 - (\langle v_\theta^2\rangle + \langle v_\phi^2\rangle)/(2 \langle v_r^2\rangle)$, and the total gravitational
potential $\Phi(r)$ as follows:
\begin{equation}
\frac{{\rm d}(\nu \langle v_r^2\rangle)}{{\rm d} r} + 2 \frac{\beta}{r} \nu \langle v_r^2\rangle = - \nu \frac{{\rm d} \Phi}{{\rm d} r}.
\label{eq:jeans_sph}
\end{equation}
An equivalent, often useful form of this equation is
\begin{equation}
\label{eq:jeans_mr}
\frac{GM(r)}{r} = \langle v_r^2\rangle (\gamma_* - 2 \beta - \alpha),
\end{equation}
where $r$ is the spherical radius, $\gamma_* = - d\log\nu/d\log r$ and $\alpha = d \log \langle v_r^2\rangle/d \log r$. 
For example, if the radial velocity and stellar density have been
measured, and we make an assumption on the velocity anisotropy
$\beta$, we may be able to derive the mass distribution (gravitational
potential) of the system. This is the most frequently used
approach. The velocity ellipsoid can be isotropic, in which case
$\beta = 0$, tangentially or radially anisotropic, when $\beta < 0$ or
$\beta > 0$ respectively, and will in the most general case, vary with
radius. In the case of $\beta = 0$, this implies that the velocity
distribution is ergodic, i.e. it is only a function of energy $f = f(E)$, while
for anisotropic systems, $f = f(E,L)$.

The above equations highlight a degeneracy between mass and anisotropy (if the stellar density is perfectly known from observations; otherwise this also enters the degeneracy). This is most easily seen if we assume that $\beta$ is constant with radius. In that case, Equation~(\ref{eq:jeans_sph}) reduces to \citep{BT2008}
\begin{equation}
\label{eq:sigma_r-beta-cst}
\langle v_r^2(r)\rangle = \frac{1}{r^{2\beta}\nu(r)} \int_r^\infty {\rm d} r' r'^{2\beta} \nu(r') \frac{{\rm d} \Phi}{{\rm d} r'}.
\end{equation}
We thus see directly that different combinations of the mass
distribution, density and anisotropy might conspire to produce the
same velocity dispersion profile in the radial direction.  The situation is worsened by the
fact that generally one deals with projected quantities, as discussed below. 

A way to reduce the degeneracy is to use higher moments, in
particular, the 4th moment equations are obtained by multiplying Eq.~(\ref{eq:boltzmann})  by
$v_r^3$ and $v_r v_t^2$ and integrating over velocity space
\citep[see][]{Merrifield1990}:
\begin{equation}
\frac{{\rm d}(\nu \langle v_r^4\rangle)}{{\rm d} r} - 3 \frac{\nu} {r} \langle v_r^2 v_t^2 \rangle + \frac{2}{r} \nu \langle v_r^4\rangle + 3
 \nu \langle v_r^2\rangle\frac{{\rm d} \Phi}{{\rm d} r} = 0,
\end{equation}
and
\begin{equation}
\frac{{\rm d}(\nu \langle v_r^2 v_t^2\rangle)}{{\rm d} r} -  \frac{\nu} {r} \langle v_t^4 \rangle + \frac{4}{r} \nu \langle v_r^2 v_t^2\rangle +  \nu \langle v_t^2\rangle\frac{{\rm d} \Phi}{{\rm d} r} = 0.
\end{equation}
If one assumes that the distribution function is of the form $f(E,L) =
f_0(E) L^{-2\beta}$, it can be shown that the anisotropy is constant, and  
these equations simplify significantly to \citep{Lokas2002} 
\begin{equation}
\frac{{\rm d}(\nu \langle v_r^4\rangle)}{{\rm d} r} + \frac{2\beta }{r} \nu \langle v_r^4\rangle + 3
 \nu \langle v_r^2\rangle\frac{{\rm d} \Phi}{{\rm d} r} = 0, 
\end{equation}
whose solution may be expressed as 
\begin{equation}
\langle v_r^4(r)\rangle = \frac{3}{r^{2\beta}\nu(r)} \int_r^\infty {\rm d} r' r'^{2\beta} \nu(r') \langle v_r^2(r')\rangle\frac{{\rm d} \Phi}{{\rm d} r'}.
\end{equation}

As discussed above, the intrinsic moments are not directly accessible
to the observer, and only projected moments of the line-of-sight
velocity distribution and stellar density profile are
measurable. Following \cite{Merrifield1990} 
these projected moments take
the form 
\begin{equation}
\mu(R) = 2 \int_R^{\infty} \nu(r) \frac{r {\rm d}r}{(r^2 - R^2)^{1/2}},
\end{equation}
\begin{equation}
\label{eq:sigma_los}
\langle v_{\rm los}^2(R)\rangle = \frac{2}{\mu} \int _R^{\infty} \nu(r) \left[\left(1 - \frac{R^2}{r^2}\right) \langle v_r^2\rangle + 
\frac{1}{2} \frac{R^2}{r^2} \langle v_t^2\rangle \right] \frac{r {\rm d}r}{(r^2 - R^2)^{1/2}}, 
\end{equation}
\begin{eqnarray}
\langle v_{\rm los}^4(R)\rangle = \frac{2}{\mu} \int _R^{\infty} \!\! \nu(r) \left[\left(1 - \frac{R^2}{r^2}\right)^2 \!\! \langle v_r^4\rangle + 
3 \frac{R^2}{r^2}(r^2 - R^2) \langle v_r^2 v_t^2\rangle + 
\frac{3}{8} \frac{R^4}{r^4} \langle v_t^4\rangle \right] \nonumber \\
\times \frac{r {\rm d}r}{(r^2 - R^2)^{1/2}}.
\end{eqnarray}
Here $R$ denotes the projected radial distance. Expressed in terms of 
the anisotropy $\beta$ these equations take the form
\begin{equation}
\label{eq:sigma_los-beta}
\langle v_{\rm los}^2(R)\rangle = \frac{2}{\mu} \int _R^{\infty} \nu(r) \left(1 - \beta \frac{R^2}{r^2}\right) \langle v_r^2\rangle  
\frac{r {\rm d}r}{(r^2 - R^2)^{1/2}}, 
\end{equation}
\begin{equation}
\label{eq:vlos4}
\langle v_{\rm los}^4(R)\rangle = \frac{2}{\mu} \int _R^{\infty} \nu(r) \langle v_r^4\rangle g(r,R,\beta)
\frac{r {\rm d}r}{(r^2 - R^2)^{1/2}}, 
\end{equation}
where
\begin{equation}
g(r,R,\beta) = 1 - 2 \beta \frac{R^2}{r^2} + \beta (1 + \beta)/2 \frac{R^4}{r^4}, 
\end{equation}
\citep{Lokas-Mamon2003}. Eq.~(\ref{eq:vlos4}) is valid for the specific form of the distribution function that
leads to a constant anisotropy, while Eq.~(\ref{eq:sigma_los-beta}) is more general. 

In the recent past, Jeans modeling has been the most frequently used
method to estimate the mass content of dSphs
\citep{Lokas2001,Kleyna2001,Koch2007,gilmore2007,Walker2007,battaglia2008}. For
simplicity, many of the works assumed a constant anisotropy, and
typically only the second moment is fit using the Jeans equation
(although see below). The first modeling attempts already showed that
mass following light models could not fit the relatively flat velocity
dispersion profiles observed, and that extended dark matter halos
were needed, for example in the case of Draco \citep{Kleyna2001}.

More recently, the focus has shifted to the type of dark matter halos
that could host dSphs. For example, \citet{gilmore2007} assumed the
velocity ellipsoid to be isotropic ($\beta = 0$), a cored light
surface density distribution and a flat (inner) l.o.s.\ velocity dispersion
profile, and found that dSphs could be embedded in cored or cuspy dark
matter halos (but shallower than the singular isothermal sphere).
\citet{Walker2007} assume NFW profiles and constant anisotropy
together with an exponentially declining surface brightness distribution. These authors fit the
total mass $M_{\rm vir}$ and (constant) anisotropy $\beta$, and assume
a particular value for the concentration from the virial mass-concentration relationship found
in cosmological N-body simulations \citep[e.g.][]{bullock2001,maccio2007}.  It
is important to stress that a quantity such as the total mass is not
well constrained, but what is better constrained is the mass within a
given radius (within the region spanned by the dataset). Such a
quantity is less sensitive to the functional form of the density
profile, and therefore preferable. Although the virial mass may be
considered just another (free) parameter of the fit, its meaning as
representing the total mass of the system is actually an
extrapolation.

\citet{walker2009uni,Walker2010} have extended the modeling of their
sample of dSphs to allow for more general forms of the density profile
of the dark matter ($\rho \propto 1/(x^{\gamma}( 1 + x^\kappa)^{(3 - \gamma)/\kappa})$, with $\gamma, \kappa \ge 0$, i.e. as in Eq.~(\ref{eq:rho_general}) with an outer slope $\alpha = 3$), 
while still assuming
constant anisotropy. They use a Monte Carlo Markov Chain (MCMC) method
to explore the space of parameters and find the best fit models. The results are
shown in Fig.~\ref{fig:walker2009}. An
interesting finding is that they can strongly constrain the
mass at the projected half-light radius $r_{\rm half}$ (the projected radius
enclosing half of the total luminosity). Therefore, they also derive
the circular velocity at $r_{\rm half}$, $V_{\rm half}$, and hence
place lower limits on $V_{\rm max}$. This is only a lower limit 
because there is a degeneracy between $V_{\rm max}$ and the scale
radius $r_s$ of NFW profiles, as there are many such profiles
consistent with a given measurement of $V_{\rm half}$
\citep{penarrubia2008}\footnote{However, this degeneracy can be broken
  by measuring the velocity dispersion profile over a large extent in
  radius, as shown by \citet{Breddels2013}. There is a second, more
  difficult to break degeneracy between the slope/functional form of
  the density profile and $r_s$.}.


\citet{lokas2009} performed Jeans modeling of Car, Fnx, Sext and Scl using the 2nd and 4th
moments of the l.o.s. velocity distribution.  As explained in Sec.~\ref{sec:secondm} an important difference with
work by other authors is the procedure to deal with interloper
removal, which leads to velocity dispersion profiles that decrease
with radius. As a consequence, \citet{lokas2009} finds systematically lower masses than 
other authors, and that models in which mass follows
light can in fact, fit the derived observables.  However, the $M/L$ derived are still much greater than expected 
from standard stellar populations (generally much greater than 10, see their Table 2). In the case of Draco,
\citet{lokas2005} found that the anisotropy was mildly tangential for
a model with a $r^{-1}$ density profile (with an exponential
cut-off). This is consistent with the Jeans model by
\citet{walker2009uni} and also with \citet{Jardel2012} orbital based Schwarzschild
model of the system (within $\sim 1 \sigma$, see below).  Although the
use of the 4th moment leads to a better constraint on the model
parameters, its effect is relatively minor, and the solutions found
are rather similar to those in which only the second moment is
used. The reason for this may be attributed to the fact that the
uncertainties on the measured kurtosis are large (samples are still
small to measure moments very reliably), and that the differences with
a Gaussian-like velocity distribution are not very big, in which
case the 2nd moment suffices to characterize the LOSVD, as discussed in Sec.~\ref{sec:higherm}.

\subsubsection{Jeans equations for axisymmetric systems}

The light distribution of dwarf spheroidals is not really spherically
symmetric, nor are the shapes of dark matter halos predicted by
$\Lambda$CDM simulations, so Jeans modeling as described in the
previous section is not necessarily justified. The next natural degree of
complexity in the modeling of dSphs is to allow for axisymmetry. Although one might
attempt to use fully triaxial models, this might not be necessary given that
there is no clear evidence of twisting in the isophotes of dSphs \citep[except for Car, which is likely to be
affected by tides, see --][and references therein]{battaglia2012car}. Furthemore, 
the analysis of the shapes of dark matter satellites in Milky Way-like simulations \citep{Kuhlen2007,
Vera2013}, indicate that these are not strongly triaxial (with axis ratios at $\sim 1$~kpc derived from the inertia tensor 
of $\langle c/a \rangle \sim 0.6$ and $\langle b/a \rangle \sim 0.75$, 
and $\langle c/a \rangle \sim 0.8$ and $\langle b/a \rangle \sim 0.9$ farther out). 

In the axisymmetric case, we take the velocity moments of Eq.(\ref{eq:boltzmann}) in the directions 
$R$, $\phi$ and $z$. This results in the following equations respectively
\begin{equation}
\frac{\partial \nu \langle v_R^2 \rangle}{\partial R} + \frac{\partial (\nu \langle v_R v_z \rangle)}{\partial z} + 
\nu \left( \frac{\langle v_R^2 \rangle - \langle v_\phi^2\rangle}{R} + \frac{\partial \Phi}{\partial R} \right) = 0,
\end{equation}

\begin{equation}
\frac{1}{R}\frac{\partial (R \nu \langle v_R v_z \rangle)}{\partial R} + \frac{\partial \nu \langle v_z^2 \rangle}{\partial z} + 
 \nu \frac{\partial \Phi}{\partial z} = 0,
\end{equation}
and
\begin{equation}
\frac{1}{R^2}\frac{\partial (R^2 \nu \langle v_R v_\phi \rangle)}{\partial R} + \frac{\partial \nu \langle v_z v_\phi \rangle}{\partial z} = 0.
\end{equation}
relating the intrinsic properties of the system.
As in the spherical case, this set of equations is not closed,
unless we make assumptions about the shape of the velocity
ellipsoid, or the form of the distribution function. A typical consideration is to assume that the distribution function
is of the form $f(E,L_z)$, in which case the mixed moments vanish, the last equation is trivially satisfied, and 
$\langle v_z^2 \rangle  = \langle v_R^2 \rangle$. In that case the other two equations reduce to 
\begin{equation}
\frac{\partial \nu \langle v_z^2 \rangle}{\partial z} +  \nu \frac{\partial \Phi}{\partial z} = 0,
\end{equation}
and 
\begin{equation}
\frac{\partial \nu \langle v_R^2 \rangle}{\partial R} + 
\nu \left( \frac{\langle v_R^2 \rangle - \langle v_\phi^2\rangle}{R} + \frac{\partial \Phi}{\partial R} \right) = 0.
\end{equation}

More generally, the velocity dispersions in the radial $R$ and
vertical $z$-directions will not be identical \citep[][and
references therein]{Cappellari2008}, but we might express $\langle
v_R^2 \rangle = b \langle v_z^2 \rangle$, where $b$ is a
constant. \citet{Vera2013} have shown the $\Lambda$CDM subhalos have
constant $\beta_z = 1 - \langle v_z^2 \rangle /\langle v_R^2 \rangle$ along the minor axis, but that there exists a weak
trend as function of distance along the major axis, with $\beta_z \sim
+0.2$ to $-0.2$ from the center to the outskirts, implying that the
simple assumption by \citet{Cappellari2008} is not significantly violated.

Naturally, these are intrinsic quantities while  we
only have projected moments for the dSphs at our disposal. The equations relating these
are given in \cite{Cappellari2008} and \cite{Hayashi2012} for the
cases discussed by these authors. Clearly one of the uncertainties in
the projection is the inclination of the object with respect to the
observer, and this is therefore generally an outcome of the modeling
procedure. Another assumption often made is that the density of the
tracer has the same orientation and symmetry as that of the dark
matter halo. For dSph galaxies, this is justifiable since this is the
dominant contributor to the gravitational potential and the stars are
simply tracers.

\cite{Hayashi2012} show how the line of sight velocity profiles vary
when objects are observed along the major and minor axis for various
density profiles and flattenings in the light distribution. They find
that the effect of a more flattened stellar system while keeping the
dark matter halo shape fixed, is to produce wavy features in the
l.o.s. velocity dispersion profile along the major axis, especially in
the central regions (see their Fig.~2), and that the shapes are also
different along the minor axis, naturally falling off faster with
radius. The LOSVD amplitude and exact shape change depending on the
dark matter density profile. These authors have also measured the
velocity dispersion profiles for Car, Fnx, Sext, Draco, Leo I and Scl,
and there is only a clear difference along the major and minor axis of
the light distribution for Car and possibly Sext at large
radii. From their modeling the authors conclude that the shapes of the
dark halos are very flattened ($Q \sim 0.3 - 0.4$) for most of their
dSphs, much more than expected in $\Lambda$CDM, and also significantly
more than the light itself ($q \sim 0.65 - 0.8$). This could be
related to the assumption of semi-isotropy ($\langle v_z^2 \rangle =
\langle v_R^2 \rangle$). There is a strong degeneracy between
$\beta_z$ and $Q$, since a flattening of the velocity ellipsoid
$\beta_z > 0$ has a very similar effect to a flattening of the halo
$Q<1$. Moreover a very small change in $\beta_z$ anisotropy can mimic
a major change in the halo shape, as shown in Figure
\ref{fig:cappellari}.  It would clearly be valuable to apply the more
general modeling by \citet{Cappellari2008} on the dSphs to establish
the reliability/confidence of the conclusions.


Another interesting result from \cite{Hayashi2012} is that 
the total mass of the dSphs enclosed within a spheroid with major-axis length of 300 pc varies from 
$10^6 - 10^7 M_\odot$, i.e. these masses are lower than those estimated from spherical models by a factor that
is roughly proportional to the flattening $Q$ \citep[estimated by][]{Hayashi2012} .

\subsubsection{Other interesting results based on the Jeans equations}

As discussed in previous sections, modeling using the spherical Jeans
equations requires assumptions on the functional form of the
anisotropy and of the density profile of the dark matter halo of the
system. The solution to e.g. Eq.~(\ref{eq:sigma_los}) then gives us
the parameters of the profile (e.g. a mass/density and scale radius,
and an anisotropy).  More generally, also the shape of the density profile
may be allowed to vary, as in e.g.  \cite{Strigari2007,walker2009uni}.

\paragraph{A common mass scale?} In two thorough studies,
\citet{Strigari2007,Strigari2008} used MCMC numerical methods to
explore a large range of models for the dark halos hosting dSphs. These
authors found that of all the parameters describing the model, the mass
within 300 pc was a robust and well determined quantity, that was
roughly independent of anisotropy or shape parameters. In an immediate
application of this result, \citet{Strigari2008} found that most dSphs must be
embedded in dark matter halos of similar mass within this radius $M_{300} \sim
1 - 2 \times 10^7 M_\odot$, despite the fact that they span several orders of magnitude in
luminosity.  Although these results have
been refined, especially for the ultrafaint dSphs, which do not even extend up to
300 pc (and hence this $M_{300}$ is an extrapolation), in general the
$M_{300}$ is confirmed to be a very weak function of luminosity $M_{300} \propto
L^{0.03 \pm 0.03}$ \citep{Rashkov2012}.

\paragraph{Robust measurement of $M(r_{1/2})$}
The virial theorem as well as the Jeans equations actually offer a
plausible explanation for why the mass at a particular radius may be
estimated reliably from the LOSVD only. The virial theorem tells us
that $M_{\rm tot} = \sigma_{\rm tot}^2 r_g/G$ where $r_g$ is the gravitational
radius of the system \citep[see Eq.~(2.42) of][]{BT2008}.  On the other
hand, \cite{Wolf2010} have shown analytically using the Jeans
equation, that at the radius at which the slope of the stellar density
profile $d\log \nu/d \log r = - \gamma_* = -3$, the mass is very well
constrained independently of the anisotropy of the system. Thus
\begin{equation}
\label{eq:mass_wolf}
M(r_{-3}) = 3 \frac{\langle \sigma_{\rm los}^2 \rangle r_{-3}}{G}
\end{equation}
for a system with a flat velocity dispersion profile. Since most of
the dSphs have such relatively flat profiles, \cite{Wolf2010} and also
\cite{walker2009uni,Walker2010} in their MCMC analysis of the Jeans
equation have been able to confirm this analytic result. In general,
however, instead of estimating the radius $r_{-3}$, \cite{Wolf2010}
use the half-light radius $r_{1/2}$ since the two are very similar for most
profiles used to model the light distribution in dSphs. Note that this
is the 3D radius containing half of the total luminosity of the
system, and not the effective radius obtained from the surface
brightness profiles nor the 2D projected radius containing half of the luminosity, $r_{\rm half}$ in \citet{Walker2010}.

These relations are also useful for ultrafaint dSphs, provided these
systems are in dynamical equilibrium. The
sample sizes for most of these systems are too sparse to warrant a full dynamical model so
general scalings as those just described may be more
useful. See also
\citet{An2011} for more information on the theory of virial mass estimators.

\paragraph{General constraints on the df}

Not every solution to the Jeans equation has an associated distribution function
that is physical, i.e. positive everywhere. This is why it is important to find 
additional conditions that can help identify when the assumptions made to solve the 
Jeans Equations will lead to plausible (physical) solutions.

\cite{An2006,An2009} and \cite{Evans2009} use the Jeans equations to
explore the asymptotic relations between the anisotropy $\beta$, the
logarithmic slope of the light distribution $\gamma_*$ and that of the
underlying dark matter density profile near the center of a spherical
system $\gamma_{DM}$. They show that, if the tracer population is
embedded in a spherical dark halo that is shallower than the singular
isothermal sphere ($\gamma_{DM} < 2$) in the center, a finite central
velocity dispersion $\sigma_{r,0}$ implies a relation between the
central value of the logarithmic slope of the tracers $\gamma_{*,0}$ and the velocity
anisotropy at the center $\beta_0$, namely $\gamma_{*,0} = 2 \beta_0$. However, it is also
possible that the system is dynamically cold at the center
(i.e. $\sigma_{r,0} = 0$), in which case the condition is
$\gamma_{*,0} > 2 \beta_0$.  This theorem highlights that care is required
in the interpretation of results based on assumptions such as isotropy
and spherical symmetry.

\cite{Ciotti2010} showed that there may be a more general relation
that should hold at all radii, which is that $\gamma_* \ge 2
\beta$. This may be seen to be related to the positivity of the mass (Eq.~\ref{eq:jeans_mr}),
as $ M \propto \gamma_* - 2 \beta - \alpha \ge 0$, where $ \alpha = d
\log \sigma_r^2/d \log r$.  \cite{Ciotti2010} have demonstrated the above
relation holds for particular forms of the distribution function
(namely those in which the augmented density is a separable function
of radius and potential, see their Eqs.(1 - 6) for more details), but the more general inequality (including
$\alpha$) should be always true.

However, it should also be born in
mind that this analysis applies to intrinsic quantities implying for example, that
even if $\sigma_{r,0} = 0$, $\sigma_{\rm los}$ can still be finite at the center,
and hence the theorem is, although correct, less powerful in
predicting the orbital behavior at the center. Furthermore,  a surface
brightness profile might have a very shallow cusp \citep[as considered in][]{Strigari2010}, 
in which case the velocity ellipsoid need not be isotropic. Another example of the limitation of projected quantities
on the power of the theorem is given example below.

Consider the following distribution function
\begin{equation}
\label{eq:df_mb}
 f(E, L) \propto (-E)^4 (-\Phi_0-(-E))^{-3} L,
\end{equation}
where $-\Phi_0$ is the potential energy at $r=0$. In this model, the
anisotropy is constant, and for this particular example we have fixed it at
$\beta=-0.5$. Let us now assume that the gravitational potential is
of the NFW form:
\begin{equation}
 \Phi(r) = - \Phi_0 \frac{\log (1+r/a)}{r/a},
\end{equation}
here $\Phi_0 = 4 \pi G \rho_0 a^3$, and for simplicity, we have 
neglected the contribution to the potential by the stars.  The light
distribution may be obtained by integrating this distribution function
with respect to velocity space. The resulting light distribution has a
logarithmic slope $\gamma_{*,0} = 0$ at the center as shown in the top
panel of Fig.~\ref{fig:df_breddels}. The 2nd moments for the radial
and tangential direction are plotted in the central panel of the
figure. Note that this model has a centrally vanishing radial velocity
dispersion.  On the other hand, the l.o.s. velocity dispersion profile
is shown in the bottom of the figure, and is relatively flat and
non-zero at all projected radii. This distribution function is
positive everywhere, it leads to a velocity dispersion profile not too
different from that observed for dSphs, but its light distribution at
the center has $\gamma_{*,0} = 0$, while the anisotropy is negative $\beta
= -0.5$ \citep[so that in this case, $\gamma_{*,0} \ne 2 \beta_0$, but the
more general result of] [does hold, as expected]{Ciotti2010}.


\subsection{Modeling through distribution functions}

As stated previously, a solution to the Jeans equation is not necessarily physical since
there is no guarantee that a distribution function will exist that is
positive definite everywhere. This is one of the reasons why several authors have
attempted to model directly the distribution function itself.

Dejonghe \& Merrit (1992) have studied the issue of how the projected
velocity distribution as a function of position
$f_{\rm los}(v_{\rm los},r_{\rm los})$ for a spherical system constrains the
distribution function and gravitational potential. They show that if the form of the
spherical potential is specified, then $f(E,L)$ is uniquely determined
by $f_{\rm los}(v_{\rm los},r_{\rm los})$.  However, if the spherical potential is
not known, they argue that there will be a family of possible
potentials, but only those that lead to a df
that is positive everywhere would be allowed, and not every potential will permit that.

\citet{Merritt-Saha1993} explore the problem of inferring the
gravitational potential of a spherical system from measurements of the
l.o.s. velocity and positions for individual stars (or galaxies, in
their case). They assume that the distribution function may
be expressed as a polynomial expansion: $f(E,L) = \sum_{m,n}
c_{m,n}f_{m,n}$ where $f_{m,n} = (-E)^{n-1/2} L^{2m}$, hence this distribution function is
non-parametric. To determine the properties of the potential, however,
a few parametric forms are considered.  Thus from a discrete set of
velocities of galaxies in the Coma cluster, they find best solutions
in a maximum likelihood sense. These authors estimate that meaningful
constraints are possible with datasets containing $\sim 1000$ objects.

\cite{wu2006} \citep[see also][]{merritt1993} take an even more
general form for the distribution function, namely they divide the
$(E, L)$ space into $N_E \times N_L$ bins, and construct a set of
top-hat basis functions, $h_{mn} = 1/V_{mn}$ where $V_{mn} = \int_{mn} d^3x
d^3v$ is the phase-space volumen associated to bin $mn$. Thus, $f(E,L)
= \sum_{mn} w_{mn} h_{mn}$, and the task consists in finding the
weights $w_{mn}$ that fit the observables after assuming a specific gravitational
potential.  \cite{wu2006} use this technique to infer the mass of M87
from the motions of its globular clusters.  These approaches are very
powerful as they use maximally the datasets, without turning to
moments to characterize the LOSVDs, and are also free of assumptions
regarding the distribution function.  It would be very valuable to apply
such methods to the modeling of dSphs in the Local Group.

\cite{wilkinson2002} introduce a family of anisotropic distribution
functions for spherical systems, in which the dominant gravitational
potential is cored and parametrized as $v_c^2 = v_0^2 r^2/(1 +
r^2)^{1 + \delta/2}$. For different values of the characteristic
parameters ($ -2 \ge \delta \ge 1$), this leads to flat or declining
rotation curves. The velocity ellipsoid is isotropic in the center,
and may become radially or tangentially anisotropic at intermediate
radii, while at large distances it is constant. The advantage of this
family of distribution functions is that the expressions for the
various moments (including the 2nd and 4th) are analytic, and depend
only on the parameters of the distribution function.  This means, that in principle, 
these characteristic parameters could be retrieved directly through comparison to observations. They also compute
the projected (observable) quantities for different values of the
parameters. The resulting l.o.s. velocity dispersion profiles (see
their Fig. 3) can be flat, rising or declining depending on the distribution function. In
\citet{Kleyna2002draco} they have applied this modeling to a dataset
for Draco with $\sim 160$ member stars, and found that the
system is best fit by a slightly tangentially anisotropic ellipsoid
and with a halo that falls off more slowly than a flat rotation curve
model ($v_c \propto r^{0.17}$), while they are also able to rule out a
mass-follows-light model and an extended harmonic core with $3\sigma$ confidence.

\subsection{Schwarzschild modeling}

Schwarzschild modeling is by now a traditional technique to derive the
mass distribution, especially in elliptical galaxies, from integrated
light spectroscopy. It was initially developed in the 1980s \citep{Schwarzschild1979,Richstone1984}, and used
extensively to derive $M/L$ and black hole masses in the 1990s and the
2000s \citep[e.g.][]{Rix1997,vanderMarel1998} where it was extended to allow for axisymmetric models, and
even triaxiality \citep{vdBosch2008}. The basic idea of the method is that the 
building blocks of galaxies are orbits, and through the right orbital
superposition it is possible to match the light and kinematic
distributions observed. 

Therefore the method consists in assuming a specific gravitational
potential, calculating the observables predicted for each orbit, and
then weighting the orbits (with non-negative weights) to obtain a
model that fits the observed data in a $\chi^2$ sense. This approach
guarantees that the distribution function obtained (which is reflected in the
orbit weights) is non- negative. The fitting procedure thus allows the determination
of the characteristic parameters of the best fit model for a
specific gravitational potential.  If one wishes to test
different functional forms for the gravitational potential, then new orbit
libraries need to be built, and the fitting procedure is
repeated.  The advantage of this method is that it does not make
assumptions about the form of the anisotropy or the distribution function (rather these are
an outcome of the model), and therefore
it is less biased than some of the modeling techniques 
described above. Naturally, it is less flexible in the sense that it
is more computationally intensive/expensive, and hence it is possible
to explore a smaller variety of gravitational potentials, than for
example, through Jeans models.

Despite the vast history, this method has not been applied
systematically to the dynamical modeling of dwarf galaxies until very
recently. For example, \citet{Breddels2012} have used this technique
to model the Scl dSph in the assumption of spherical
symmetry. They fit the light, 2nd and 4th projected moments. They have found, in
agreement with other authors, similar estimates for the mass of Scl
within 1 kpc, but perhaps more interestingly, and for the first time,
they measured the velocity anisotropy of the system to be tangential
and relatively flat with radius. Furthermore, they are able to rule out very steep
density cusps ($\gamma_{DM} > 1.5$), although they cannot distinguish
statistically an NFW (or shallower cusp) from a $\gamma_{DM} = 0$ profile  for the
dark matter.

Still considering spherical models, \citet{Breddels2013} also fitted
the velocity dispersion and kurtosis profiles of Fnx, Sext and Car
using this technique. They performed a Bayesian comparison of a suite
of different density profiles (2 Einasto, NFW, 4 cored, with different
slopes). They found that most models are statistically
indistinguishable. However, they show that it is unlikely (with odds
of 1:10) that all these dSphs are all embedded in cored profiles where the
density falls off steeply as $\rho(r) \propto 1/(1 +
(r/r_s)^2)^{\kappa/2}$, where $\kappa = 3,4$. What is also very
interesting from their work, is that they show that for each of the
systems, the mass distribution from a radius slightly below $\sim
r_{1/2}$ up to the last measured kinematic data point is the same for
all models. This means that even though one cannot reliably distinguish
the inner shape or slope of the dark matter halos, they can certainly
state what the mass distribution is over a large range of radii, and therefore also
derive the slope of the density profile at some intermediate point, as
shown in Fig.~\ref{fig:breddels_fnx} for the Fnx dSph. Furthermore,
since all these models effectively have the same mass distribution,
they also have the same anisotropy profile, and hence these authors
have essentially measured the anisotropies for these systems. These
are found to be relatively flat and mildly tangential 
($\beta \gtrsim -0.5$). This must be telling us something about the
formation and evolution of these systems since it must be indicative
of some amount of circular motions present in the system (otherwise,
for a radial collapse, one would expect a radially anisotropic
ellipsoid).


\cite{Jardel2012} presented three-integral, Schwarzschild models of
Fnx that take into account the non-spherical light distribution of
this galaxy, although embedded in a spherical dark matter halo. These
authors have tested a cored profile $\rho \propto (3 r_c^2 +
r^2)/(r_c^2 + r^2)^2$ and the NFW model. They find that the cored model is
strongly favored, and that the velocity ellipsoid is mildly radially
anisotropic. Their mass for Fnx $M(R_e) = 3.9 ^{+0.46}_{-0.11} \times
10^7 M_\odot$ is somewhat smaller than what the estimators by
\citet{Wolf2010} or \citet{walker2009uni,Walker2010} would
predict. \cite{Jardel2012} argue that this might be related to the
fact that those estimators have been established (and shown to be
independent of anisotropy) for spherical models. Another difference
might lie in that the amplitude of the line-of-sight velocity
dispersion profile they derive for Fnx is somewhat lower than that
shown, for example, in Fig.~\ref{fig:breddels_fnx}.

\cite{Jardel2013}, return to spherical models, but assume
that the density profile for the dark matter is non-parametric. They model Draco in this way, and
find that the preferred model is a power-law, with a slope quite similar to the NFW, that is
$\gamma_{DM} = 1$ for $ 20 \le r \le 700$ pc, and that the velocity
ellipsoid is radial. Note that, in comparison to \citet{wilkinson2002}, \citet{Jardel2013}
have allowed greater freedom in the form of the density profile
(and have not forced cored models), and hence their results are potentially more robust.

\subsection{Made to measure}

The Made-to-Measure (M2M) is a numerical method that integrates the
orbits of test particles in a gravitational potential in order to
reproduce a given set of observables \citep{Syer1996}. Particles have
associated weights, which themselves follow equations of motion. The
system is evolved in time until a satisfactory solution has been found.
The gravitational potential may be specified or determined
self-consistently, and the resulting distribution function is
completely non-parametric, and determined by the final particle's
configuration that satisfies the observational constraints. The method
can be used to model individual measurements or moments of a LOSVD
\citep[as with N-MAGIC in][]{delorenzi2007}.

\citet{Long2010} have
modeled Draco using the data from \citet{Kleyna2002draco} and assumed
an isotropic velocity ellipsoid, and the same type of cored potentials
as \citet{wilkinson2002}. The best fit model has asymptotic slope for
the squared circular velocity $v_c^2$ of $\delta =
-0.90^{+0.36}_{-0.35}$, while for the mass within three core radii
they find $9.7 \pm 2.3 \times 10^7 M_\odot$, in comparison to
\citet{Kleyna2002draco} who obtain $\delta \sim -0.34$ and a somewhat
smaller mass. \citet{Long2010} attribute this difference to their
assumption of isotropy.

\subsection{Modeling dSphs with composite stellar components}
\label{sec:dyn-multiple}

As discussed in Sec.~\ref{sec:multiple} several dSphs host multiple stellar
chemo-dynamical components.  Since these components are embedded in the same
gravitational potential, they allow one to place more stringent
constraints on the properties of this potential, since e.g. each of
the component has to satisfy the Jeans equations independently. In
practice this means that there are fewer free parameters since each
component will follow its own distribution function entering the
left-hand-side of Eq.~(\ref{eq:jeans_sph}), but the right-hand-side will be the same,
thereby effectively leading to a reduction in the number of degrees of
freedom. 

This idea was first exploited by \citet{battaglia2008}, who
modeled Scl using two components, a metal-rich centrally
concentrated, and a metal-poor hot and extended, both embedded in a
dark matter halo. These authors found, using Jeans models, that the
metal-poor component was better fit with a nearly flat anisotropy profile,
while the metal-rich one, because of its rapidly falling velocity dispersion
profile (see Fig.~\ref{fig:sigmascl}), required a radially anisotropic ellipsoid. They found that
cored models provided better fits but that NFW models could not be
ruled out.

Another use of the composite stellar components was put forward by
\citet{walker2011} to infer the slope of the dark matter profile. These
authors argue that one might consider Eq.~(\ref{eq:mass_wolf}) for
each component separately, so that the mass of the host halo is
constrained at the half-mass radius of each component
independently. This then leads to two measurements of the mass at two
different radii, and hence to a slope. They have performed many tests
of their method, whose basic assumption is that the l.o.s. velocity
profiles are flat, and found that their results are relatively robust
to such (and other) assumptions, although systematic uncertainties
affect the masses at $r_{\rm half}$ which depend on the density
profile of the halo and the degree of embedding of the stars). These
authors define
\begin{equation}
\Gamma = \Delta \log M/\Delta \log r =  \frac{\log(M_{h,2}/M_{h,1})}{\log(r_{h,2}/r_{h,1})} \sim 1 + 
\frac{\log(\sigma^2_{2}/\sigma^2_{1})}{\log(r_{h,2}/r_{h,1})},
\end{equation}
where $r_{h,{\rm pop}}$ and $M_{h,{\rm pop}}$ refer to the projected
half-light radius and the mass at this point, while $\sigma^2_{\rm
  pop}$ is the global velocity dispersion that characterizes the
population, and where ${\rm pop} = 1, 2$, i.e. metal-rich or
metal-poor components. In the limit of $r \rightarrow 0$, then $d \log M/d
\log r = 3 - \gamma_{DM}$ where $\gamma_{DM}$ is the central value of
the slope of the dark matter density profile. Since $d \log M/d \log
r$ decreases as $r$ increases for any reasonable density profile, this implies
that $3 - \gamma_{DM}> \Gamma$, or alternatively that $\gamma_{DM} < 3
- \Gamma$, as the slope $\Gamma$ is measured at a finite distance from
the center.  \citet{walker2011} find $\Gamma =
2.61^{+0.43}_{-0.37}$ for Fnx, while for Scl $\Gamma =
2.95^{+0.51}_{-0.39}$. This thus implies that NFW-like profiles
($\gamma_{DM} = 1$) would be ruled out at significance levels $\gtrsim
96\%$ and $\gtrsim 99\%$ respectively for these systems. These results
are much more stringent than any of the previously reported findings
by other authors, where typically both profiles are consistent with
the data.

More recently, \cite{Amorisco2012} have modeled the two populations in
Scl using Michie-King models.  These are isotropic in the center
and become radially anisotropic in the outskirts. The validity of
these assumptions for the velocity ellipsoid is taken from their
analysis of the shape of the l.o.s. velocity distributions of Scl
in \citet{amorisco2012lp}, whose estimates of the 4th moment would
suggest a radially anisotropic ellipsoid \citep[see however][who find
a kurtosis profile that is consistent with tangential anisotropy. As discussed in Sec.~\ref{sec:higherm}, the
difference is possibly related to the treatment of
foreground/membership determination]{Breddels2012}. Under these
assumptions for the velocity ellipsoid (or the distribution function)
these authors find that cored mass distributions are preferred over
cusped ones. Given the uncertainties in the measurements of the 4th
moments, this result could be related to the assumed shape of
$\beta(r)$ rather than necessarily reflect the underlying mass
distribution.

\cite{Agnello2012} use the projected virial theorem and argue
that the two populations in Scl should satisfy independently the virial
theorem,
\begin{equation}
\frac{K_{los,1}}{K_{los,2}} = \frac{W_{los,1}}{W_{los,2}} 
\end{equation}
from which they obtain the relation
\begin{equation}
\left(\frac{\sigma_{0,1}}{\sigma_{0,2}}\right)^2 > 2 \left(\frac{R_{h,1}}{R_{h,2}}\right),
\end{equation}
if the stars follow Plummer profiles and are embedded in NFW halos. Given their estimates of
these various observables, \cite{Agnello2012} conclude that no NFW halo can be compatible with the energetics 
of the two populations. Because the two populations should co-exist in virial equilibrium, the authors argue that 
this implies that the
dark halo must be cored, and they estimate its size to be $\sim 120$~pc.

The results presented in this section all argue that the modeling using two (or multiple)
components disfavor NFW/cuspy profiles for dSphs, at least
for Fnx and Scl.  It is striking that the consideration of two
components in dynamical equilibrium point all in the same
direction. It would be important to confirm these results using
fully-fleshed non-parametric methods, such as Schwarzschild or
Made-to-Measure, that explore the presence of multiple populations and
remove some of the (systematic) uncertainties in the use of global
scaling relations. It would also be desirable to understand the extent down to 
which these systems' properties are better described using a few independent 
components, rather than to assume that the properties of
the stars change gradually  throughout the system, and specifically how these assumptions affect the dynamical models and their conclusions.

Figure \ref{fig:breddels_scl} compares the results of various modeling
approaches on Scl. In this figure we have plotted the mass
distribution derived using Schwarzschild models by
\citet{Breddels2013}. We have included here estimates of $M_{300}$ by
\citet{Strigari2008} and \citet{walker2009uni} and at the 3D half-light
radius by \citet{Wolf2010}. These estimates are all consistent with
those obtained by \citet{Breddels2013} which is reassuring. The mass
estimated at 1.8 kpc obtained by \citet{battaglia2008} assuming a
cored density profile but modeling simultaneously metal-rich and
metal-poor populations is also shown (open black circle). It is on the
upper side of the curves, but is consistent within error bars, and is
beyond the region where the mass profiles are indistinguishable, so
this mass estimate is likely to be more model dependent.  Finally, the
two estimates of the mass derived by \citet{walker2011} are shown as
diamonds in this Figure. These two estimates of the mass at the
projected half-light radius of the metal-rich and metal-poor
components of Scl, appear to be somewhat larger than what is found by
\citet{Breddels2013}.  This is consistent with the systematic
uncertainties that \citet{walker2011} reported from their Monte Carlo
simulations. However, we notice that the mass at the half-light radius
of the metal-poor component is more overestimated than that of the
metal-rich one (and even higher than $M_{300}$ or $M_{1/2}$ for
example).  In view of this, it seems plausible that the slope $\Gamma$
that \citet{walker2011} derived could be overestimated, in which case,
cuspy profiles with $\gamma_{DM} > 0$ could still be allowed.

\section{Future directions}
\label{sec:disc}

In the previous sections we have discussed the status of the field,
and have begun to identify directions where more research would be
desirable to understand the properties and dynamics of dSphs. In the
case of the dynamical modeling, as this review reflects, much of this
work has been done assuming that the dSphs are embedded in spherical
dark matter halos (and often, even assuming their light distribution
is approximately spherical). First attempts to veer from this
assumption have been made using the Jeans equations, but as discussed
earlier, these have the limitation of exploring parametric models, and
unfortunately, the results sometimes reflect the hypotheses made. It
is therefore desirable to apply non-parametric modeling, for example
along the lines of \citet[][although these authors still assume the
dark halos are spherical]{Jardel2012}, who have used Schwarzschild
models assuming non-spherical light distributions. In fact, \citet{vdBosch2008} have performed triaxial modeling
of elliptical galaxies, implying that the tools needed for the dSphs
may already have been largely developed \citep[see also][for the
modeling of OmegaCen in the limit of axisymmetry]{vdVen2006}. Another
example, is the M2M modeling of the Galactic bar by
\citet{Long2013}. Unfortunately the sample sizes for the most
classical dwarfs may be still too small to warrant such sophisticated
approaches, with the possible exceptions of Sculptor and
Fornax. Nonetheless, it is worthwhile establishing what are
the degeneracies/limitations in the modeling, and to what extent 
they can be broken by different datasets. This can be addressed by applying
dynamical models on Mock datasets, for example extracted from
N-body simulations or generated from known distribution functions.

Another aspect is the consideration of the discrete nature of the
datasets, which has not been exploited in full. Most works so far 
fit the full LOSVD or its moments, both as function of projected radius $R$, rather than attempting a
full likelihood analysis using the individual measurements (position
on the sky, and l.o.s.\ velocity) for each star in the dataset.  This
is a direction that needs to be exploited further, since binning
always leads to loss of information. First steps have been taken in
the context of Schwarzschild models by \citet{Chaname2008}, or when using
distribution functions to model the dynamics of planetary nebulae or
globular cluster systems around elliptical galaxies by
\cite{Merritt-Saha1993} and \cite{wu2006}. There is also recent work employing
the Jeans equations for modeling the dynamics of galaxies in a cluster
environment \citep{Mamon2013}.

The use of
proper motion measurements of stars in dSphs is another unexplored aspect of the dynamical modeling. The reason is, of course,
that this has been beyond the capabilities of current
instrumentation. However, the situation is likely to change in the
coming years.  For example, it is now possible to constrain the mean
tangential motions of dSphs using the Hubble Space Telescope
\citep[see][and references therein]{Piatek2008}, and these
measurements are likely to be significantly more accurate with the
advent of Gaia\footnote{See {\sf http:\/\/www.rssd.esa.int\/index.php?project=GAIA\&page=Science\_Performance}
for the latest estimates of its performance.}. The internal motions 
may still be just about beyond
reach for individual stars in dSphs. For example for a star at 70 kpc, 
an internal tangential velocity of $v_t \sim 10$~\kms
translates into proper motion of $\mu \sim 30 \mu$as/yr.  For a star
of magnitude $G \sim 17$, the accuracy expected for the Gaia mission
is $\sigma_\mu \sim 36 \mu$as/yr, and hence the internal velocity and
its error will be of comparable magnitude. This implies, however, that
one ought to be able to bin the data to obtain a tangential velocity
curves with reasonable accuracy, as the error on the dispersion is
inversely proportional to the square root of the sample
size. For the UFDs the situation is less clear, as these objects have faint and sparsely populated red giant branches. At these characteristic faint magnitudes Gaia's proper motion accuracies degrade quickly, from $\sim 80 \mu$as/yr at $G = 18$, to $140 \mu$as/yr at $G = 20$. For an object at
a distance of $\sim 40$~kpc, this implies an error in the tangential velocity of
$\sim 15 - 25$~\kms. Although in principle one can reduce this error by binning, 
the sparsely populated RGBs prevent from obtaining the significant gains needed to characterize the internal kinematics of the UFDs. Nonetheless, these measurements clearly will allow the determination of their orbit, as well as
aid in establishing membership and potentially finding extra-tidal stars and streamers. On the other hand, for the brightest stars in the LMC (those with $G
\lesssim 15$), the expected accuracies are in the range $\sim 4 - 14
\mu$as/yr, which at a distance of $\sim 50$~kpc, corresponds to a
tangential velocity error of $\sim 0.1 - 3.3$~km/s, comparable to what can
be obtained nowadays for the l.o.s.\ velocities routinely from the
ground. 

\cite{wilkinson2002} have studied the impact of proper
motion information following the specifications planned for the former
SIM mission, namely 3 -- 6 $\mu$as/yr, which translates into 1-2 km/s
for stars in Draco of magnitude $V \sim 19 - 20$. These authors show
that by adding proper motion information for samples as small as 160
stars, it is possible to obtain accurate estimations of both the
velocity anisotropy and mass slope, and thereby break modeling
degeneracies unambiguously. This is also confirmed by
\citet{Strigari2007b}, who show that, for general dark matter density
and anisotropy profiles, the log slope of the dark matter profile at
about $\sim 2 r_c$ can be measured to within $ \pm 0.2$ if the proper
motions of 200 stars (with tangential velocity errors of $\sim 5$
km/s) are added to the l.o.s.\ velocity measurements. This would allow
to place tighter constraints on the type of dark matter halos hosting
dSphs, and hence possibly also on the nature of dark matter.

Dynamical modeling of dwarf galaxies is not only interesting to
understand the mass distribution of these systems, but it is also
important in terms of evolutionary paths. Numerical simulations of
isolated dwarf galaxies \citep[e.g.][]{Valcke2008,revaz2009} model
these systems as the result of the radial collapse of a gas cloud
embedded in a dark matter halo. As a consequence of the collapse, the
orbital distribution is likely biased to radial orbits. On the other
hand, if there is any amount of angular momentum, as expected in the
context of $\Lambda$CDM, the gas might collapse towards a disk-like
distribution \citep[see][]{Governato2010,Sawala2010,Schroyen2011}, but
this might depend on the amount of merger activity at the time of
formation of most of the stars \citep{Gnedin2006}. The spheroidal-like
morphologies might thus have to be the result of external mechanisms,
and it has been proposed by \cite{mayer2001} that the tidal force
field of the host galaxy (in this case, the Milky Way) can be the
driver of morphological transformations. These authors, and more
recently, \cite{kazantzidis2013} have shown, that the dSphs could be
tidally stirred disky dwarf galaxies. The morphologies, because of the stirring, would
become triaxial, and the systems could lose most of their angular
momentum and end up being largely pressure supported, with nearly
isotropic velocity ellipsoids \citep{Lokas2010stirring}.
\cite{Helmi2012} have proposed another scenario to transform disk-like
dwarf galaxies into spheroidal systems through a major merger (with a
dark satellite) \citep[see also][for a merger between
dwarfs]{Kazantzidis2011merg}. While such events are not necessarily very
common, they might lead to a spheroidal morphology. In either case,
one might expect the orbital distribution to be biased to high-angular
momentum orbits as a reflection of the initial conditions.

Hydrodynamical simulations are also important to understand the interplay between baryons and dark matter, and to place firmer predictions on the expected properties of dark matter halos, such as their shape and density profile \citep{DiCintio2012,Zolotov2012}. Baryonic processes such as feedback are poorly understood, but it is on the scale of dwarf galaxies that we are likely to obtain the largest insights because of the shallowness of their potential wells, and also because  these systems are the easiest to resolve numerically. 

The dynamics of the dwarf galaxies in the Local Group are also of
interest, as it has been proposed that they may be tidal dwarf
systems, and in that case, expected to be devoid of dark matter. This
proposal has been put forward to explain the somewhat peculiar
relatively planar distribution of satellites around the Milky Way
\citep{Metz2009}. More recently, \cite{Ibata2013} has even shown that 
about half of the satellites of M31 appear to be located in thin vast
co-rotating plane, which \citet{Hammer2013} explain in the context of
a gas-rich merger experienced by M31 $\sim 5$ Gyr ago. However, it is
unclear whether the internal dynamics of the dSphs, their extended star
formation histories and complex chemical enrichment evolutionary
paths, and the fact that they have survived until the present day
would support such a scenario. Therefore it is important to do a detailed
comparison between the simulations and the rich datasets that
are already available. Clearly, the impact of the Gaia mission will also be enormous here
because thanks to the accurate proper motion  measurements it will possible to reconstruct the full
orbital history of the satellites around the Milky Way.

The Milky Way system of satellite galaxies is likely to remain the place where 
we will be able to gather the most accurate kinematic data-sets for resolved stars. However it would be desirable to 
expand the sample of dwarf galaxies to explore different environments. For isolated 
dwarf galaxies, we can exclude environmental effects such as tidal and ram-pressure stripping as 
major drivers of their present stellar and DM properties. On the other hand, probing the 
satellite system of other large spirals such as M~31 would highlight possible differences due to mass of the host 
and assembly history of the whole system, and provide larger statistics. 

There are some works that have started to gather large samples of RGB stars for kinematic studies in 
Local Group dwarf galaxies, beyond the MW system \citep[e.g. the isolated WLM at $\sim$1~Mpc 
observed both with VLT/FORS and Keck/DEIMOS][]{leaman2012}, although this 
implies a significant investment of telescope time per galaxy. For the closest ones 
(for example, the M~31 satellites found on the close side to the MW), if populous enough, it is possible to 
gather kinematic samples of comparable size and quality as those for MW dSphs \citep[e.g. as for And~II, 
observed with Keck/DEIMOS][]{ho2012}, suitable for a full dynamical modeling. For the most remote and intrinsically faint 
Local Group dwarfs, it is already challenging to gather samples of a few dozens l.o.s. velocities \citep[e.g. VV124][]{kirby2012}, 
which permits to apply the scaling relations discussed in Sect.~3. The challenge can be easily understood 
if we consider that the tip of the RGB at the distance of VV~124 (1.1Mpc) is at about $I \sim 21.2$, about 5.5mag fainter 
than for the typical dSph at $\sim$80~kpc. The situation will be greatly improved by the next generation of 30m-40m class 
telescopes, in particular if equipped with spectrographs with multiplex capabilities over a field-of-view of one to a few arcmin 
to encompass a large enough portion of these galaxies. This would also allow targetting much fainter stars in the MW UFDs, and significantly expanding their kinematical datasets. Accurate l.o.s. velocities of large samples of evolved red giant 
stars would become a reality even up to $\sim$4~Mpc \citep[e.g.][and E-ELT Design Reference Mission report\footnote{http://www.eso.org/sci/facilities/eelt/science/drm/drm\_report.pdf}]{evans2013}, opening the exploration of the satellite 
systems of other large spirals such those in the Sculptor group.

\section{Conclusions}
\label{sec:concl}

In the last decade, we have experienced the vast increase in the number and extent of datasets 
with kinematic measurements of stars in the dSphs of the MW. 

The leap forward in sample size and spatial coverage, coupled to exquisite velocity accuracy, due to advent of 
wide-field multi-object spectrographs on the largest telescopes world-wide allowed us to uncover 
velocity gradients of a few km s$^{-1}$ deg$^{-1}$  
\citep[e.g.][]{munoz2006, battaglia2008, walker2008, ho2012} and to establish that the classical dSphs have nearly flat \dispr around 
6-10 \kms. In combination with the  
metallicity information, this has unveiled an unexpected degree of complexity in these small galaxies, such as 
the presence of multiple stellar chemo-dynamical components \citep{tolstoy2004, battaglia2008, battaglia2011}. 
 
These superb datasets have been modeled using a variety of techniques, mostly assuming spherical symmetry. They have confirmed that dSphs are the most dark matter dominated objects known to date, with $M/L \sim 10 - 1000$ in the regions where the stars are located. Their masses at their half-light radii have been accurately measured, and are in the range of $\sim 4 \times 10^{5}$\sm\, for Wilman I to $7.5 \times 10^{7}$\sm\, for Fnx \citep{Wolf2010,walker2009uni,Walker2010}.  These estimates are robust as they do not suffer from uncertainties in the anisotropy of the velocity ellipsoid. Thanks to sophisticated modeling techniques such as Schwarzschild's method, we have begun to discover that the internal motions of dSphs do not deviate strongly from being isotropic, with a slight preference for constant slightly tangentially anisotropic ellipsoids for some of the dSphs \citep{Breddels2013}. This would be in contrast to what has recently been found for some round giant ellipticals, which have radially aligned ellipsoids \citep[e.g.][]{delorenzi2009,morganti2013}. This would suggest that the dwarf and such giant spheroids could have had different formation paths. 

Although the debate concerning whether the central density profile is cusped or cored still remains for these systems, we would like to argue that this dichotomy is {\it pass\'e}, and that the exact value of the slope might depend on the specific history of the baryons in these systems. Furthermore, from an observational perspective, it is very hard to obtain the kinematics of stars truly located near the center of the dSph and not just in projection, to use these to constrain the slope. Nevertheless, we have made substantial progress in demonstrating that it is possible to measure robustly (in a model independent fashion) the shape of the mass distribution $M(r)$ over a distance range of $\sim 1$ kpc in projection \citep{Breddels2013}. These measurements are likely to be more constraining for cosmological simulations that the measurement of the very central slope. 

The next years promise to be exciting in the field because of advances
both on the theory as well as the observational
front. Theoretically, this will be driven by more sophisticated
dynamical modeling techniques, and by the large improvements in
numerical simulations of galaxy formation on the dwarf galaxies
scale. From an observational perspective, larger radial velocity
surveys, and the advent of proper motion information should enable us
to understand the dynamics and evolution of the dSphs in much better
depth.

\section*{Acknowledgments} 
\begin{figure}
\centerline{\includegraphics[scale = 0.75,clip=true]{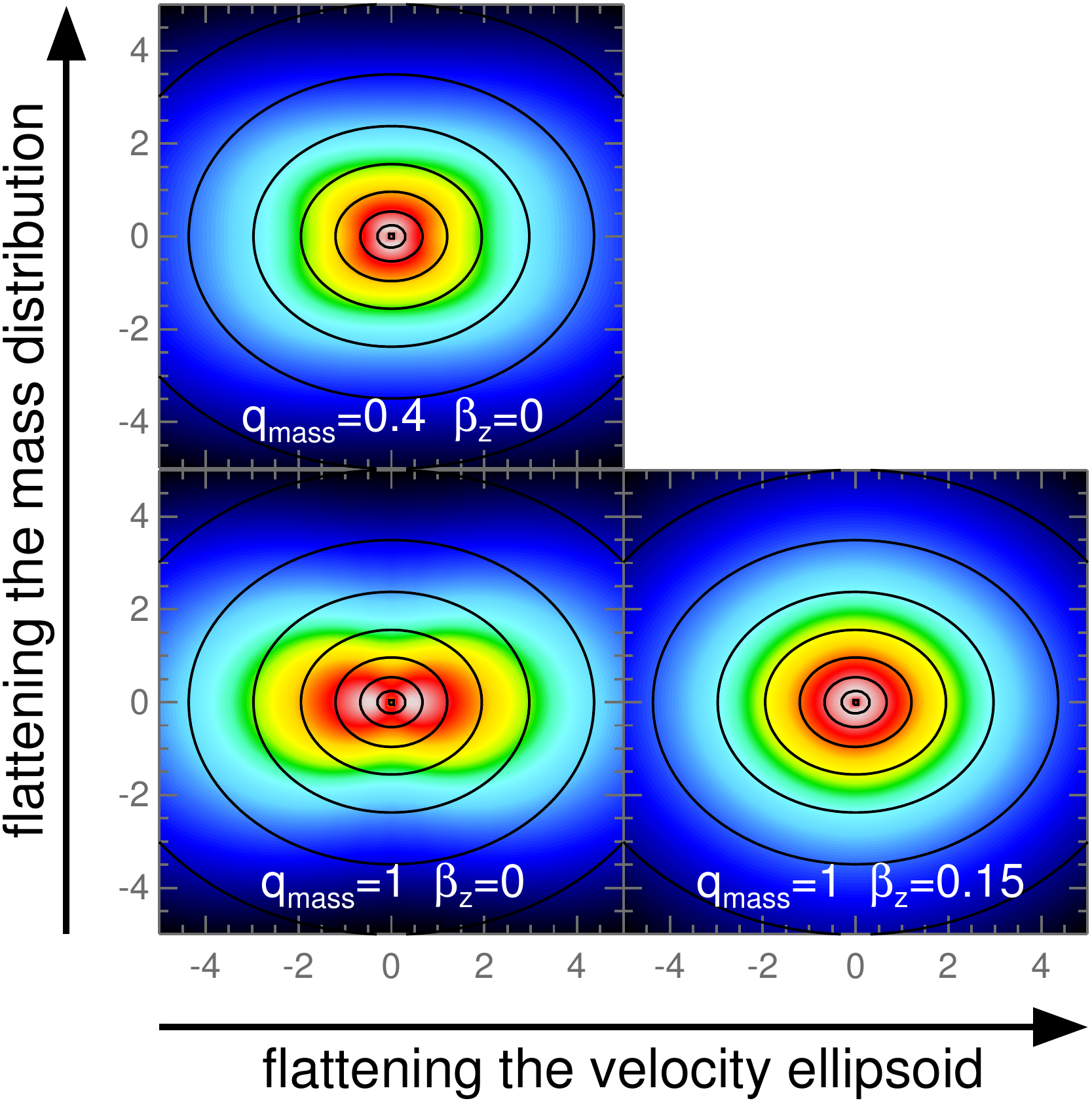}}
\caption{JAM models by \citet{Cappellari2008} of \citet{Hernquist1990} profiles for
  both the light and mass distribution, where for simplicity the stars
  have been treated as a massless tracers in the given
  potential. Plotted is the true second moment
  $V_{RMS}=<v^2_{los}>^{1/2} = \sqrt{V^2+\sigma^2}$ of the models as
  given in Eq.~(28) of \citet{Cappellari2008}.  The spherical Hernquist mass
  distribution was flattened while keeping the surface density profile
  unchanged along circularized isophotes $r=\sqrt{ab}$. By definition
  $\beta_z=1-\sigma_z^2/\sigma_R^2$. In all cases the light
  distribution was kept fixed with $q=0.8$, as shown by the isophotes
  (in 0.5 mag steps). The color scale in the plots is not the same,
  but it is adapted to the ranges of values in each panel. The axis
  are in units of the Hernquist scale radius $a$.
\label{fig:cappellari}}
\end{figure}

\begin{figure}
\centering
\includegraphics[scale=.55]{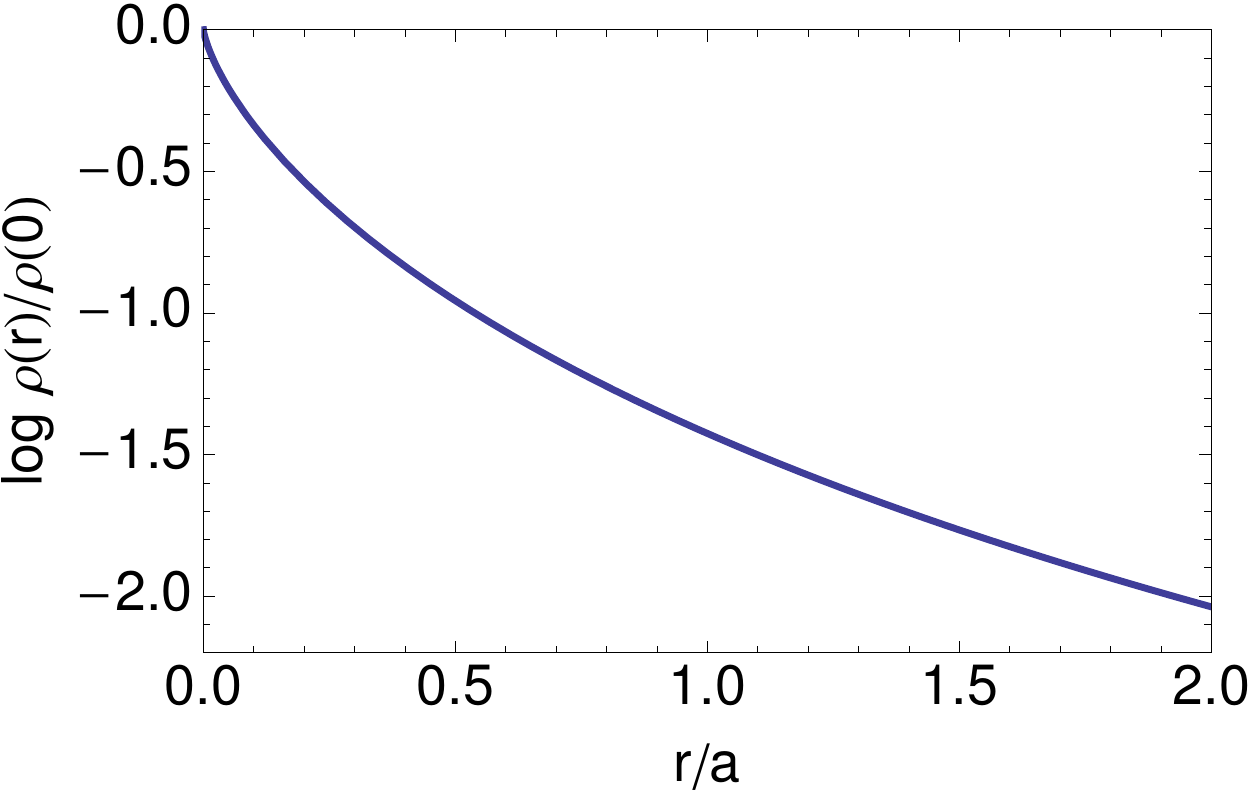}
\includegraphics[scale=.55]{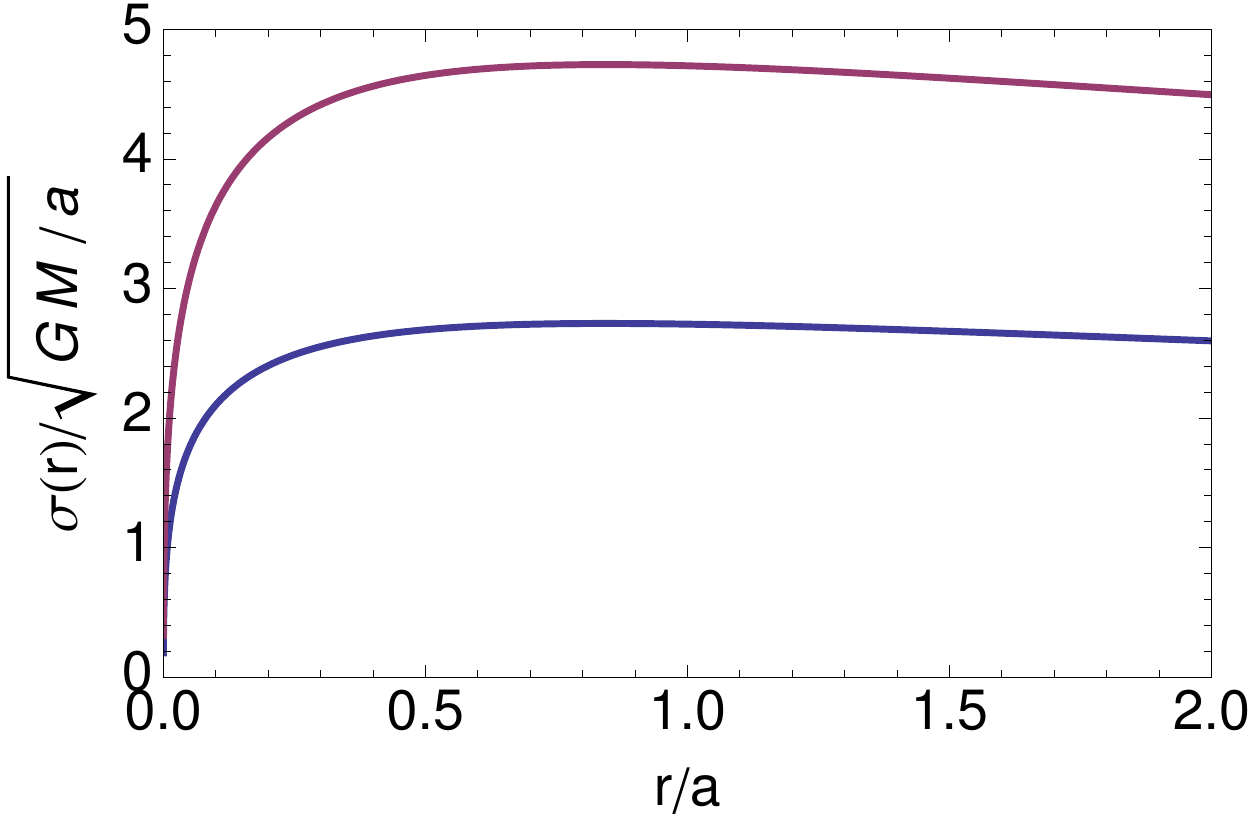}
\includegraphics[scale=.55]{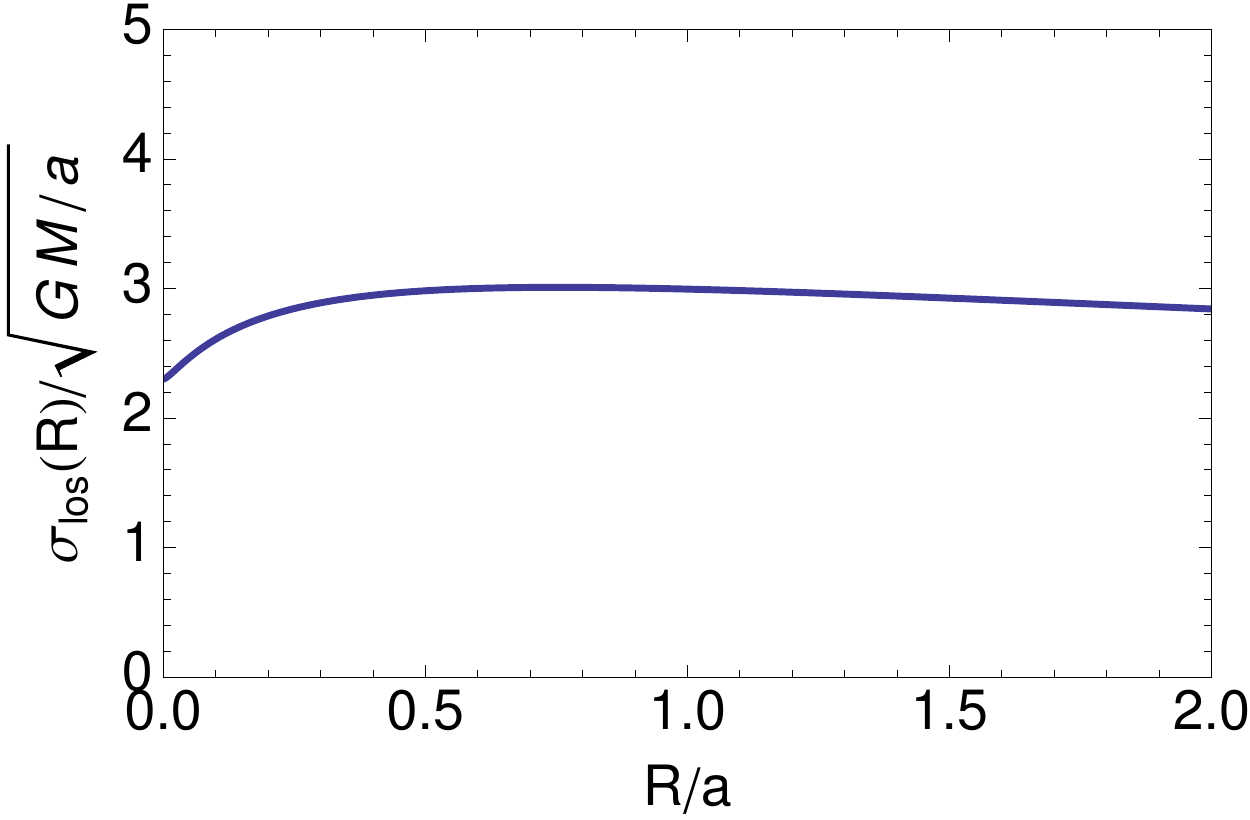}
\caption{Various moments of the distribution function described in
  Eq.~(\ref{eq:df_mb}). The top panel shows the predicted stellar
  density $\nu(r)$ versus distance (in units of the scale $a$). The
  central panel shows the 2nd velocity moments in the radial (blue)
  and tangential (red) directions. In the bottom panel we plot the
  $\sigma_{\rm los}$ profile. Note that this example shows that it is
  possible for the system to have a relatively flat velocity
  dispersion profile, even if $\sigma_r = 0$ at the centre. Moreover,
  a $\gamma_{*,0} = 0$ stellar profile can be embedded in a cuspy NFW
  halo, without having an isotropic velocity dispersion at the centre, since here
  $\beta = -0.5$.}
\label{fig:df_breddels}
\end{figure}

\begin{figure}
\centering
\includegraphics[scale=0.5]{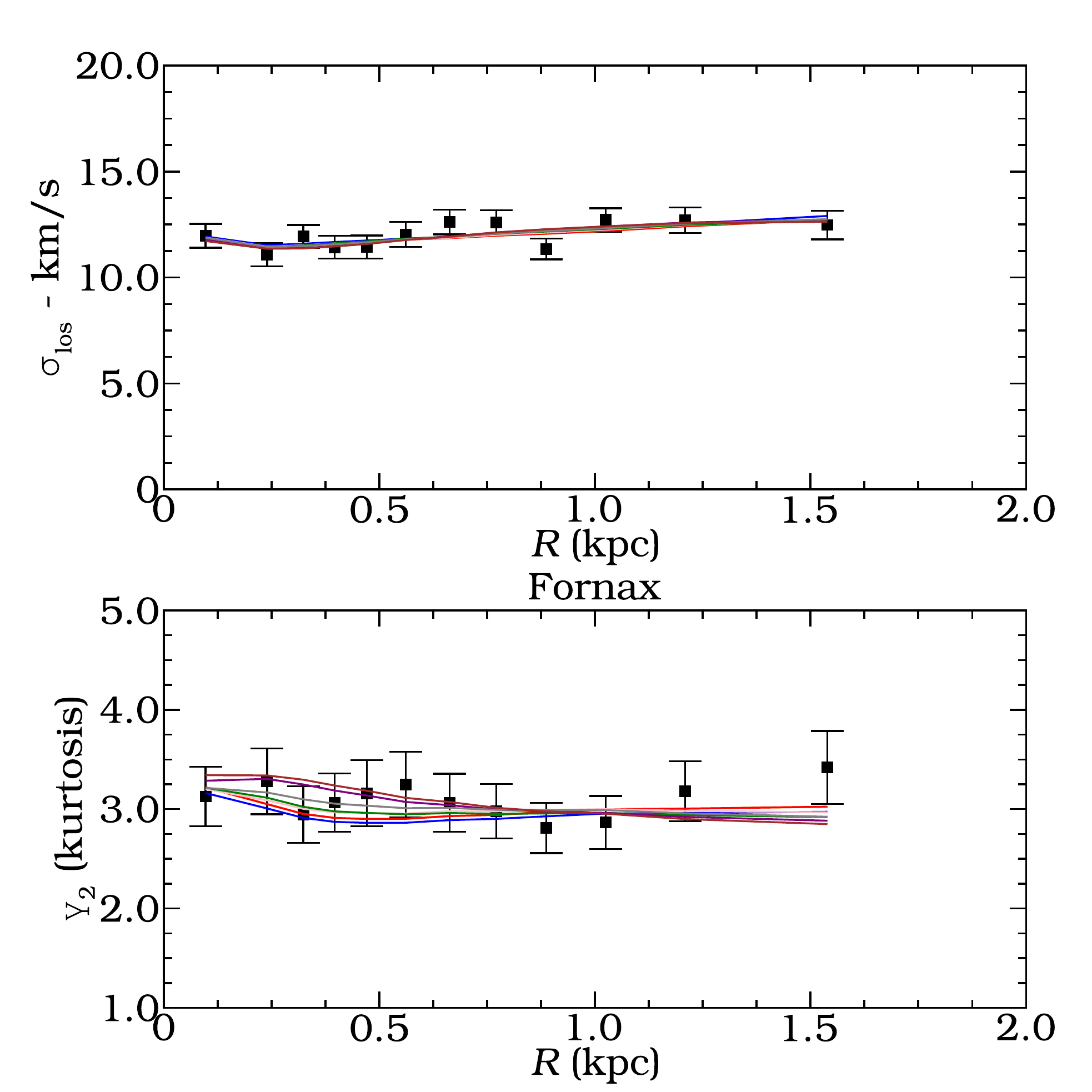}
\includegraphics[scale=0.7]{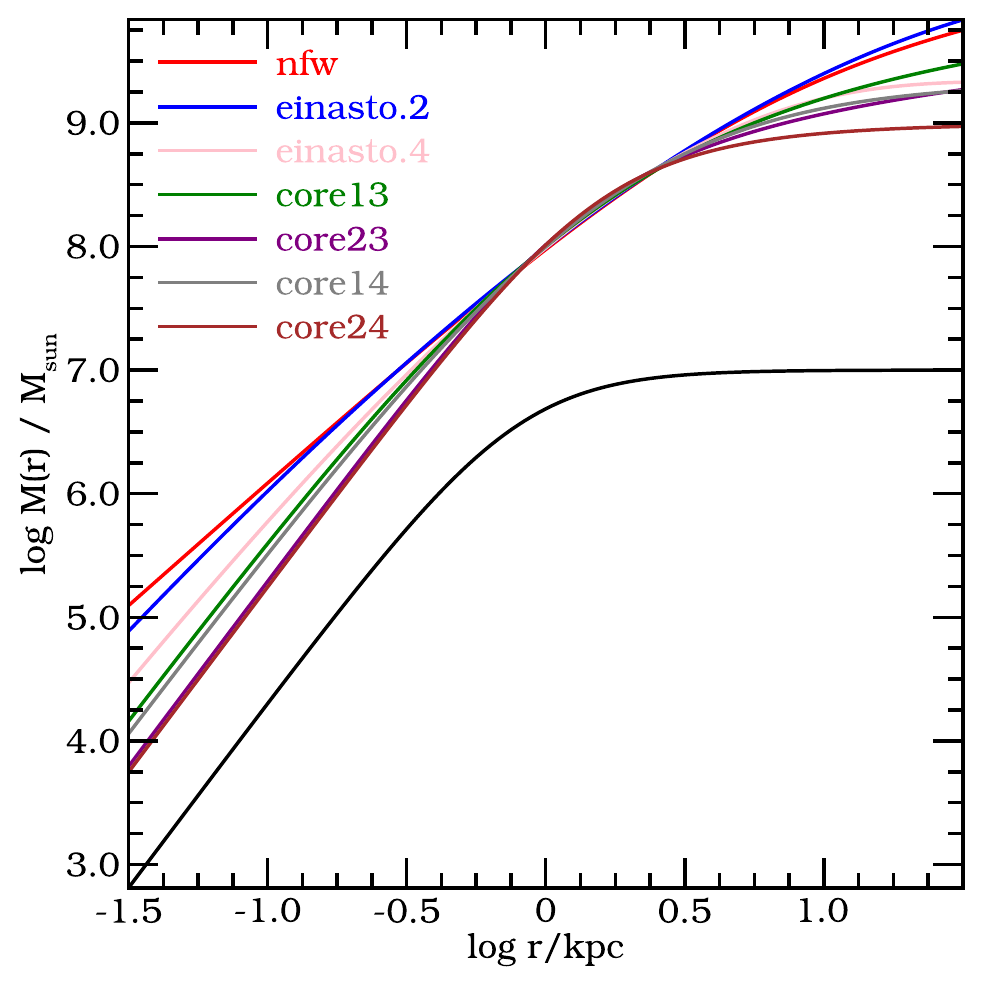}
\caption{The top and middle panels show the velocity dispersion
  profiles and kurtosis for Fornax (squares with error bars). The
  various curves correspond to the best fit models obtained using
  Schwarzschild's method of orbit superposition. The mass distribution
  of these models is shown in the bottom panel, and clearly shows that
  the various models are effectively indistinguishable within a
  distance range that goes from slightly less than $r_{1/2}$ up-to the
  last measured data point. \citep[From][]{Breddels2013}.}
\label{fig:breddels_fnx}
\end{figure}

\begin{figure}
\centering
\includegraphics{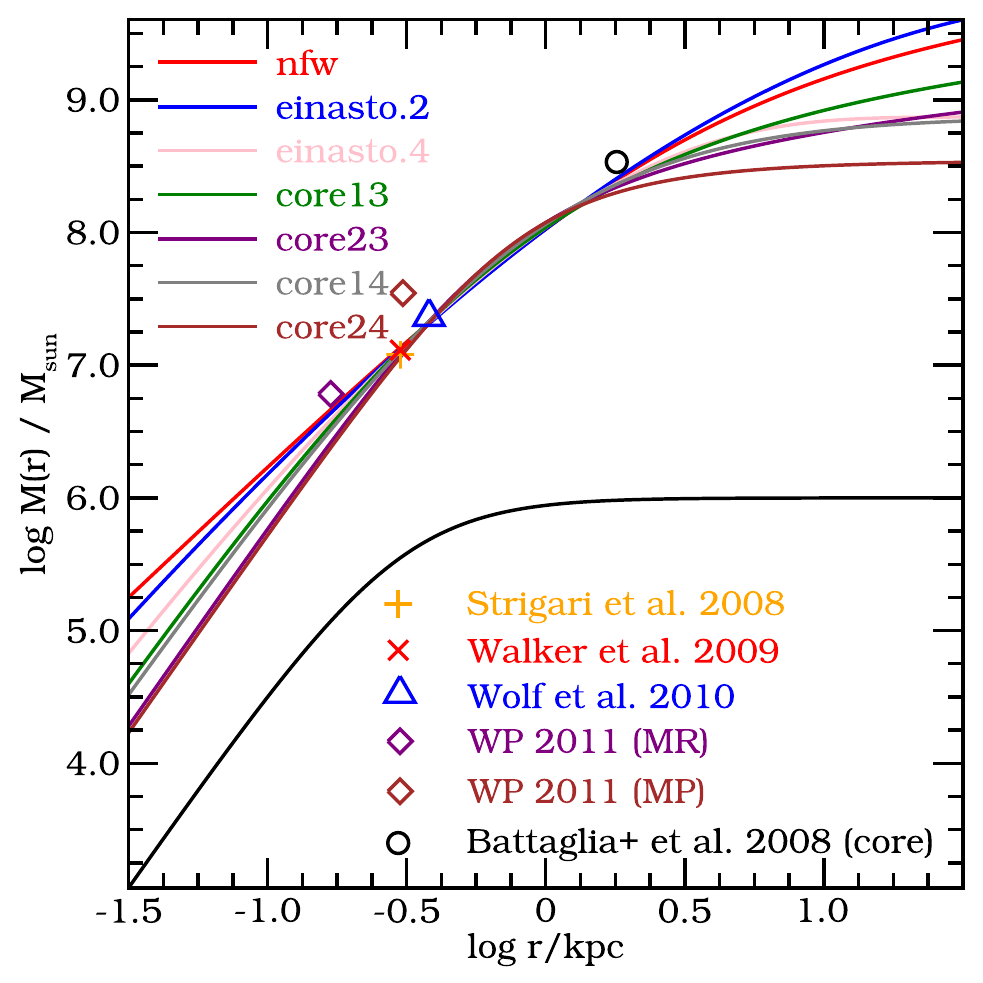}
\caption{Mass distribution $M(r)$ for Scl derived by
  \citet{Breddels2013}. The various curves represent the best fit
  models obtained using Schwarzschild's orbit-based method. As in the
  case of Fnx, the various density profiles appear indistinguishable
  from one another. We have overlaid various measurements for the mass
  at different radii, including those obtained considering separately
  the metal-rich and metal-poor populations by \citet{battaglia2008}, and by estimating 
separately $M_{1/2}$ for the metal-rich (MR) and metal-poor (MP) component in \citet{walker2011}.}
\label{fig:breddels_scl}
\end{figure}

G.~Battaglia gratefully acknowledges support through a Marie-Curie
action Intra European Fellowship, funded from the European Union
Seventh Framework Program (FP7/2007-2013) under grant agreement number
PIEF-GA-2010-274151. A.~Helmi and M.A.~Breddels acknowledge financial support from NOVA (the Netherlands Research School for Astronomy), and 
the European Research Council under ERC-StG grant GALACTICA-24027. We are also grateful to numerous enjoyable interactions throughout the past few years with members of the GALACTICA group: H.~Tian, T.~Starkenburg, H.~Buist, G.~Monari, C.~Vera-Ciro, E.~Starkenburg, T.~Antoja, R.~Sanderson, S.~Jin and K.~Westfall. Finally, we would like to thank M.~Cappellari for his generosity in making Fig.~\ref{fig:cappellari}. 
We acknowledge discussions with S.~Koposov, E.~Lokas, J.~Pe\~{n}arrubia, and M.~Walker.

\begin{sidewaystable}
\begin{tabular}{c|c|c|c|c|c|c}
\hline
\hline
\multirow{2}{*}{dwarf galaxy} & \multicolumn{2}{|c|}{Jeans}  & \multicolumn{2}{|c|}{Schwarzschild} & \multirow{2}{*}{DF} & \multirow{2}{*}{M2M} \\
\cline{2-5}
 & Spherical & Axis. & Spherical & Axis. & \\
\hline
Carina & \gilmore,\walkerseven,\lokasinter,\walkeruni & \hayashi & \bredfour & &  \\
\hline
Draco  & \lokasdraco,\gilmore,\walkerseven,\walkeruni & \hayashi & \jardeldraco & & \kleynaone,\kleynatwo & \longdraco \\
\hline
Fornax & \lokasfnx,\walkerseven,\lokasinter,\walkeruni & \hayashi & \bredfour & \jardfnx$^\dagger$ &  \\
\hline
LeoI & \kochleoone,\gilmore,\walkerseven,\mateoleoone,\walkeruni & \hayashi & & & \\
\hline
LeoII & \kochleotwo,\gilmore,\walkerseven,\walkeruni &  & & & \\
\hline
Sculptor & \walkerseven,\giusscl,\lokasinter,\walkeruni  & \hayashi & \bred,\bredfour & & \amevscl &  \\
\hline
Sextans & \gilmore,\walkerseven,\walkeruni,\lokasinter,\battsxt & \hayashi & \bredfour & & \\
\hline
Ursa Minor & \gilmore,\walkeruni &  & & & \\
\hline
\end{tabular}
\caption{Overview of various modeling techniques applied to Local Group dSph galaxies. {\bf \amevscl:} \citet{Amorisco2012},
{\bf \giusscl:} \citet{battaglia2008}, {\bf \battsxt:} \citet{battaglia2011},
{\bf \bred:} \citet{Breddels2012}, {\bf \bredfour:} \citet{Breddels2013},
{\bf \gilmore:} \citet{gilmore2007}, {\bf \hayashi:} \citet{Hayashi2012},
{\bf \jardfnx:} \citet{Jardel2012}, {\bf \jardeldraco:} \citet{Jardel2013},
{\bf \kleynaone:} \citet{Kleyna2001}, {\bf \kleynatwo:} \citet{Kleyna2002draco},
{\bf \kochleotwo:} \citet{Koch2007}, {\bf \kochleoone:} \citet{koch2007leo1},
{\bf \lokasfnx:} \citet{Lokas2001}, {\bf \lokasfnx:} \citet{lokas2005}, {\bf \lokasinter:} \citet{lokas2009},
{\bf \longdraco:} \citet{Long2010}, {\bf \mateoleoone:} \citet{mateo2008}, {\bf \walkerseven:} \citet{Walker2007},
{\bf \walkeruni:} \citet{walker2009uni}, $^\dagger$ Note that their dark matter halo is still spherical}\label{tab:models}
\end{sidewaystable}

\pagebreak

\pagebreak

\pagebreak

\pagebreak

\pagebreak

\pagebreak

\newpage
\pagebreak

\pagebreak

\pagebreak

\pagebreak

\pagebreak

\vspace*{2cm}

\noindent

\bibliographystyle{elsarticle-harv}
\bibliography{review_all_3b,refs-ah.v3c}

\end{document}